\documentclass[a4paper,onecolumn,11pt,accepted=2024-08-03]{quantumarticle}
\pdfoutput=1
\usepackage[utf8]{inputenc}
\usepackage[margin=1in]{geometry}
\usepackage[dvipsnames]{xcolor}
\usepackage{amsmath}
\usepackage{physics}
\usepackage{graphicx}
\usepackage{amssymb}
\usepackage{enumitem}
\usepackage{mathtools}
\usepackage{authblk}
\usepackage{siunitx}
\usepackage{hyperref}
\hypersetup{colorlinks=true, citecolor=red, urlcolor=blue, linkcolor=blue}
\usepackage{caption}

\newcommand{\tfd}{\text{TFD}}

\newcommand{\magn}[1]{\left|#1\right|}

\newcommand{\id}{\mathbb{I}}
\newcommand{\bracket}[2]{\left\langle#1|#2\right\rangle}
\newcommand{\adj}[1]{#1^{\dagger}}
\usepackage{datetime}
\newdate{date}{09}{08}{2024}

\title{Approximate Quantum Codes From Long Wormholes}
\author{Gregory Bentsen, Phuc Nguyen, Brian Swingle}
\affil{Department of Physics, Brandeis University, Waltham, MA}
\date{\displaydate{date}}

\begin{document}

\maketitle

\begin{abstract}
    We discuss families of approximate quantum error correcting codes which arise as the nearly-degenerate ground states of certain quantum many-body Hamiltonians composed of non-commuting terms. For exact codes, the conditions for error correction can be formulated in terms of the vanishing of a two-sided mutual information in a low-temperature thermofield double state. We consider a notion of distance for approximate codes obtained by demanding that this mutual information instead be small, and we evaluate this mutual information for the SYK model and for a family of low-rank SYK models. After an extrapolation to nearly zero temperature, we find that both kinds of models produce fermionic codes with constant rate as the number, $N$, of fermions goes to infinity. For SYK, the distance scales as $N^{1/2}$, and for low-rank SYK, the distance can be arbitrarily close to linear scaling, e.g. $N^{.99}$, while maintaining a constant rate. We also consider an analog of the no low-energy trivial states property which we dub the no low-energy adiabatically accessible states property and show that these models do have low-energy states that can be prepared adiabatically in a time that does not scale with system size $N$. We discuss a holographic model of these codes in which the large code distance is a consequence of the emergence of a long wormhole geometry in a simple model of quantum gravity.

\end{abstract}

\tableofcontents

\section{Introduction}
\label{sec:intro}

Quantum error correction is a foundational ingredient in fault tolerant quantum information processing. Stabilizer codes \cite{gottesman1997stabilizer,aaronson2004improved} are an important and widely studied subclass of quantum error correcting codes with the property that they can be defined as the exact ground space of a ``code Hamiltonian'' built from commuting terms. Studying families of such codes, indexed by the number of physical qubits, has been very fruitful across physics and quantum information. For example, it has led to connections to topological phases of matter~\cite{Kitaev_2003} and quantum gravity~\cite{almheiri_holo_code,happy_2015}, among many other subjects.

One of most exciting recent developments in quantum error correction is the discovery of asymptotically ``good'' quantum codes of ``low-density parity check'' (LDPC) type \cite{panteleev2022asymptotically,dinur_qldpc_decoder}, with many important related works, e.g.~\cite{leverrier2022quantum,tang_decoder,zemor_decoder}. As the number of physical qubits grows large, an asymptotically good code is one in which the code distance and the number of encoded qubits are both proportional to the number of physical qubits. Similarly, a quantum LDPC code is one in which the code Hamiltonian can be constructed from terms that each act non-trivially on a bounded number of qubits with each qubit participating in a bounded number of terms \cite{gallager1963lowdensity,gottesman2014faulttolerant,breuckmann2021quantum}. As with other stabilizer codes, these good quantum LDPC codes have revealed new physical phenomena, including the existence of robust forms of quantum entanglement \cite{freedman2013quantum,anshu2023nlts,herasymenko2023fermionic}. Hence, it is interesting to understand these good quantum LDPC codes from a physics perspective and to  generalize and refine the existing constructions.

\begin{figure}[h!]
    \centering
    \includegraphics[width=0.5\columnwidth]{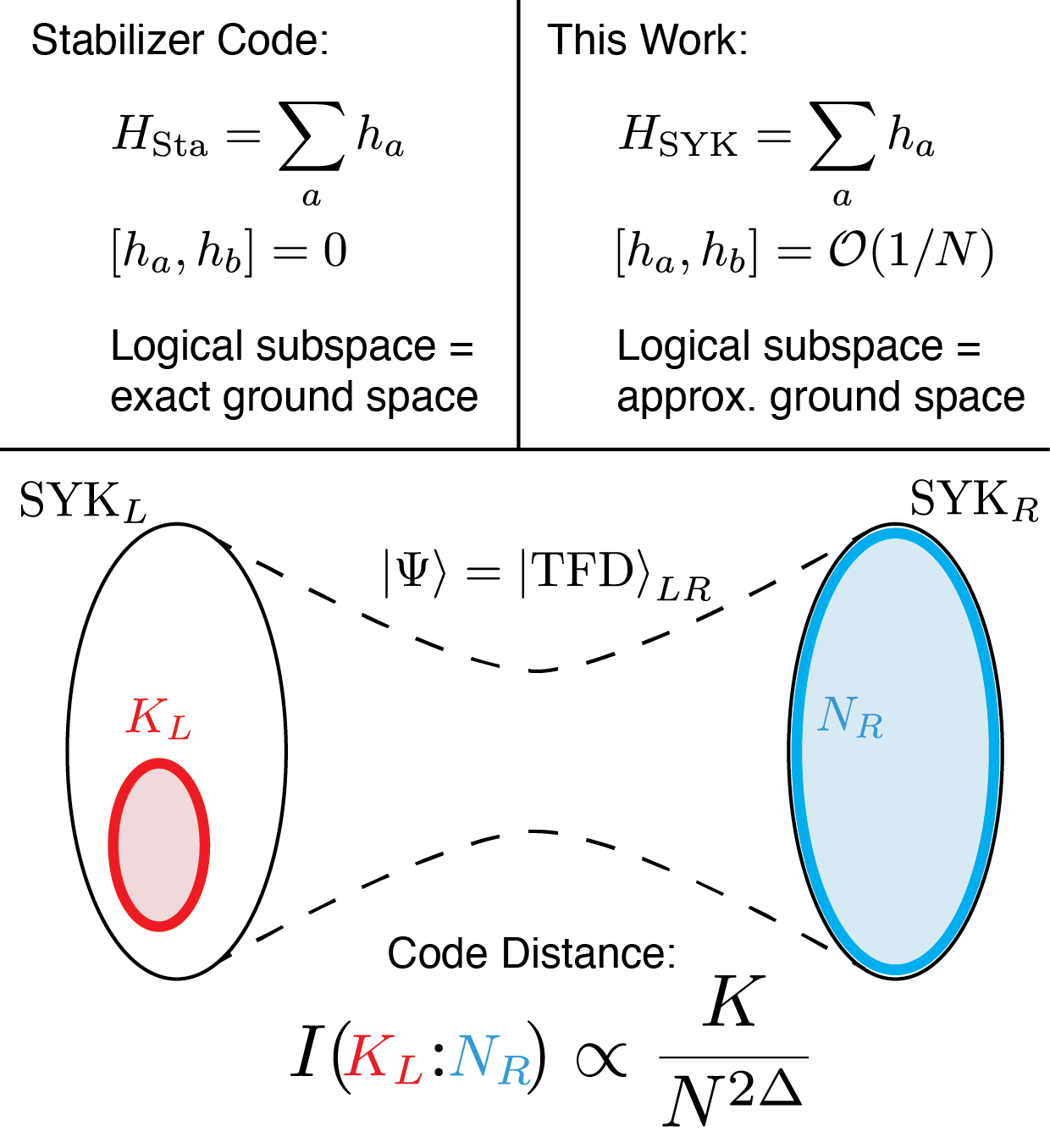}
    \caption{In this work we view the ground space of the SYK model as an approximate quantum error-correcting code. For a stabilizer quantum error-correcting code (top left), the logical subspace can always be viewed as the ground subspace of a Hamiltonian with completely commuting terms. In the SYK model (top right) the terms in the Hamiltonian only approximately commute. Nevertheless, one can still view the ground space of the SYK model as an approximate logical subspace, where the ground-space entropy density $s_0$ measures the rate of the code. To analyze the distance of the code, we prepare two copies of the SYK model in a thermofield double (TFD) state at low temperature $\beta \sim O(N)$ (bottom), and measure the mutual information $I(K_L:N_R)$ between a subset $K$ on the left and the entire right. Vanishing mutual information $I(K_L:N_R) = 0$ indicates that errors acting on $K_L$ cannot perturb the logical information stored in $R$, so the mutual information thereby provides a measure of the code distance. As discussed in the main text, the extrapolation to $\beta \sim O(N)$ does push our methods of analysis to their limits, nevertheless we argue at length in favor of the validity of the extrapolation (see Appendix~\ref{app:extrap} and~\ref{app:lr_qes}).}
    \label{fig:sykcode}
\end{figure}

Towards this end, in this paper we describe a phenomenon of approximate codes arising from the nearly degenerate low-lying states of families of Hamiltonians composed of non-commuting terms as illustrated in Fig. \ref{fig:sykcode}. We consider both the Sachdev-Ye-Kitaev (SYK) model \cite{sachdev1993gapless,kitaev2015simple,maldacena2016remarks,rosenhaus_syk,suh_softmode} and a related family of ``low-rank'' SYK models \cite{kim2020lowrank,kim2021dirac}. These models are built from $N$ Majorana fermions and yield a class of approximate fermionic codes. Fermionic codes have been studied previously, motivated by topological quantum computing and simple models of holography~\cite{kitaev_maj_wires,bravyi_fqc,bravyi2010majorana,vijay_fqecc,jahn_maj_holo_qecc}, but these prior examples are significantly different from the codes we consider. The SYK and low-rank SYK codes turn out to have constant rate and a distance that scales as $N^c$ as $N$ goes to infinity. For SYK, we find $c=1/2$, while for the low-rank models, we find that $c$ can be tuned arbitrarily close to $1$, e.g. $c=.99$, while preserving a non-vanishing rate. Hence, certain of these codes are almost good. They are also roughly analogous to LDPC codes in that the terms in the Hamiltonian are all low-weight; however, they are ``mean-field'' in form, meaning every fermion participates in many interactions. The setup is somewhat analogous to the approximate good LDPC codes considered in~\cite{yuen_approximate_qldpc}, although our Hamiltonians are not frustration free. One can consider sparsified versions of SYK and low-rank SYK which might have similar properties \cite{xu2020sparse}, but we do not study them here. Existing constructions of approximate quantum codes include approximate good LDPC codes~\cite{yuen_approximate_qldpc}, spin chain energy eigenstate codes~\cite{brandao_eth_qecc}, covariant codes~\cite{hayden_qecc_frame,faist_qecc_cont_sym,woods_qecc_clock_frame,zhou_qecc_cov,kong_qecc_sym}, and quasi-exact codes~\cite{wang_quasiexact}. The error correcting properties of some large $N$ models have been studied in \cite{milekhin_largeN_qecc,cheng_largeN_qecc}, and those of SYK specifically have been studied in \cite{syk_code_cl}; our findings accord with theirs where there is overlap.

In our setup, determining the rate of the code maps to a well-understood calculation of ground state entropy which can be carried out in the large $N$ limit using path integral techniques. Determining the distance is more challenging. As illustrated in Fig. \ref{fig:sykcode}, we map the distance calculation to a mutual information calculation between two sides of a low-temperature thermofield double (TFD) state, with one side representing the physical system (left, $L$) and the other side representing a reference which purifies the maximally mixed state of the code (right, $R$). The code distance can then be defined as the largest number of fermions on the left side which have less than one bit (or some small constant $\delta$) of mutual information with the code space represented by the right side. We demonstrate how this mutual information criterion can be rigorously connected to standard notions of approximate error correction in terms of recoverability as measured by the trace norm. These mutual information results are obtained numerically after an extrapolation procedure: we compute the mutual information for a general TFD state as a function of inverse temperature $\beta$ and then argue that the result is reliable even if $\beta$ scales with $N$. The mutual information becomes small in this limit, suppressed by an inverse power of $N$, when considering sufficiently small subsets of the fermions. It is important to note that this extrapolation procedure pushes our methods of analysis to their limits, but we argue extensively in favor of the validity of the extrapolation (see Appendix~\ref{app:extrap} and~\ref{app:lr_qes}).

The smallness of the mutual information and the resulting expression of the code distance is the main result as far as the code properties are concerned. In addition, we describe a physical picture which explains the code properties in terms of an emergent geometry (emergent because the SYK model has no obvious geometry---it features all-to-all interactions). In particular, using the holographic description of the low temperature physics of SYK (reviews include \cite{sarosi2018ads,rosenhaus_syk_review,trunin_syk_review}), the smallness of the mutual information can be interpreted as a manifestation of the weak correlations between two sides of a long wormhole. In a precise sense, the errors and the logical information are separated by a large spatial distance at large $N$, and this separation in turn leads to a large code distance.

Our study of approximate error-correcting codes in the SYK model and 1+1$D$ gravity is motivated by several overlapping interests. The approximate codes we study here are fundamentally different from the exact stabilizer codes that are frequently studied in the literature, and therefore broaden the toolset available and provide new sources of codes with potential experimental relevance. Moreover, our understanding of approximate codes is limited to a handful of specific examples, without a unifying framework. The codes we find here extend that set of examples to the gravitational setting and could aid in developing a more comprehensive understanding of approximate error-correction. Importantly, approximate error-correcting codes have been shown to outperform exact codes in various ways \cite{leung1997approximate,crepeau2005approximate,bergamaschi2022approaching} and could therefore lead to better performance in principle and in practice. 
We are also motivated by the new perspectives and tools that the language of quantum error-correction provide for understanding fundamental issues in many-body physics. For example, the realization \cite{almheiri_holo_code} that holographic states in the AdS/CFT correspondence can be understood as quantum error-correcting codes was a valuable insight and led to progress both in holographic quantum gravity and to the development of new quantum error-correcting codes \cite{happy_2015}. In this paper, we are providing that same connection explicitly for the SYK model and its low-rank generalizations, with one explicit goal being an improved understanding of recently proposed tensor network models of SYK~\cite{nora}. We are also interested in questions of Hamiltonian complexity, such as the NLTS conjecture~\cite{freedman2013quantum}, in the context of mean-field models where they have not been well studied.

The rest of this paper is organized as follows. In Section \ref{sec:tools}, we provide a technical overview of our results and discuss the tools we will need for our analysis, including a formulation of the code distance in terms of a ``two-sided'' mutual information in the zero temperature TFD state.
In Section \ref{sec:syk_mi}, we present our results for the code rate and the mutual information in SYK and low-rank SYK models. In Section \ref{sec:gravity_mi}, we give a holographic interpretation of the code results. Finally, in Section \ref{sec:outlook} we give an outlook and discuss related issues, including the encoding and decoding process and (the absence of) an adiabatic analog of the no low-energy trivial states (NLTS) property~\cite{freedman2013quantum,eldar2017local,anshu2023nlts}. A number of appendices contain technical details.

\section{Tools and technical overview}
\label{sec:tools}

The Sachdev-Ye-Kitaev (SYK) model is a quantum many-body model of $N$ strongly-interacting fermions, which we review in more detail below. The purpose of this paper is to discuss an interpretation of the SYK ground space as an approximate quantum error correcting code. Here we give a high level overview of the parameters of this code and how they relate to the usual specification of codes.

One typically labels exact codes by three numbers, $n$, the number of physical qubits, $k$, the number of logical qubits, and $d$, the distance, which is a measure of the minimum number of operations needed to move from one code state to another orthogonal one. The rate of the code is $k/n$.

For the approximate codes we consider in this paper, the role of $n$ is played by $N$, the number of Majorana fermions in the SYK model. For the rate and distance, we argue that the SYK ground space forms a fermionic quantum code $C_{\text{SYK}}$ with 
\begin{itemize}
    \item rate $ = s_0$ and 
    \item distance $\sim  \delta N^{1/2}$,
\end{itemize}
where $s_0$ is an $O(1)$ constant measuring the ground-space entropy density, such that $\Omega \equiv \dim C_{\text{SYK}} \propto e^{s_0 N}$. 
Here the distance is defined as the largest number of fermions which have at most $\delta$ mutual information with the code space, as discussed in detail in Section~\ref{sec:syk_mi}. We refer to this code as the SYK ground space code or just the SYK code. The SYK code is loosely analogous to a low-density parity check code (LDPC), in that it is specified by many low-weight terms in a Hamiltonian. However, we emphasize that the SYK model is ``dense'' in that every fermion participates in a large number of interaction terms.

We also extend these results to a class of ``low-rank'' SYK models which generalize the usual SYK model and are indexed by a continuous parameter $\gamma \in (0,\infty)$, with the SYK case recovered as $\gamma \rightarrow \infty$. This class of models also has an approximately degenerate ground space of size $e^{s_0(\gamma) N}$, and we argue that this space forms a code with 
\begin{itemize}
    \item rate $= s_0(\gamma)$ and
    \item distance $\sim \delta N^{2 \Delta(\gamma)}$,
\end{itemize}
where, for small $\gamma$, $s_0(\gamma) \propto c_1 \gamma + O(\gamma^2) $ and $ 2 \Delta \propto 1 - c_2 \gamma + O(\gamma^2)$. Hence, this family of codes achieves constant rate and nearly constant relative distance for small fixed $\gamma$. This class of ground space codes is also loosely analogous to LDPC codes, in the same sense as for the SYK model (and with the same caveat about the model being ``dense'').

We emphasize that these codes are not stabilizer codes. They are approximate codes which are loosely analogous to stabilizer codes in that they can be specified as the approximate ground space of some Hamiltonian. In the stabilizer case, the relevant Hamiltonian is a commuting projector Hamiltonian built from the stabilizer group; in our case, the Hamiltonian consists of many non-commuting terms. We also emphasize that these codes lack two properties that one typically expects from stabilizer codes: the ground code space is not separated from the excited states by a gap; and the ground state is not frustration free. Nevertheless, we demonstrate that these codes are still well-defined in the sense of being defined by a projector onto the code subspace, which we define as the collection of the $\Omega$ lowest-lying energy eigenstates, and are approximately recoverable as measured by the trace norm \cite{schumacher2002approximate,hayden2020approximate,yi2023complexity}. It is also not clear if these codes are useful for practical information processing. They may be more akin to random codes, e.g.~\cite{brown_random_code} , but with a relatively succinct definition via the SYK or low-rank SYK Hamiltonians.

\subsection{Ground space codes}

In this paper, our approach to defining a code is to consider it as the exact or approximate ground space of some Hamiltonian $H$. The number of encoded qubits is then equal to the ground state entropy, and the corresponding rate is the ground state entropy density (entropy per qubit or per particle). For Hamiltonians with an exactly degenerate ground space and a gap to excited states, the maximally mixed state on the code is the zero-temperature Gibbs state, and the purification of this state is the zero-temperature thermofield double. Later in the paper we will relax these conditions and consider instead low-temperature Gibbs ensembles and low-temperature thermofield double states with $J \beta \rightarrow N$ (see discussion in subsection 2.2), but for the moment let us first discuss the exactly-degenerate, zero-temperature case to build intuition.

Let us begin with stabilizer codes, which can always be formulated as a ground space code. A stabilizer code on qubits is a subspace $C$ specified by a stabilizer group $S$. This group is generated by commuting generators $G_a$ where each $G_a$ is a Pauli string that squares to the identity, $G_a^2=1$. The code is defined as the common $+1$ eigenspace of all the $G_a$. Note that the choice of generators is not unique and the code does not depend on the choice.

Given a set of generators of the stabilizer group, the code Hamiltonian is defined as 
\begin{equation}
    H_C = \sum_a \frac{I - G_a}{2}.
\end{equation}
Each term in $H_C$ commutes with all others, and the ground space of $H_C$ is identical to $C$. The excited states depend on the precise choice of generating set, but the ground space does not. A code family is LDPC if, as the number of physical qubits goes to infinity, (1) the set $\{G_a\}$ can be chosen to have bounded Pauli weight and (2) no qubit participates in more than a bounded number of the $G_a$.\footnote{The rough analog of the LDPC property (1) in SYK is that each term in \eqref{eq:syk} consists of $q$ fermion operators. However, LDPC property (2) is violated by SYK.}

Let $\Pi_C$ be the projector onto the code space. Clearly we have $H_C \Pi_C = 0$. The zero temperature Gibbs state is 
\begin{equation}
   \rho_C =  \lim_{\beta \to \infty} \frac{e^{- \beta H_C}}{\mathrm{tr} e^{-\beta H_C}} = \frac{\Pi_C}{\mathrm{tr} \Pi_C}.
\end{equation}
We can also introduce a reference which purifies $\rho_C$, and it will be convenient to take this reference to be another copy of the code. To synergize with the wormhole language used below, we denote the original copy by $L$ (left) and the purifying copy by $R$ (right). 

For a Hamiltonian with energy levels $E_n$, the thermofield double state (TFD) purifying the canonical thermal state at inverse temperature $\beta$ is
\begin{equation}
|\text{TFD},\beta\rangle = \sum_n \sqrt{\frac{e^{-\beta E_n}}{Z}} |E_n \rangle_L |E_n \rangle_R,
\end{equation}
where $Z$ is the thermal partition function. Now, for the code Hamiltonian above, this TFD state at general $\beta$ depends on the precise choice of stabilizer generators since it depends on the full spectrum of the Hamiltonian.

However, in the limit $\beta \to \infty$ of interest here, the TFD turns out to depend just on the code itself, i.e. the ground space. Given a basis $|\phi_i \rangle$ of $C$, the purified zero temperature Gibbs state is
\begin{equation}
    |\text{TFD},\beta\to \infty\rangle = |\Phi \rangle = \frac{1}{\sqrt{\dim C}} \sum_i |\phi_i \rangle_L \otimes |\phi_i \rangle_R.
\end{equation}
We refer to this state as a zero-temperature TFD.

\subsection{Approximate ground space codes}

Since the ground space degeneracy in SYK is not exact\footnote{It turns out the degeneracy becomes exact in some supersymmetric SYK models, which leads to somewhat different physics as we discuss in Section~\ref{sec:outlook}.}, we must specify the code space more carefully. In the many-body fermion models we study, we define the code subspace as the span of the $\Omega$ lowest energy eigenstates of the Hamiltonian. We emphasize that this definition yields an ensemble of quantum codes, one for each realization of the random couplings. The maximally mixed state on the code is then purified to a microcanonical TFD state in which the amplitudes are equal within an energy window and zero outside of the window.

However, for the purposes of computation, it is more convenient to use a canonical TFD state at a specially chosen value of $\beta=1/T$. Because the mutual information diagnostic we discuss below depends on correlations that change slowly with respect to temperature, i.e. are intensive variables at large $N$, we should obtain approximately the same result either way.

Quantitatively, the probability to find the system in some energy eigenstate with energy $E$ (not the probability of a single eigenstate) in the canonical ensemble is
\begin{equation}
    p(E) \propto e^{- E/T + S(E)}.
\end{equation}
We want to compare this to the microcanonical probabilities in the case where $1/T = a N$ for some constant $a$. Writing $E = E_0 + \Delta E$ and using the formula $S(E) = s_0 N + b \sqrt{N (E-E_0)} $, which is accurate at the not-exponentially-fine energy scale of interest \cite{maldacena2016remarks}, we find
\begin{equation}
    p(\Delta E) \propto e^{- a N \Delta E + b \sqrt{ N \Delta E}}.
\end{equation}
This is peaked around $\Delta E_{\text{peak}} = \frac{b^2}{4 a^2 N} \propto \frac{1}{N}$, so it is very unlikely for the energy to exceed $E_0$ by even a constant amount at large $N$. However, the value of $p(E)$ at $\Delta E=0$ is only suppressed relative to the peak value by a finite amount at large $N$. Hence, all energies less than $E_0 + \Delta E_{\text{peak}}$ have similar probabilities at large $N$. In other words, the distribution $p(E)$ is really only peaked in the positive $\Delta E- \Delta E_{\text{peak}}$ direction.

We can use the fact that $p(\Delta E)$ is peaked to argue that the microcanonical correlation functions are well-approximated by the canonical correlation functions. As a warmup, we review the situation for energies that sit above the ground state by a low but extensive amount. Let the left-right correlator in the microcanonical TFD be $G_{\text{mc}}(E)$. This translates into a correlation in the canonical TFD of order 
\begin{equation}
    \int dE p(E) G_{\text{mc}}(\Delta E/N) \sim G_{\text{mc}}(\Delta E_{\text{peak}}/N) + O(G_{\text{mc}}''(\Delta E_{\text{peak}}/N)/N) 
\end{equation}
where the first derivative term vanishes since $p(\Delta E)$ is approximately Gaussian around the peak. Formally, this is shown by Taylor expanding $G_{\text{mc}}$ around $\Delta E_{\text{peak}}$ and using $\int dE p(\Delta E) (\Delta E - \Delta E_{\text{peak}}) = 0 $ and $\int dE p(\Delta E) (\Delta E - \Delta E_{\text{peak}})^2 \sim O(N)$. Thus, we can obtain the microcanonical correlations from the canonical correlations via the correspondence between $\Delta E_{\text{peak}}$ and $\beta$.

In the case of smaller $\Delta E$, the analysis is complicated by the fact that $p(E)$ is now only peaked in the positive $\Delta E - \Delta E_{\text{peak}}$ direction, as discussed just above. This means we cannot set to zero the first derivative term since the $p(E)$ distribution is significantly asymmetric. Nevertheless, because $G_{\text{mc}}$ is still slowly varying with respect to $\Delta E$, it will be the case that the canonical correlators still track the microcanonical correlators up to multiplicative corrections of order $O(1/\sqrt{N})$.

Before proceeding, we wish to emphasize that the SYK code depends on the choice of $\Omega$. In this paper, we are going to focus on the choice $\Omega \sim e^{s_0 N}$ in which we include the entirety of the approximate ground space. This choice corresponds to $\beta \sim N$ as can be seen from the above analysis of $\Delta E_{\text{peak}}$. This case is naturally interesting for a variety of reasons highlighted throughout the text, but it is also the current limit of our ability to control the calculation of the distance. It is plausible that significantly smaller values of $\Omega$ would lead to improvements in the distance, but we cannot reliably calculate in this regime. Nevertheless, we offer some speculations on this point in Section~\ref{sec:outlook}.

\subsection{Mutual information diagnostic}

For such ground space codes, the distance can also be formulated in terms of the TFD state. The idea is to consider subregions of the left and study their correlation with the entire right. When the correlation vanishes, then the left subregion has no information about the code. We first discuss the connection between the mutual information and the Knill-Laflamme (KL) error correction conditions \cite{kl} in the exact case before discussing the generalization to approximate error correction.

Starting from the Knill-Laflamme (KL) error correction conditions~\cite{kl}, an exact error-correcting code with distance $d$ has
\begin{equation}
    \langle \phi_i | E_a^\dagger E_b | \phi_j \rangle = c_{ab} \delta_{ij}
\end{equation}
for all errors (Pauli strings) $E_a, E_b$ of weight less than or equal to $(d-1)/2$. Intuitively, this means that one needs a Pauli string of length at least $d$ in order to have a non-trivial matrix element between code states. As we show in Appendix~\ref{app:wormhole-kl}, the KL conditions can be reformulated in terms of a left-right mutual information in the zero temperature TFD. A code having distance $d$ is equivalent to the statement that the mutual information between any subregion $A$ of the left of size less than $d$, $|A_L| < d$, and the entire right is zero,
\begin{equation}
    |A_L | < d \to I(A_L : R) = 0.
\end{equation}

Thus, we can determine the distance of the code by constructing the TFD state and studying the left-right mutual information as a function of region $A$. The distance is the size of the smallest region $A$ on the left such that $I(A_L:R)\neq 0$. This is an exact notion which is well suited to exact codes.

Below we will consider approximate codes \cite{leung1997approximate,crepeau2005approximate,Flammia2017limitsstorageof} and define the distance as the size of the largest subregion (subset of fermions) on the left which has mutual information at most $\delta$ with the right, where $\delta \ll 1$ is some small fixed constant. Depending on the context, we interchangeably use $A$, $K$, or $K_L$ to denote this set of $K$ fermions on the left. We show in Appendix~\ref{app:approxrecovery} that the smallness of the mutual information is rigorously related to standard notions of approximate error correction as measured by the trace norm. In particular, we demonstrate that $I(A_L:R) < \delta$ implies that there exists a decoding map $\mathcal{D}$ acting on $L$ such that
\begin{equation}
    || \left(\mathbb{I} \otimes \mathcal{D} \circ \mathcal{N}\right) \psi_{LR} - \psi_{LR} ||_1 < c' \delta^{1/4}
    \label{eq:appcodedefmain}
\end{equation}
where $\mathcal{N}$ is an arbitrary noise channel acting on the subset $A_L$, $\psi_{LR}$ is the TFD code state, and $c'$ is an $O(1)$ constant. Thus our mutual information criterion has a precise operational meaning in terms of the recoverability of the code state.

In terms of the code state, it is important to highlight a difference relative to the stabilizer code case. For SYK and low-rank SYK, we cannot send $\beta \to \infty$ at fixed $N$ since the ground space is not exactly degenerate. Instead, we must choose $\beta$ to scale with $N$, the number of fermions, in order to obtain a sensible analog of the zero-temperature TFD as discussed in detail in the previous subsection. At a technical level, our calculations are fully controlled for any finite $\beta$ in the $N\to \infty$ limit. However, to obtain the distance results quoted in the introduction, we rely on an extrapolation to $\beta$'s that scale with $N$. In particular, we will consider taking $\beta = N^a$ for some $a$ with $a \rightarrow 1$. The reason for this is that we are motivated to find codes with constant rate and extensive distance, if they exist. Specifically, one of our primary results is that the mutual information scales like
\begin{equation}
    I(A_L:R) \propto \frac{K}{\beta^{2\Delta}}
\end{equation}
where the scaling exponent is $\Delta = 1/4$ for the $q = 4$ SYK model and $\Delta \rightarrow 1/2$ for the low-rank SYK model. We obtain a $\delta$-recoverable code when $I(A_L:R) < \delta$ or, equivalently, when $K < \delta \beta^{2\Delta}$. Thus, if we are allowed to extrapolate $\beta = N^a$ for $a \rightarrow 1$, we therefore expect to find a nearly extensive code distance $K \propto N$ for the low-rank SYK model. This is one of our primary motivations for considering the extrapolation to $\beta \rightarrow N$. 
We discuss and justify this extrapolation in Appendix~\ref{app:extrap}.

As an aside, it is worth noting that the left-right mutual information in the thermofield double state can be exactly recast as a ``one-sided'' quantity. The mutual information is
\begin{equation}
    I(K_L:R) = S(K_L) + S(R) - S(K_L \cup R)
\end{equation}
evaluated in the $LR$ TFD state. Each term can be equivalently obtained from the corresponding Gibbs state $\hat{\rho}$ on, say, $L$. Specifically, $S(K_L)$ is the entropy of $K$ fermions in the $N$ fermion state $\hat{\rho}$, $S(R)$ is the full entropy of $\hat{\rho}$, and $S(K_L \cup R)$ is the entropy of $N-K$ fermions in $\hat{\rho}$. The first and second statements are immediate while the third follows from the purity of the TFD. We make use of both formulations as convenient.

\subsection{Models}
\label{sec:models}

With the notion of ground space code and our mutual information diagnostic defined, we now turn to the models we will analyze. We describe first the SYK model and then the low-rank generalizations.

The Sachdev-Ye-Kitaev (SYK) model refers to an ensemble of Hamiltonians defined by an even number $N$ of Majorana fermions and an even integer $q$. The fermions obey the standard algebra
\begin{equation}
    \{ \chi_a,\chi_b \} = \delta_{ab},
\end{equation}
which implies that $\chi_a^2 = 1/2$. An instance of the Hamiltonian is written as
\begin{equation}
    H = i^{q/2} \sum_{a_1 \cdots a_q} J_{a_1 \cdots a_q} \chi_{a_1} \cdots \chi_{a_q}.
    \label{eq:syk}
\end{equation}
We use a multi-index $A$ to refer to the $q$-tuple $a_1 \cdots a_q$. The coefficients $\{J_A\}$ are Gaussian random variables with mean zero and covariance
\begin{equation}
    \mathbb{E}(J_A J_B) = \delta_{AB} \frac{J^2(q-1)!}{N^{q-1}}.
\end{equation}
An important symmetry shared by all instances of the model is fermion parity, given by
\begin{equation}
    (-1)^F = (2i)^{N/2} \chi_1 \cdots \chi_N.
\end{equation}

This model is interesting for a variety of reasons. Of particular relevance for this work is the model's low temperature properties. For $q>2$, the ground space of the model consists of approximately $e^{s_0(q) N}$ approximately degenerate states. The number is approximate in the sense that, while the logarithm of the dimension of the ground space per particle is equal to $s_0$ up to $1/N$ corrections, the precise number of states is not exactly $e^{s_0 N}$ (which is not an integer in general). Similarly, these states are approximately degenerate in that they are split in energy by an amount of order $e^{-s_0 N}$ at finite $N$.

At very low temperature but still far above the energy splitting in the ground space, the model exhibits an emergent nearly conformal dynamics. The fermions have a definite scaling exponent $\Delta$ under the scaling transformation which is a part of this conformal symmetry, and this exponent plays a crucial role in our analysis. The nearly conformal dynamics turns out to be related to a simple model of quantum gravity known as Jackiw-Teitelboim (JT) gravity in $1+1$d Anti-de-Sitter space (AdS). Among other features, this gravitational description controls the leading-in-$N$ thermal entropy and energy, and is responsible for many of the special dynamical properties of SYK at low energy (reviews include \cite{sarosi2018ads,rosenhaus_syk_review,trunin_syk_review}).

In the SYK model the integer $q$ affords some control over the scaling exponent $\Delta = 1/q$ simply by changing the degree of interactions.
A closely related family of models, called low-rank SYK models, gives us access to a continuous spectrum of scaling exponents $\Delta$ interpolating between $\Delta = 1/2$ and $\Delta = 1/4$ \cite{kim2020lowrank,kim2021dirac}. The low-rank SYK model has the same form as a $q = 4$ SYK model, but the couplings $J_{ijkl}$ now form a $N^2 \times N^2$ matrix of decreased rank $R = \gamma N$:
\begin{equation}
    J_{ijkl} = \frac{1}{2} \sum_{n=1}^R \lambda_n u_{ij}^{(n)} u_{kl}^{(n)}
\end{equation}
with eigenvalues $\lambda_n$, where $u_{ij}^{(n)}$ are independent Gaussian random variables with zero mean and variance $\mathbb{E} [ u_{ij}^{(n)} u_{kl}^{(m)}] = \delta_{ik} \delta_{jl} \delta_{nm} / N^2$. We assume the eigenvalues form a Class III distribution, according to the classification of Ref. \cite{kim2020lowrank}, meaning that the largest eigenvalue $\lambda \geq \lambda_n$ is supported on a delta-function peak in the eigenvalue distribution:
\begin{equation}
    \rho(\lambda_n) = c_0 \delta(\lambda_n - \lambda) + \ldots
\end{equation}
The rank is controlled by the parameter $\gamma = R/N$, and allows us to tune between an essentially free-fermion theory at $\gamma \rightarrow 0$ and the usual strongly-interacting SYK limit as $\gamma \rightarrow \infty$. The rank parameter also controls the scaling exponent $\Delta$, which in turn determines the ground-space code properties. As we demonstrate, the low-rank SYK models therefore serve as examples of ground-space codes with continuously tunable rate and distance.

In the next Section, we define the analog of the maximally mixed state on the code in terms of a certain low-temperature Gibbs state and present a direct numerical analysis of the SYK (and low-rank SYK) path integrals to compute the ground state entropy and code distance via the mutual information diagnostic. Then in Sec.~\ref{sec:gravity_mi} we discuss the physics again from the perspective of the duality with JT gravity.

\section{Microscopic analysis}
\label{sec:syk_mi}

Following the discussion in Section~\ref{sec:tools}, we will define and investigate an approximate ground space code for several interrelated models. We focus on two main tasks, (1) determining the rate of the code by computing the ground state entropy and (2) determining the distance of the code by computing the left-right mutual information. All of our results will be averaged over the code ensemble.

Consider a single instance $H$ of the SYK Hamiltonian. Let $\rho_C$ denote the Gibbs state at inverse temperature $\beta \sim N/J$. This is our proxy for the maximally mixed state on the code state discussed in Section~\ref{sec:tools}. The entropy of the Gibbs state in this regime is $s_0 N$ to leading order in $N$. One should ask why this particular scaling of $\beta$ with $N$? This choice is special because it isolates the approximate ground space without giving up the use of saddle point methods when analyzing the path integral (or at least is on the edge of the validity of the saddle point approach, see App.~\ref{app:extrap}). One can consider lower temperatures as well, however, this qualitatively changes the physics (fermion correlators behave diffently and we would eventually probe the splittings in the the approximate ground space) and we lose control of the calculation (saddle point methods no longer apply). Regarding the first task, (1), the rate is thus $s_0$. The computation of the ground state entropy of SYK and low-rank SYK models has been reported before, but we repeat the calculations here for completeness and consider some values of $\gamma$ in the low-rank case that were not previously considered. We compute $s_0$ for SYK in subsection \ref{subsec:syk_num} and for low-rank SYK in subsection \ref{subsec:syk_lowrank_num}, finding agreement with prior results where applicable, e.g. the explicit SYK formulas in \cite{maldacena2016remarks}.

To address the second task, (2), we need the left-right purification of $\rho_C$. This is the code TFD state, and we discuss its physical properties below. As our analog of qubit subsystems, we consider subsets of $K$ fermions on the left and study their mutual information with the entire right in the code TFD. Because of the statistical permutation symmetry between the fermions in the disorder-averaged SYK ensemble, we do not expect the mutual information to depend strongly on precisely which fermions are chosen on the left. So by convention we consider just the first $K$ fermions on the left. To determine the distance, our goal is to evaluate 
\begin{equation}
    I(K) = I(K_L : R)_{\text{TFD}} = S(K_L)_{\text{TFD}}+S(R)_{\text{TFD}} - S(K_L \cup R)_{\text{TFD}}.
\end{equation}
A convenient notion of fermion entanglement using twist fields is specified in Appendix~\ref{app:fermion}. Using this notion, we will show that
\begin{equation}
    I(K) \propto \frac{K}{N^{2\Delta}} + \cdots
\end{equation}
provided $K \ll N^{2 \Delta}$ and where $\cdots$ indicates corrections that are further suppressed in powers of $N$. Requiring that the mutual information be less than some small constant $\delta$ then determines a maximum $K$, the distance, thus addressing the second task, (2).

We will carry out these two tasks first in the microscopic models of interest using path integral techniques. The SYK model is discussed in subsection \ref{subsec:syk_num} and the low-rank SYK model in subsection \ref{subsec:syk_lowrank_num}. We obtain the code rate as well as the mutual information. For technical reasons, our initial approach to calculating the mutual information in Section \ref{sec:syk_mi} is limited to $K \ll N$; but more advanced techniques described in Appendix~\ref{app:numericalmethods} allow us to extend the range of validity to any value of $K$. While our numerical results only directly access Renyi versions of the mutual information, we obtain estimates of the von Neumann mutual information by utilizing a model of the relevant density matrix. This calculation also agrees with a direct twist field calculation~\cite{syk_code_cl}.

\subsection{Structure of the TFD state}
\label{subsec:tfd}

We first describe the stucture of the TFD state for general $\beta$. From the point of view of individual fermion operators $\chi_{iL}$ and $\chi_{jR}$, the TFD state exhibits weak $LR$ correlations of the form
\begin{equation}
    \langle \chi_{iL} \chi_{jR} \rangle_{\tfd} = i G(\beta/2) \delta_{ij} + \mathcal{O}(1/N) \label{eq:lr_syk}
\end{equation}
where $G(\beta/2)$ is the thermal correlator
\begin{equation}
    G(\tau) = \frac{1}{N Z(\beta)} \sum_i \mathrm{tr}( e^{-(\beta - \tau) H} \chi_i e^{- \tau H} \chi_i ) 
\end{equation}
evaluated at $\tau = \beta/2$ \cite{maldacena2018eternal}. Note that $G$ is an average over all the fermion correlations, so each fermion has average correlations up to $1/N$ corrections.

The leading $\delta_{ij}$ factor in \eqref{eq:lr_syk} is an important and standard feature of large $N$ mean-field models. Physically, it arises because the ensemble of SYK Hamiltonians has a statistical symmetry, the orthogonal group $O(N)$, under which fermions transform as $\chi_i \to \sum_k O_{ij} \chi_j$. Hence, any average over the ensemble will have an exact $O(N)$ symmetry acting on fermion indices. Furthermore, the mean-field nature of the SYK Hamiltonian implies that correlators of intensive variables are close to their average value at large $N$. Hence, the leading order term in \eqref{eq:lr_syk} is the $O(N)$ symmetric Kronecker delta whereas the sample-dependent pieces are subleading in $N$. See~\cite{maldacena2016remarks,rosenhaus_syk,suh_softmode} for further discussion of this point.

In the low-temperature regime where $N \gg \beta J \gg 1$, the thermal correlator takes the universal form 
\begin{equation}
    G(\tau) \propto \frac{1}{\left(\beta \sin \frac{\pi \tau}{\beta}\right)^{2\Delta}},
    \label{eq:g_conf}
\end{equation}
where $\Delta = 1/4$ is the fermion scaling dimension in $q=4$ SYK \cite{maldacena2016remarks}. This form is determined by the emergent conformal symmetry mentioned above. In particular, it can be obtained from the zero temperature correlation, $G(\tau) \propto \tau^{-2\Delta}$, via a time reparameterization. Technically, this expression is only fully valid in the regime where $\beta J \ll N$, whereas we are primarily interested in $\beta J \sim N$. Nevertheless, we may obtain the $N$-dependence of the LR correlation in this regime by extrapolating the above formula in the limit $\beta J \to N$. Hence, in the regime of interest we shall take
\begin{equation}
    G(\beta/2)_{\beta \rightarrow N/J} \sim \frac{1}{N^{2\Delta}}.
	\label{eq:LRcorrs}
\end{equation}
As discussed above, further increasing $\beta$ qualitatively changes the physics. For the SYK model, the correlator $G(\tau)$ can have different behavior thanks to the Schwarzian mode becoming strongly coupled~\cite{bagrets_syk_liouville,bagrets_syk_otoc}; in SUSY SYK models, the correlator instead saturates~\cite{lin2023looking,lin2023holography}. Nevertheless, even when properly accounting for the Schwarzian mode at low temperatures, the LR correlator is known to take the form Eq.~\eqref{eq:LRcorrs}  \cite{bagrets_syk_liouville}, so long as $J\beta < N$. Furthermore, we shall find that this form of the LR correlations is well-supported by both our numerical results in Secs. \ref{subsec:syk_num} and \ref{subsec:syk_lowrank_num} and by our holographic analysis in Sec. \ref{sec:gravity_mi}. The weak correlations captured by Eq. \eqref{eq:LRcorrs} will turn out to control the relevant mutual information $I(K)$ and therefore also the recoverability of our code state.

\subsection{Numerical results: SYK}
\label{subsec:syk_num}

Having previewed the code state above, we now compute its properties in the SYK model. We numerically solve the Schwinger-Dyson equations and use these solutions to extract the relevant information. Due to its all-to-all interactions, the SYK model is strongly mean-field, with fluctuations supressed by $1/N$. As a result, the dynamics of the model are governed by semiclassical saddle-point equations of motion called the Schwinger-Dyson equations. By summing melon diagrams \cite{maldacena2016remarks} or equivalently by deriving the semi-classical Euler-Lagrange equations at large $N$ \cite{suh_softmode}, one arrives at the imaginary time Schwinger-Dyson equations for the SYK model:
\begin{align}
    G(\omega_f) &= \left(-i \omega_f - \Sigma(\omega_f) \right)^{-1} \nonumber \\
    \Sigma(\tau) &= J^2 \left[ G(\tau) \right]^{q-1}
\end{align}
where the Green's function $G(\tau)$ is antiperiodic on the thermal circle $G(\tau+\beta) = - G(\tau)$ and $\omega_f = (2m+1) \pi / \beta$ are the fermionic Matsubara frequencies. These equations can be exactly solved in the conformal limit $\beta J \gg 1$ or in the limit of large $q$ \cite{maldacena2016remarks}. Here we numerically solve these equations by iteration, using a weighted update \cite{maldacena2016remarks,maldacena2018eternal}, which we explain in more detail in Appendix~\ref{app:numericalmethods}.

\begin{figure}[h!]
    \centering
    \includegraphics[width=0.8\textwidth]{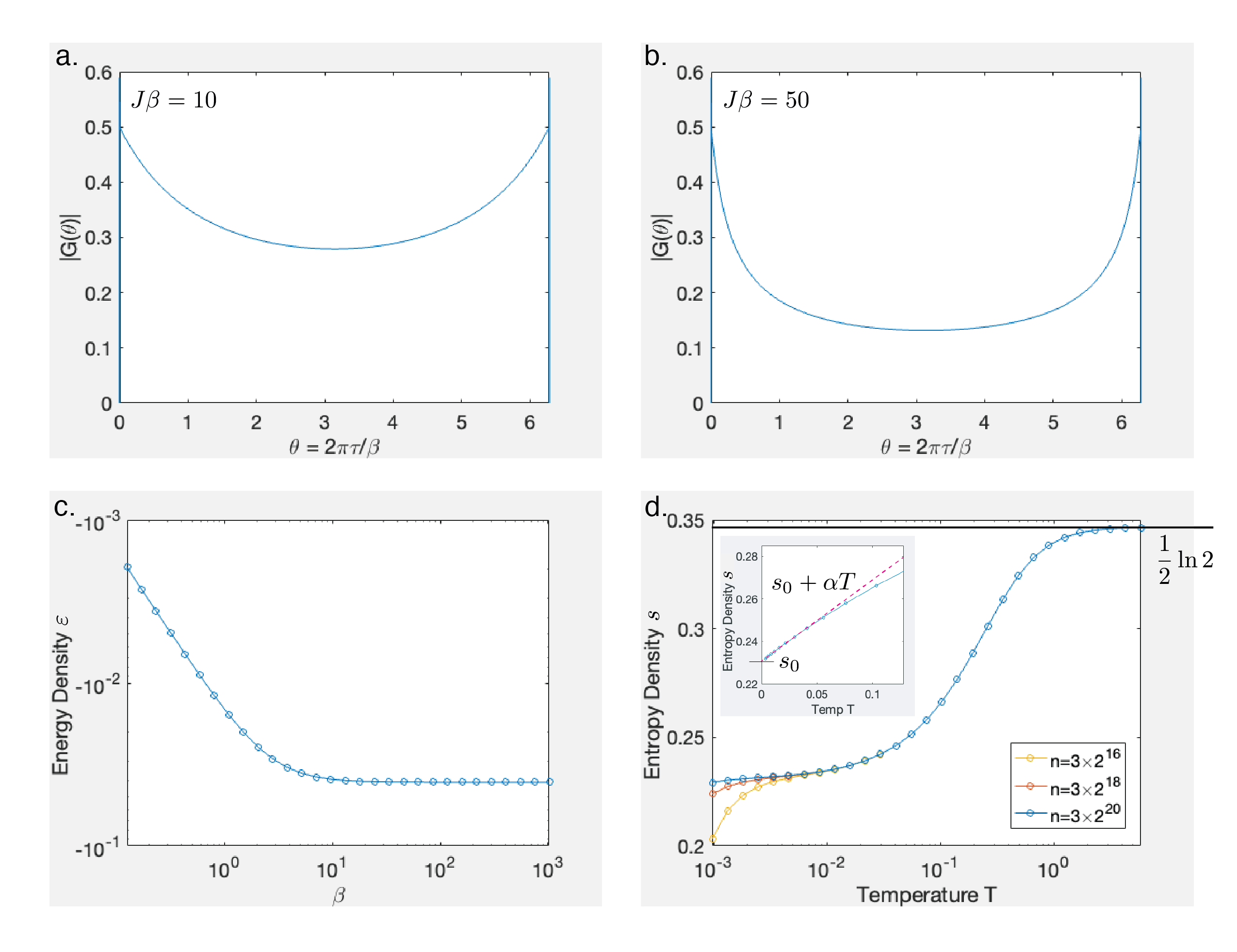}
    \caption{Numerical solutions of the imaginary-time Schwinger-Dyson equations for the SYK model. Top row shows the absolute magnitude $\magn{G(\theta)}$ of the Green's function versus $\theta = 2 \pi \tau / \beta$, calculated at $J \beta = 10$ (a) and $J \beta = 50$ (b). Bottom row shows energy density $\varepsilon$ as a function of $\beta$ (c) and entropy density $s$ as a function of temperature $T = 1/\beta$ (d). The left-most `droop' in $s$ is a numerical artifact of using a finite number of timepoints $n$ in $G(\tau)$, which effectively introduces an artificial UV cutoff; using more timepoints corrects this. Inset shows approximately linear growth of entropy with $T$ at low temperatures, with a ground space entropy density around $s_0 \approx 0.23$.}
    \label{fig:numericalsyk}
\end{figure}

We use the numerical solution $G(\tau)$ to the Schwinger-Dyson equation to compute the properties of the ground space code. We first compute the model's ground space entropy density $s_0$. This quantity tells us how many degrees of freedom comprise the ground space, and is therefore a measure of the code $\emph{rate}$, i.e. how many logical qubits are encoded in the ground subspace. We can obtain the ground space entropy density $s_0$ by calculating the system's energy density \cite{maldacena2016remarks}):
\begin{equation}
    \label{eq:sykenergydensity}
    \varepsilon = - \frac{J^2}{4} \int_0^{\beta} d \tau \ G(\tau)^4.
\end{equation}
over a range of temperatures $T = 1/\beta$. From this data we can compute the system's heat capacity $C_V$, and then integrate down the heat capacity to find the entropy as described in Appendix~\ref{app:numericalmethods}. The results of these numerical simulations are shown in Fig. \ref{fig:numericalsyk}. We conclude that the entropy density grows like
\begin{equation}
    s = s_0 + \alpha T + \ldots
\end{equation}
at low temperatures $T$, with a ground space entropy density of roughly $s_0 \approx 0.23$. Taking the ratio of the ground state entropy to the logarithm of the total Hilbert space, we obtain a rate of $\frac{s_0}{ \frac{1}{2} \ln 2} \approx 2/3$.

Next we probe the code distance of the ground space. Following the arguments in Section \ref{sec:tools}, we characterize the code distance by preparing a thermofield double state at inverse temperature $\beta$ and calculating the mutual information $I(A:R) \equiv I(K_L:R)$ between $K$ fermions on the left side (which we call subregion $A$) and the entire right side $R$. We can estimate the mutual information using our numerical solutions to the Schwinger-Dyson equations as follows. The $k$th-order Renyi mutual information
\begin{equation}
    I^{(k)}(A:R) = S^{(k)}_{A} + S^{(k)}_{R} - S^{(k)}_{A \cup R}
\end{equation}
has contributions from subregion $A$ alone, subregion $R$ alone, and the two subregions together. For small subregions $A \ll R$, we know that $S^{(k)}_{A} \approx \frac{1}{2} K \ln 2$, where $K = \magn{A}$ is the number of fermions in subregion $A$. We similarly know that $S^{(k)}_{R} = S^{(k)}_{L}$ is given by the ground-space entropy computed above. So the interesting piece is the contribution $S^{(k)}_{A \cup R}$ from both subregions together. Because a thermofield-double state is pure, we can equivalently compute the entropy of the complement $S^{(k)}_{A \cup R} \equiv S^{(k)}_{\overline{A}}$.

The $k$th-order Renyi entropy
\begin{equation}
	S^{(k)}_{\overline{A}} = \frac{1}{1-k} \log \mathrm{tr}(\rho_{\overline{A}}^k)
\end{equation}
can be calculated by introducing $k$ copies of the system and performing a SWAP test on the fermions in region $\overline{A}$ as illustrated in Fig. \ref{fig:tfddet} (see Appendix~\ref{app:fermion} for a brief review of the replica trick and fermion twist / SWAP operators). Performing the trace and rearranging terms, we find that the result can be written as a `flagpole' diagram with $k$ blocks of bulk dynamics $e^{-\beta H}$ connected by fermion wires. The presence of the SWAP imposes different boundary conditions in regions $A,\overline{A}$: the wires in region $A$ form $k$ small loops each of length $\beta$, while the wires in region $\overline{A}$ form a single large loop of total length $k \beta$. The fermions have antiperiodic conditions on each of these loops.

\begin{figure}[h!]
    \centering
    \includegraphics[width=0.8\textwidth]{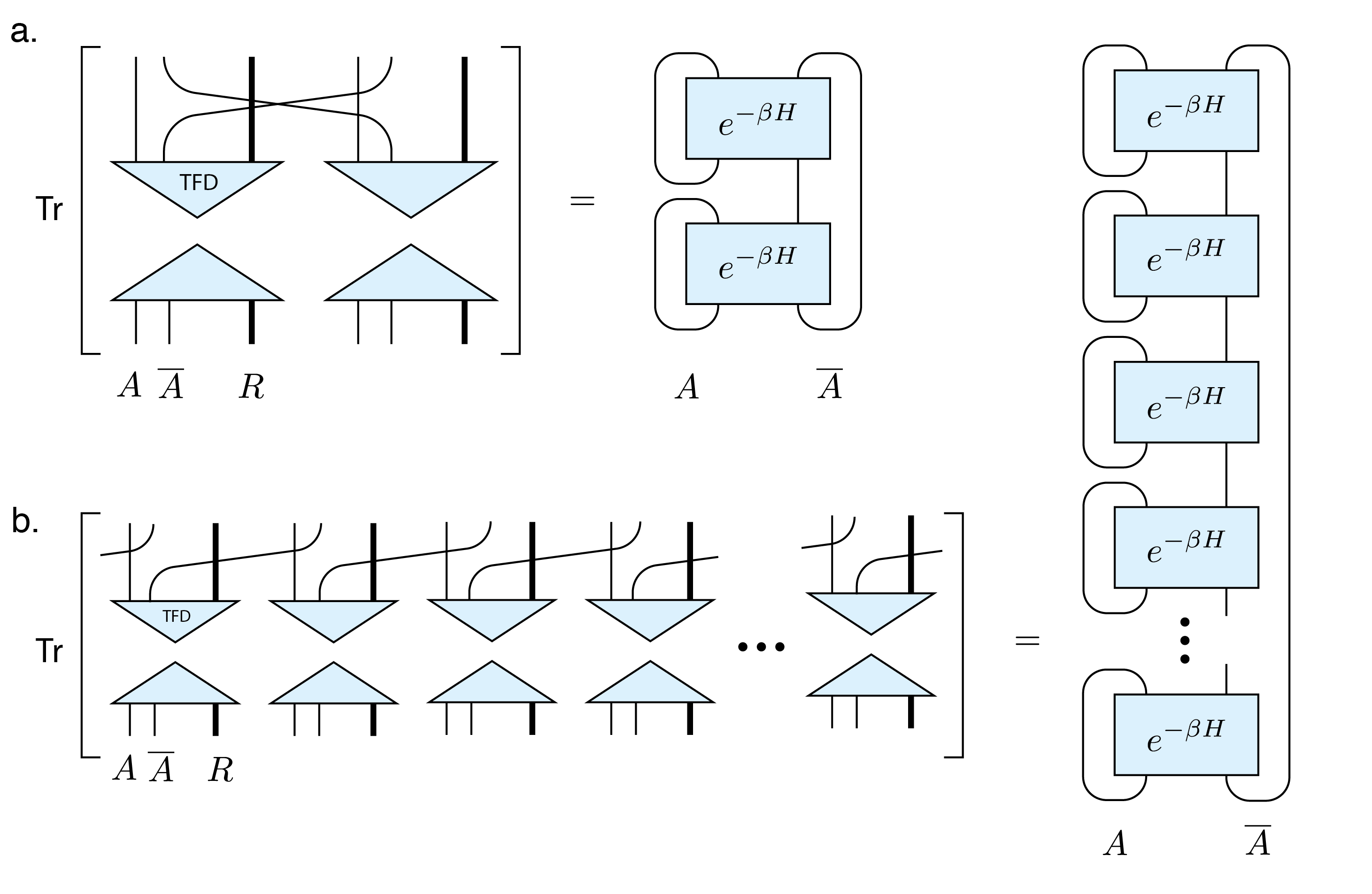}
    \caption{`Flagpole' diagrams for the $k$th Renyi entropy $S_{\overline{A}}^{(k)}$, shown for $k = 2$ (a) and $k > 2$ (b).}
    \label{fig:tfddet}
\end{figure}

To calculate $S^{(k)}_{\overline{A}}$ from our numerical solutions to the Schwinger-Dyson equations, consider starting from the limit where $A = \emptyset$. In this case we only have a single large loop of length $k \beta$:
\begin{equation}
    \mathrm{tr}(\rho^k_{\overline{A}}) = \mathrm{tr}(\rho^k_{L}) = \mathrm{tr} [ e^{- k \beta H} ]
\end{equation}
(where for the moment we are ignoring the normalization of the density matrices $\rho$). The trace can be written as a $G$-$\Sigma$ path integral whose action is dominated by the solutions to the Schwinger-Dyson equations on a large circle of length $k \beta$. These are precisely the same solutions that we found numerically above when computing the ground-space entropy. Now suppose we add a small number $K \ll N$ of fermions to $A$. We now have a very similar $G$-$\Sigma$ path integral but where $K$ fermions have modified boundary conditions. We argue that for sufficiently small $K \ll N$, these boundary conditions do not substantially change the saddle point solutions $G,\Sigma$. So we may simply plug in our numerical solutions to the Schwinger-Dyson equations, and the only remaining change to the path integral comes from the modified boundary conditions on $K$ fermions.

\begin{figure}[h!]
    \centering
    \includegraphics[width=0.8\columnwidth]{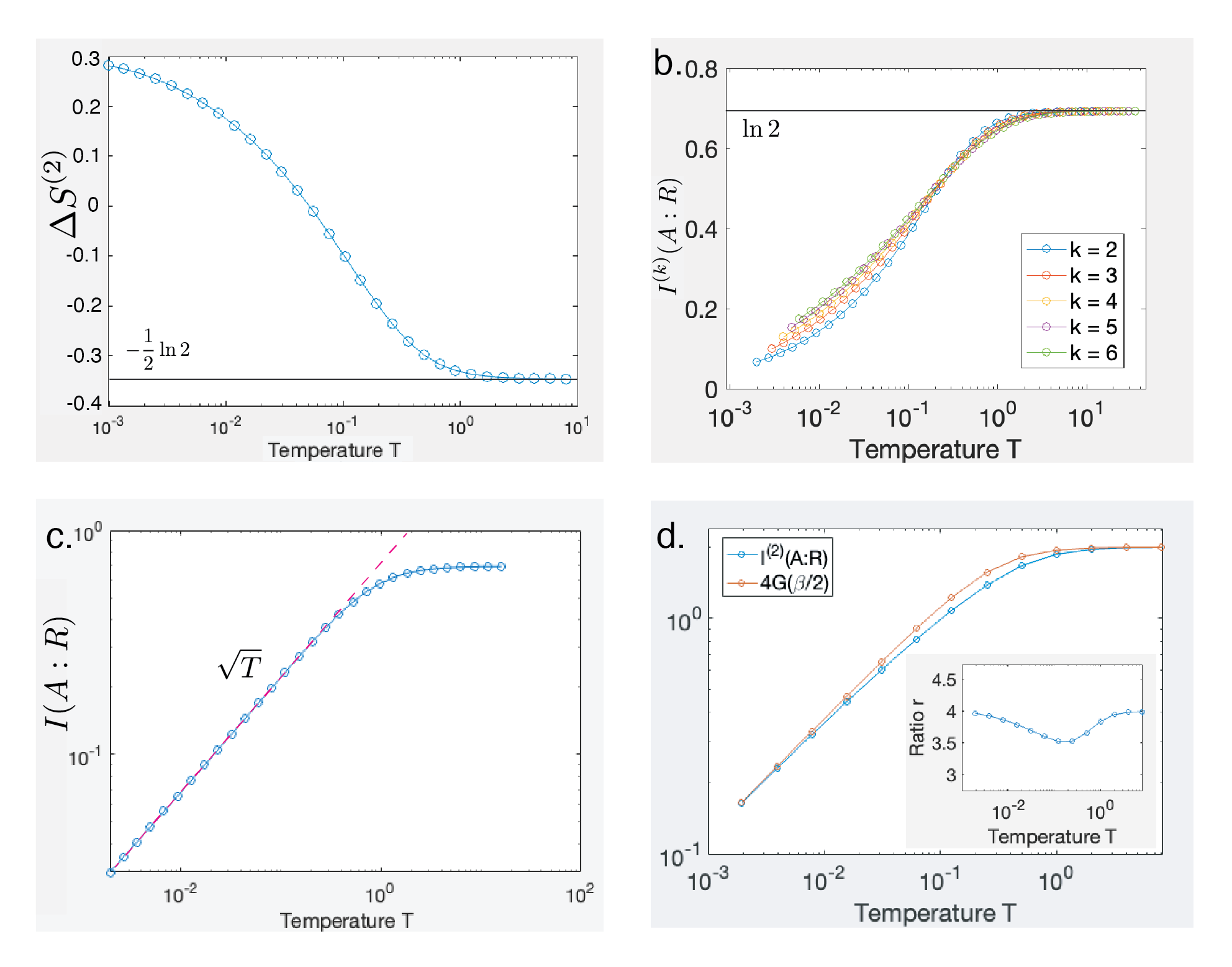}
    \caption{Numerical calculation of mutual information $I(A:R)$ by way of fermion determinants in the $q = 4$ SYK model. We first compute the Renyi entropy difference $\Delta S^{(k)}$ (see main text) using fermion determinants ($k = 2$ shown in a), and use this to compute the mutual information $I^{(k)}(A:R)$ (b). We also extract the von Neumann mutual information $I(A:R)$ as a function of temperature (c) and find that the mutual information scales like $I(A:R) \sim \sqrt{T} = T^{2 \Delta}$ as expected for $\Delta = 1/q = 1/4$. We can also directly compare the mutual information to the Green's function (d) and find good agreement across a range of temperatures (inset plots the ratio $r = I^{(2)}(A:R) / G(\beta/2)$ as a function of temperature).}
    \label{fig:deltasmisyk}
\end{figure}

We discuss in Appendix~\ref{app:numericalmethods} how the thermal propagator with these modified boundary conditions can be explicitly calculated using the technology of fermion determinants. In particular, we show how to compute the change in entropy $K \times \Delta S^{(k)}$ in going from $A = \emptyset$ to $\magn{A} = K$, such that we have
\begin{equation}
    S^{(k)}_{\overline{A}} = S^{(k)}_{L} + K \times \Delta S^{(k)}
\end{equation}
where we refer to $\Delta S^{(k)}$ as the Renyi entropy difference. With these definitions, the mutual information therefore goes as
\begin{equation}
    I^{(k)}(A:R) = K \left( \frac{1}{2} \ln 2 - \Delta S^{(k)} \right)
\end{equation}
We plot the results of these calculations in Fig. \ref{fig:deltasmisyk}. We see that the 2nd-order Renyi entropy difference $\Delta S^{(2)}$ changes sign as the temperature is decreased. This may appear surprising at first but makes sense: at high temperature the thermofield double state consists of EPR pairs between $L,R$, so adding a single fermion to $A$ causes the entropy of $\overline{A}$ to drop by half a bit, $\Delta S^{(k)} = - 1/2 \ln2$. At low temperature, we are projected onto the ground subspace consisting of highly entangled states, so removing a single fermion from $\overline{A}$ and adding it to $A$ causes the entropy of $\overline{A}$ to increase by half a bit, $\Delta S^{(k)} = + 1/2 \ln2$. Computing higher-order fermion determinants allows us to calculate the mutual information $I^{(k)}(A:R)$ up to order $k = 6$ as illustrated in Fig. \ref{fig:deltasmisyk}(b). We see that the mutual information approaches zero as the temperature is decreased, indicating a protected code subspace whose logical information stored in $R$ is decoupled from errors acting on subregion $A$.

Using the simplified model for fermion entanglement described in Appendix~\ref{app:fermion}, we can use our calculation of $I^{(k)}$ to extract an estimate of the von Neumann mutual information $I(A:R)$, which is the truly meaningful quantity when evaluating a code's performance. We show in Fig. \ref{fig:deltasmisyk}(c) that the von Neumann mutual information follows a power law at low temperatures $I(A:R) \sim \sqrt{T}$, which is exactly in agreement with our expectation that $I(A:R) \approx 4 G(\beta/2) \sim T^{2 \Delta}$ for $\Delta = 1/q = 1/4$ (see the discussion around Eq. \eqref{eq:migbulk} in Sec.~\ref{sec:gravity_mi}). In fact, we find good agreement between the mutual information and the Green's function across a range of temperatures as shown in Fig. \ref{fig:deltasmisyk}(d), where we show the ratio $r = I^{(2)}(A:R) / G(\beta/2)$ in an inset.

Our fermion determinant methods are limited to small subregion sizes $K \ll N$, but we can extend our results to any finite value of $K$ by explicitly computing the `flagpole' diagrams shown in Fig. \ref{fig:tfddet} using more advanced numerical techniques. We discuss these numerical calculations in detail in Appendix \ref{app:numericalmethods}, but we present some of the main conclusions here for completeness. First, when $K/N = 1$ we find excellent agreement with the entropy density shown in Fig. \ref{fig:numericalsyk}.d., which gives us an independent check on the heat capacity method described above. Second, we verify that the entropy $S^{(2)}_{A} \approx \frac{1}{2} K \ln 2$ has little dependence on temperature for small $K/N$. Third, we are able to compute the mutual information $I^{(2)}(A:R)$ at any arbitrary value of $K/N$, and we find good agreement between the mutual information computed using fermion determinants and the mutual information computed using the complete `flagpole' numerics when $K/N \ll 1$ (see Fig. \ref{fig:gagbnumerics}.d.). Fourth, calculations at general $K/N$ are also possible holographically and we can attempt to compare them with the microscopic SYK calculations; this comparison is limited because of numerical issues that make it difficult access low temperatures at general $K/N$ but we we see some common features begin to emerge (see Appendix~\ref{app:lr_qes} for further discussion). Finally, we characterize the fluctuations in our ensemble calculations of the Renyi entropy $S_{\overline{A}}^{(2)}$ by comparing the mean of the square to the square of the mean. The close agreement between these two quantities (see Fig. \ref{fig:typicalitykAkB}) implies that our results are highly typical in the ensemble, with small fluctuations.

\subsubsection*{Summary}

For the SYK$_4$ model, the code rate is $s_0$, where the ground state entropy per fermion is $s_0 \approx .23$. We directly access a Renyi version of the mutual information and estimate the von Neumann mutual information from a model of the density matrix. We find that the mutual information at large $\beta$ is $I(K) \sim K/\beta^{1/2}$, and, upon extrapolating $\beta \to N$, the mutual information is less than $\delta$ provided $K < \delta N^{1/2}$. See Appendix~\ref{app:extrap} for a discussion of the extrapolation.

\subsection{Numerical results: low-rank SYK}
\label{subsec:syk_lowrank_num}

\begin{figure}[h!]
    \centering
    \includegraphics[width=0.8\textwidth]{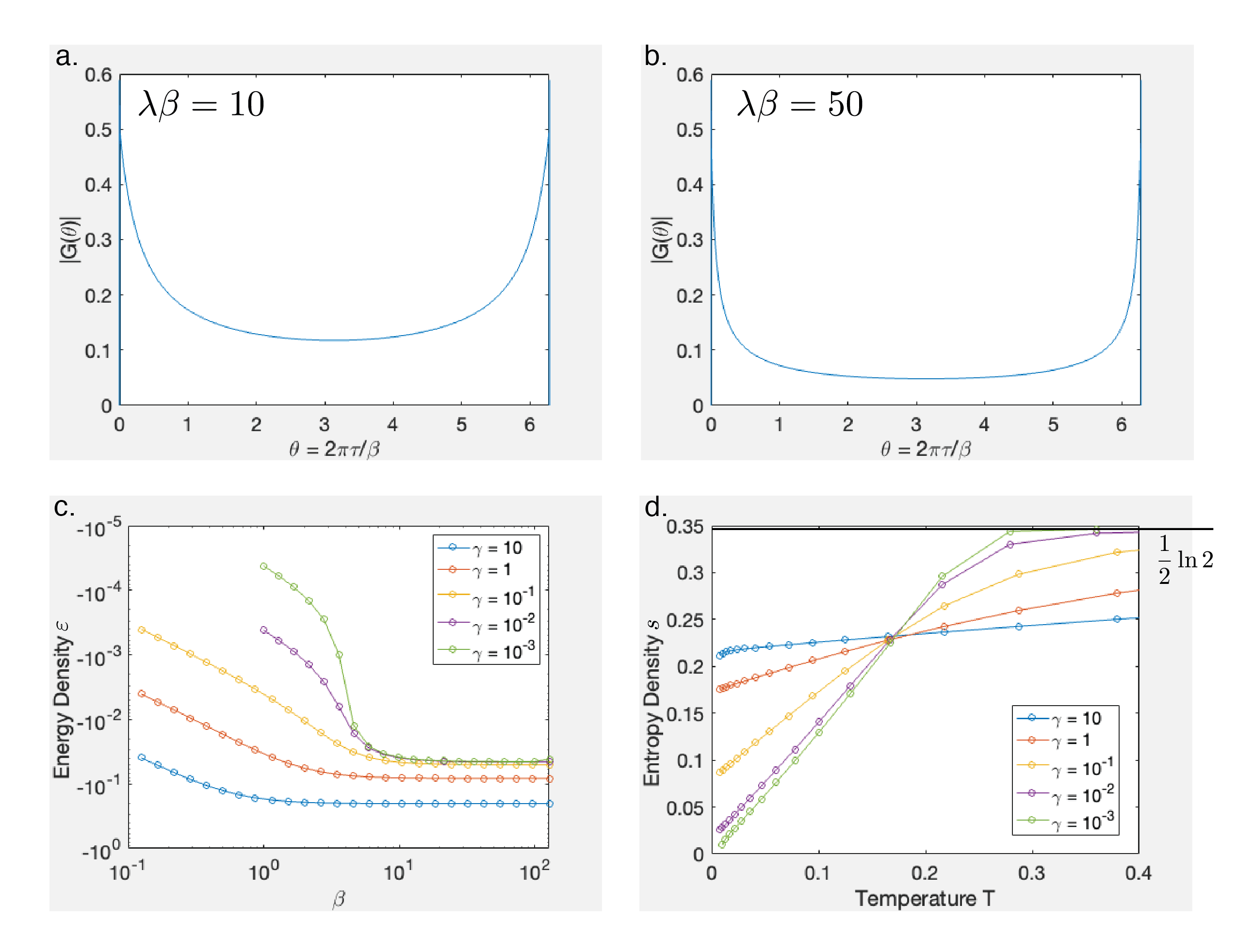}
    \caption{Numerical results for the low-rank SYK model. Top row shows the absolute magnitude $\magn{G(\theta)}$ of the Green's function versus $\theta = 2 \pi \tau / \beta$, calculated at $J \beta = 10$ (a) and $J \beta = 50$ (b). Bottom row shows energy density $\varepsilon$ as a function of $\beta$ (c) and entropy density $s$ as a function of temperature $T = 1/\beta$ (d). We see that the ground-state entropy density $s_0$ vanishes as $\gamma \rightarrow 0$.}
    \label{fig:lowranksykresults}
\end{figure}

We perform a similar analysis for the low-rank SYK models introduced in Section \ref{sec:tools}. The Schwinger-Dyson equations for the Class III low-rank SYK model are \cite{kim2020lowrank}
\begin{align}
    G(\omega_f) &= \left( - i \omega_f - \Sigma(\omega_f) \right)^{-1} \nonumber \\
    \Sigma(\tau) &= 2 \gamma G(\tau) F(\tau) \nonumber \\
    F(\omega_b) &= \lambda \left( 1 - \lambda [G^2](\omega_b) \right)^{-1}
\end{align}
where $G(\tau)$ is antiperiodic on the thermal circle $G(\tau+\beta) = -G(\tau)$, $\omega_f = (2m+1) \pi / \beta$ are the fermionic Matsubara frequencies, and $\omega_b = 2m \pi / \beta$ are the bosonic Matsubara frequencies. Here $[G^2](\omega_b)$ is the Fourier transform of $G^2(\tau)$. As above, we find self-consistent solutions to these equations using a weighted iteration method. In numerics we set $c_0 = \lambda = 1$, which puts everything in units of $\lambda^{-1}$. During the weighted iteration procedure, we also check the condition $1 - \lambda [G^2](0) > 0$ which corresponds to boson condensation \cite{kim2020lowrank}.

Similar to the SYK model, we can estimate the ground space code rate in the low-rank SYK model by computing the ground space entropy. The energy density for the model is \cite{kim2020lowrank}:
\begin{equation}
    \varepsilon = - \frac{\gamma}{2 \beta} \sum_{\omega_b} [G^2](\omega_b) F(\omega_b).
\end{equation}
From this, we compute the heat capacity and integrate down to find the entropy density $s$ as discussed in Appendix~\ref{app:numericalmethods}. The results of these numerical calculations are shown in Fig. \ref{fig:lowranksykresults}. We again find that the entropy density grows linearly with $T$ at small temperatures $s = s_0 + \alpha T$, where $s_0$ is the ground-space entropy density. For large rank $\gamma = 10$ we obtain ground space entropy densities as large as $s_0 \approx 0.22$, similar to the regular SYK model. The ground-space entropy vanishes $s_0 \rightarrow 0$ as we take $\gamma \rightarrow 0$. Nevertheless, at any small but fixed $\gamma$, we find error-correcting codes with a small non-zero rate. The ground space of low-rank SYK models thus serves as a tunable approximate error-correcting code, whose rate is tuned by the rank parameter $\gamma$.

\begin{figure}[h!]
    \centering
    \includegraphics[width=0.8\columnwidth]{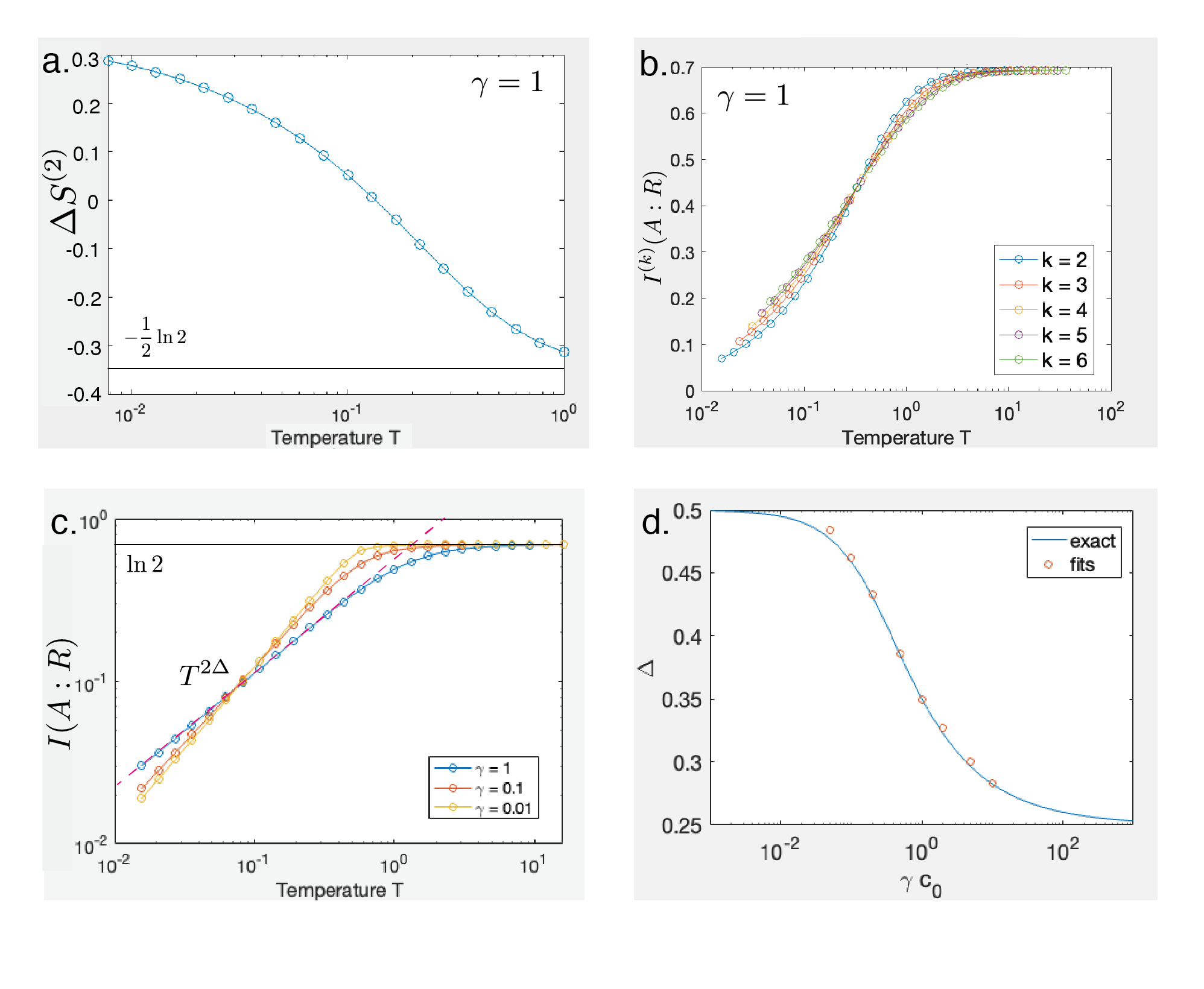}
    \caption{Numerical calculation of mutual information $I(A:R)$ by way of fermion determinants in the low-rank SYK model. We first compute the Renyi entropy $\Delta S^{(k)}$ (a, see main text for definition), and use this to extract the Renyi mutual information $I^{(k)}(A:R)$ as a function of temperature (b). Using the model in Appendix~\ref{app:fermion}, we can then extract the von Neumann mutual information $I(A:R)$ as a function of temperature (c). We find that the mutual information scales like $I(A:R) \sim T^{2 \Delta}$ as expected (dotted pink shows linear fit on a log-log scale), where $\Delta$ varies between 1/4 and 1/2, determined by the rank parameter $\gamma$. The scaling exponent $\Delta$ obtained from these fits (d, orange dots) agrees with the exact expression for $\Delta(\gamma)$ (solid blue) over a wide range of values of $\gamma c_0$.}
    \label{fig:deltasmilowranksyk}
\end{figure}

We can also probe the code distance in the low-rank SYK model using the same fermion determinant methods we introduced above. We plot the results of these calculations in Fig. \ref{fig:deltasmilowranksyk}. We again find that the entropy difference $\Delta S^{(k)}$ changes sign from $\pm 1/2 \ln 2$ as we lower the temperature. Using the entropy difference we can immediately calculate the Renyi mutual information, plotted in Fig. \ref{fig:deltasmilowranksyk}(b) for $\gamma = 1$. We see that the mutual information monotonically decreases toward zero as we decrease the temperature. Using the model in Appendix~\ref{app:fermion} we can extract an estimate for the von Neumann mutual information. Plotting on a log-log scale versus temperature, we find that the mutual information follows a scaling law $I(A:R) \sim T^{2 \Delta}$ that is tunable with the rank $\gamma$ as shown in Fig \ref{fig:deltasmilowranksyk}(c). Fitting these scaling laws and plotting versus $c_0 \gamma$, we see that these fits agree with the exact expression \cite{kim2020lowrank}
\begin{equation}
    \gamma c_0=\frac{(2 \Delta-1)(\sec (2 \pi \Delta)-1)}{8 \Delta-2}, \quad \quad  \Delta \in(1 / 4,1 / 2)
\end{equation}
as shown in Fig. \ref{fig:deltasmilowranksyk}(d). This establishes direct numerical evidence that the mutual information in the low-rank SYK models vanishes at low temperatures and is controlled by the tunable scaling exponent $\Delta$.

\subsubsection*{Summary}

For the low-rank SYK, the physics depends on the rank of the coupling matrix, which is $\gamma N$. As $\gamma \to \infty$, we recover the SYK physics discussed in the previous subsection. As $\gamma \to 0$, new features emerge. For any fixed $\gamma$, the rate of the code is given by the ground state entropy per particle, $s_0(\gamma)$. As $\gamma \to 0$, $s_0(\gamma)$ also vanished linearly with $\gamma$. Hence, for any small but fixed $\gamma$ as $N \to \infty$, the code has a small but constant rate as $N \to \infty$. We again directly access a Renyi version of the mutual information and estimate the von Neumann mutual information from a model of the density matrix. We find that the mutual information at large $\beta$ is $I \sim K/\beta^{2 \Delta(\gamma)}$, where $\Delta(\gamma)$ is a tunable scaling exponent. As $\gamma \to 0$, this exponent is $1/2$ minus a term that vanishes linearly with $\gamma \to 0$. Upon extrapolating $\beta \to N$, the mutual information is less than $\delta$ provided $K < \delta N^{2 \Delta(\gamma)}$. Again, see Appendix~\ref{app:extrap} for a discussion of the extrapolation. Since $\Delta$ can come arbitrarily close to $1/2$ by taking $\gamma \to 0$, we can tune the distance of the code arbitrarily close to linear scaling with $N$. 

\section{Holographic perspective}
\label{sec:gravity_mi}

We will now discuss how the above microscopic results can be obtained from the dual gravity picture. The model is JT gravity coupled to $N$ bulk fermion fields with tunable mass $m$ which encodes a variable scaling dimension $\Delta$. Each $\chi_i$ in the microscopic model has its own dual ``bulk'' field in the gravity model. These bulk fermion fields are weakly interacting, coupling only to gravity and not directly to each other (this is an avatar of the $\delta_{ij}$ in \eqref{eq:lr_syk}). The JT field content is the metric and a scalar ``dilaton''. The relevant spacetime turns out to be a piece of two-dimensional Anti de Sitter space (AdS$_2$). We refer to this as the holographic model or the gravity model, and via the tunable mass, it serves as a model of both SYK and low-rank SYK.

The key feature of the holographic model is that the low-temperature thermofield double state of SYK is dual to a particular wormhole solution of the gravity theory. In this gravitational model, the code properties are tunable parameters. The rate of the code is set by the entropy of a zero-temperature black hole and the distance is determined by the structure of correlations in the black hole background. The central virtues of this holographic perspective are that it offers a clear physical interpretation of the smallness of the mutual information, it allows direct access to the von Neumann mutual information, and it allows us to extend the calculation beyond the regime of small $K$ (we consider that extension in Appendix~\ref{app:lr_qes}). All the results will be consistent with the direct microscopic calculations in Sec.~\ref{sec:syk_mi}.

\subsection{Structure of the holographic TFD state}

We begin by discussing how the correlation \eqref{eq:lr_syk} in the low-temperature TFD state is encoded holographically. This small correlation turns out to arise from a large spatial separation between the degrees of freedom. In other words, the fermions represented by $\chi_{iL}$ and $\chi_{iR}$ can be viewed as being physically separated by a large but finite distance in the low temperature TFD state. This separation is not in any physical space defined by the model but rather in an emergent space which encodes the entanglement structure of the low temperature TFD.

\begin{figure}
    \centering
    \includegraphics[width=.5\textwidth]{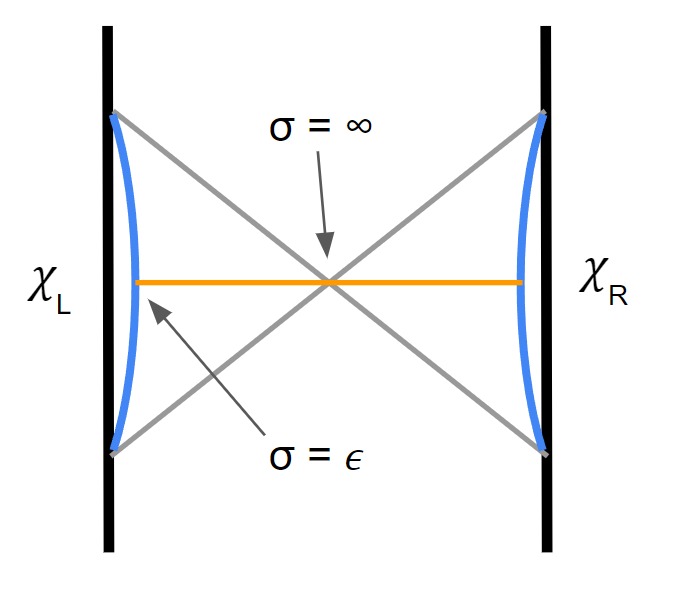}
    \caption{A depiction of the emergent wormhole geometry describing the low temperature TFD state. The two solid black lines are the boundaries of a mathematical space known as two-dimensional Anti de Sitter space (AdS$_2$). The two blue lines represent dynamical boundaries of a cut off AdS$_2$ space. Because of the simplicity of low-dimensional gravity, these boundaries are the only dynamical degrees of freedom of the geometry. We may think of the left and right SYK models living on the left and right blue lines, respectively, with the vertical direction corresponding to time. The horizontal direction corresponds to an emergent space direction, with the orange line giving one geodesic between the left and right boundaries. It is the length of this orange line which is large at low temperature, thus suppressing the correlations between the left and right fermions living at the blue boundaries.  }
    \label{fig:wormhole}
\end{figure}

The basic setup is illustrated in Figure~\ref{fig:wormhole}, which shows a piece of AdS$_2$ spacetime that we can interpret as a wormhole connecting two asymptotic boundaries on the left (L) and right (R). One convenient choice of coordinates covers half the relevant geometry corresponding to the interior of the left grey triangle in Figure~\ref{fig:wormhole}. The metric is 
\begin{equation}
    ds^2 = \frac{4\pi^2}{\beta^2} \frac{-dt^2 + d\sigma^2}{\sinh^2 \frac{2\pi \sigma}{\beta}}, \label{eq:bh_metric}
\end{equation}
where $\sigma=0$ is the left asymptotic boundary (black line in Figure~\ref{fig:wormhole}) and $\sigma \to  \infty$ is a black hole horizon (grey cross in the middle of Figure~\ref{fig:wormhole}). The black hole has a temperature set by $1/\beta$. The regulated blue boundary is specified by $\sigma=\epsilon$ and is where the left SYK can be viewed as living. As a technical aside, all bulk distances are measured in units of the AdS radius which is thus set to unity.

The coordinates above describe the exterior of a left black hole. The interior of the right grey triangle has an essentially identical metric and corresponds to the exterior of a right black hole. It is also possible to choose coordinates which cover the whole spacetime and these describe a wormhole connecting the left and right black holes. The left and right exteriors, viewed separately, are analogous to the left and right Gibbs states. Meanwhile, the global coordinates describing the wormhole correspond to the TFD state.

Now we can return to discuss the left-right correlations in equation~\eqref{eq:lr_syk}. According to the AdS/CFT dictionary, each boundary fermion is dual to a bulk field (a fermion field) such that the correlation in \eqref{eq:lr_syk} is obtained as a limit of the bulk field propagator between two points. Moreover, the mass $m$ of the bulk field is related to the dimension $\Delta$ of the boundary operator. In the large mass limit, the bulk propagator is given by a geodesic approximation,
\begin{equation}
    G_{\text{bulk}}(p,p') \sim e^{ - m (\text{geodesic distance from $p$ to $p'$})}.
\end{equation}
(see Ref \cite{hartnoll2018holographic} for a derivation of this fact). We first discuss the geodesic approximation to give intution then describe the general result. It is also important to keep in mind that $G_{\text{bulk}}$ depends on the points $p$ and $p'$, and it is only simply related to the boundary correlator, $G(\tau)$, for special points. 

Hence, to understand the left-right correlation we should find the length of a geodesic from the left boundary to the right boundary. From the $d\sigma^2$ part of \eqref{eq:bh_metric}, we learn that the proper distance from the boundary to the horizon is
\begin{equation}
   \frac{2\pi}{\beta} \int_\epsilon^\infty d\sigma \frac{1}{\sinh \frac{2\pi\sigma}{\beta}} = \int_{2 \pi \epsilon/\beta} ^\infty d\sigma' \frac{1}{\sinh \sigma'} = - \ln \tanh \frac{\pi \epsilon}{\beta} \sim \ln \frac{\beta}{\pi \epsilon}.
\end{equation}
Hence, for fixed cutoff $\epsilon$ the distance across the wormhole grows long as $\beta \to \infty$. The full left-right distance is twice this value, so the left-right correlation in the geodesic approximation is
\begin{equation}
    e^{ - m ( 2 \ln \frac{\beta}{\pi \epsilon})} \sim \frac{1}{\beta^{2m}}.
\end{equation}
This has the form of \eqref{eq:g_conf} with $m = \Delta$. In fact, from the holographic dictionary the relationship between $m$ and $\Delta$ is given by 
\begin{equation}
	\Delta(\Delta-1) = m^2
\end{equation}
in units where the AdS radius is equal to 1 \cite{hartnoll2018holographic}, so $m$ is technically only proportional to $\Delta$ in the limit of large $m$ and $\Delta$, but the form \eqref{eq:g_conf} turns out to be the right generalization. The virtue of the geodesic approximation is that it clearly shows how the smallness of the correlation arises from the long wormhole.

\begin{figure}
    \centering
    \includegraphics[width=.7\textwidth]{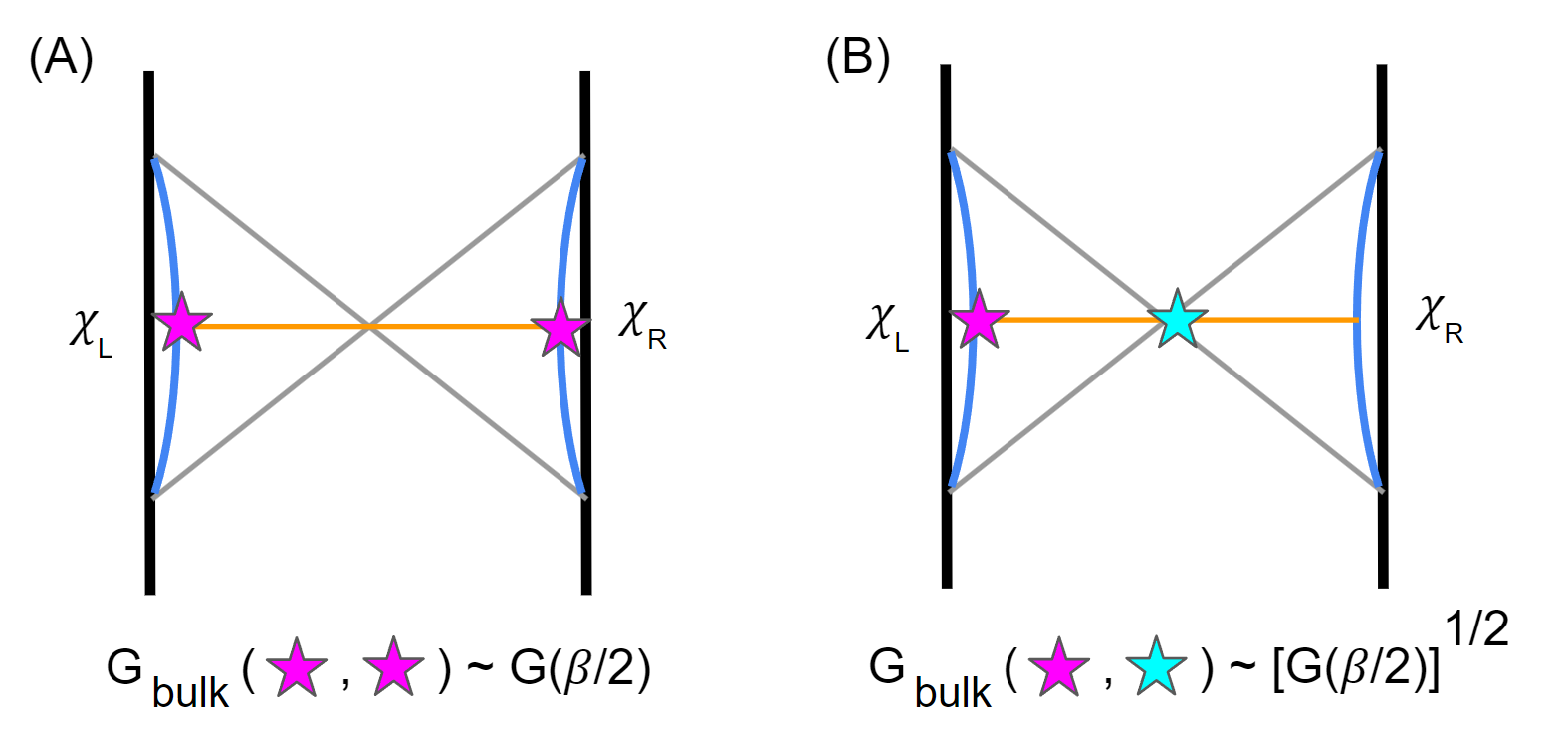}
    \caption{Illustrating the correlation structure in the wormhole geometry. One can show that the correlation between different fields (purple stars) is controlled by the separation between the corresponding points in the bulk spacetime. Panel (A): Here we are studying the correlation between a fermion on the left and a fermion on the right. We do not have access to any other degrees of freedom on the left or right. As a result, the correlation is controlled by the full length of the wormhole since our fields are inserted at the left and right boundaries. Panel (B): Here we are studying the correlation between a fermion on the left and the entire right. According to the rules of holography (entanglement wedge reconstruction), access to the full right gives us access to fields that live in the bulk between the left-right midpoint and the right boundary. As such, the correlation is now controlled by half the wormhole length instead of the full wormhole length. The calculation culminating in \eqref{eq:migbulk} then gives a mutual information proportional to $G(\beta/2)^2$ in case (A), whereas it is proportional to $G(\beta/2)$ in case (B). This enhancement of the correlation is the result of having access to the entire right algebra, (B), instead of single fermion, (A). }
    \label{fig:wormhole_corr}
\end{figure}

The calculation we just discussed captures the correlation between $\chi_{iL}$ and $\chi_{iR}$, but what we actually need is the correlation between $\chi_{iL}$ and the entire right as encoded in the mutual information. According to the AdS/CFT dictionary, given access to the entire right (and not just $\chi_{iR}$), we have access to fields deeper in the bulk (not just at the right blue boundary). These two cases, access to a single right fermion and access to the entire right, are contrasted in the two panels of Figure~\ref{fig:wormhole_corr}. The crux is that the left-right correlation is now larger since we can access a bulk field just to the right of the horizon (instead of only at the right boundary). The geodesic distance is now $\ln \frac{\beta}{\pi \epsilon}$ instead of $2 \ln \frac{\beta}{\pi \epsilon}$. Plugging into the geodesic approximation then gives a propagator of the form $\frac{1}{\beta^m}$ instead of $\frac{1}{\beta^{2m}}$. At large $\beta$, the correlation is thus significantly enhanced when we have access to the entire right. Removing the large mass approximation, the result is $\frac{1}{\beta^\Delta}$. We now turn to a description of the formal calculation of the mutual information. At large $\beta$, it will turn out to be controlled by the correlations we have just described.

\subsection{Holographic calculation of the mutual information}
\label{subsec:holo}

The left-right mutual information we need to calculate is composed of three entropy terms. The general prescription for computing these entropies is known as the quantum extremal surface (QES) prescription~\cite{qes}. We will only need to consider time-symmetric situations, which simplifies the discussion. The first step is to select the boundary degrees of freedom whose entropy we wish to compute. Then we consider candidate points\footnote{Generally, these are spacetime codimension $2$ surfaces; here they are just points since the bulk spacetime dimension is $2$.}, each of which defines a candidate entropy. The true QES and associated true entropy is obtained from the candidate point whose associated entropy is minimal under variations of the candidate point. We use a 2d QES prescription developed in \cite{qglab_2,syk_code_cl,antonini_holo_meas_2d}. We also show that this prescription can be obtained from a dimensional reduction in Appendix~\ref{app:holography}; this dimensional reduction approach is not the same as the SYK model but it yields similar low energy physics (many fermions coupled to JT gravity) and is useful for clarifying some subtleties in the 2d QES prescription.

\begin{figure}
    \centering
    \includegraphics[width=.9\textwidth]{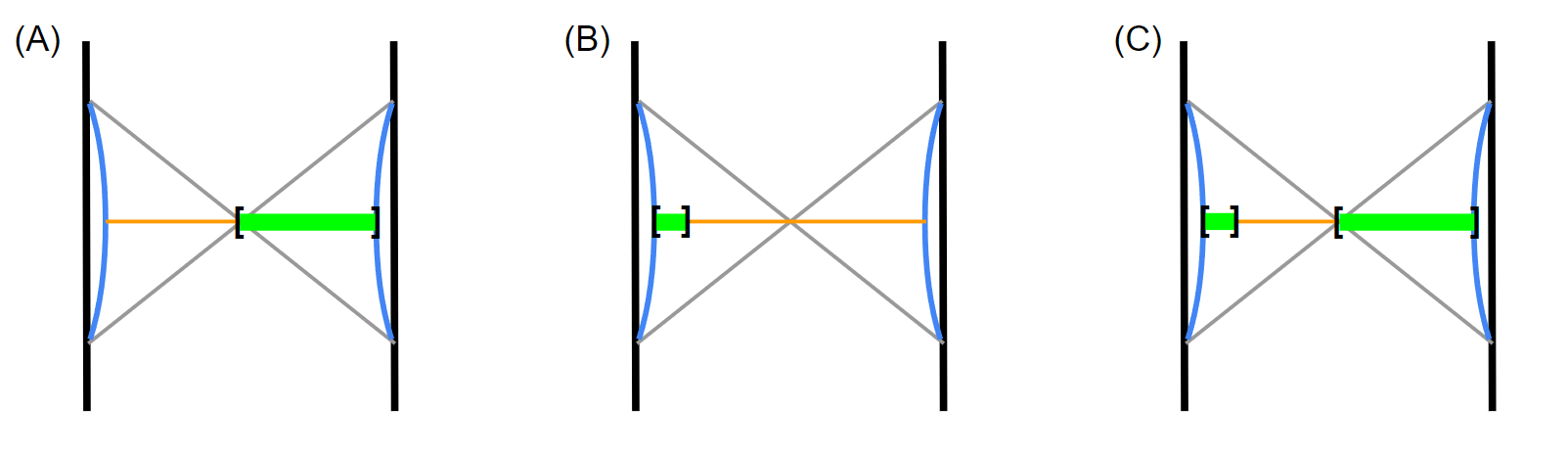}
    \caption{Bulk regions (in green) for the three terms in $I(K_L:N_R)$: (A) $S(R)$, (B) $S(K_L)$, (C) $S(K_L \cup R)$. }
    \label{fig:wormhole_entanglement}
\end{figure}

To illustrate the main idea, let us first consider the entropy of the entire right (or the entire left, since the left-right state is pure). The orange curve in Figure~\ref{fig:wormhole} represents a bulk spatial slice that corresponds to the TFD state. A candidate point is a point on that orange curve. The entropy associated with such a candidate point $p$ is
\begin{equation}
    S_{\text{gen}} = \frac{\phi(p)}{4 G_N} + S_{\text{bulk}}(p),
    \label{eq:qes}
\end{equation}
where $\phi(p)$ is the ``dilaton'' field evaluated at point $p$ and $S_{\text{bulk}}(p)$ is the entropy of bulk fields, i.e. the $N$ fermion fields, on the interval from $p$ to the right boundary (panel (A) of Fig.~\ref{fig:wormhole_entanglement}). The dilaton term is analogous to the area contribution in the perhaps more familiar Ryu-Takayanagi formula~\cite{rt}, while the bulk term arises because the bulk fields can also be entangled across the bulk point $p$. 

In the same coordinates as above, the dilaton profile is
\begin{equation}
    \phi = \phi_0 + \phi_r \frac{2\pi}{\beta} \frac{1}{\tanh \frac{2\pi \sigma}{\beta}},
    \label{eq:dil_profile}
\end{equation}
where $\phi_0$ is a constant that controls the zero-temperature entropy (and hence the code rate) and $\phi_r$ is related to the boundary value of the dilaton, 
\begin{equation}
    \phi(\sigma=\epsilon) = \frac{\phi_r}{\epsilon}.
\end{equation}
There is a corresponding profile on the right, such that the dilaton is smallest at the horizon and increases symmetrically towards either boundary. The bulk entropy is more complicated to evaluate, but fortunately we will only need some general properties of it.

To be precise, at the current state of development, we cannot analytically compute the bulk entropy for general mass and general choice of bulk regions. This makes the argument below necessarily less concrete than would be ideal. Nevertheless, we can reliably obtain certain asymptotics of the bulk entropy in the limit of a long wormhole, and as we carefully argue in the rest of this section, these asymptotics plus general properties of the bulk entropy are enough to establish our main result.

Our main goal is to compute $I(K)$, which means evaluating the formula
\begin{equation}
    I(K) = I(K_L:R) = S(K_L) + S(R) - S(K_L \cup  R).
\end{equation}
Each of these entropies has a QES formulation. The relevant bulk regions are shown in green in Figure~\ref{fig:wormhole_entanglement}. We begin by discussing the entropy $S(R)$ of the entire right, corresponding to panel (A). In this case, the minimal point $p$ is simply the left-right midpoint, also called the bifurcation surface. The dilaton profile is manifestly minimal at the left-right midpoint. Moreover, even without an explicit calculation, one can argue that the bulk entropy term is symmetric about the left-right midpoint since the total bulk state is pure. In fact, for conformal matter, the bulk entanglement is actually independent of the candidate point (for a single interval) and more generally we expect only weak dependence on the endpoint. From this we conclude that the midpoint is indeed minimal for the full generalized entropy. The associated entropy is the thermal entropy of the right SYK (or the left SYK) at inverse temperature $\beta$ since the left-right entanglement in the TFD state is the thermal entropy of a single side.

Consider next the term $S(K_L)$. When $K \ll N$, the relevant bulk regions turns out to be a small interval on the left (panel (B) of Figure~\ref{fig:wormhole_entanglement}). Intuitively, this is because we have access to only a small number of fermions on the left, so most of the bulk should be inaccessible. Concretely, these fermions are all essentially in a maximally mixed state and hence know nothing about the state of the full system. This is represented in the QES formalism with a UV-scale region as in panel (B). We should of course minimize over the location of the endpoint of the bulk region using a generalization of \eqref{eq:qes} in which the dilaton contribution is reduced by a factor of $K/N$, with the result being the indicated small region near the left boundary.

Consider finally the term $S(K_L \cup R)$, which contains the interesting physics. The relevant bulk region is shown in panel (C) of Figure~\ref{fig:wormhole_entanglement}. In detail, we consider $K$ of the bulk fermions on the union of the bulk regions in panels (A) and (B) (these are the fermions we have access to on the left and right) and $N-K$ of the bulk fermions on just the bulk region in panel (B) (these are the fermions we only have access to on the right). As before, we should really minimize the entropy over the choice of endpoints, however, provided $K \ll N$, the result is close to the locations obtained in panels (A) and (B).

In fact, the only difference between panel (C) and the union of panels (A) and (B) is that the $K$ fermions on the union of the bulk regions now have a more complicated bulk state. If the distance between the two bulk intervals in panels (A) and (B) were very large, then we might expect that the bulk intervals are barely correlated leading to a bulk entropy which is additive, i.e. the sum of the entropies of each region on its own. If this were the case, then the full entropy in panel (C) would simply be the sum of the entropies in panels (A) and (B) and the left-right mutual information would be zero. 

However, the entropy of the union of bulk regions is not exactly additive because there is some weak correlation between the two, i.e. the correlation we discussed in the previous subsection. This weak correlation implies the \emph{bulk mutual information} is non-vanishing and hence the bulk entropy is not exactly additive. In the approximation that the bulk endpoints don't change in going from panels (A) and (B) to panel (C), the bulk entropy in panel (C) is thus the bulk entropy in panels (A) and (B) plus $K$ times the bulk mutual information. The approximation of ignoring the change in the bulk endpoints is a good one because $K$ is small and the bulk mutual information is very small (if it were zero, then the minimization problem for the endpoints in panel (C) would reduce to a separate minimization of the end points in panels (A) and (B)).

We finally reach our main conclusion: up to subleading corrections, the left-right mutual information is simply $K$ times the bulk mutual information of a single fermion field. This is so because all the dilaton terms cancel and all the bulk entropy terms combine to give the bulk mutual information. It remains then to determine the bulk mutual information. This is still a challenging problem in full generality, but we again find a simplication in the limit of large $\beta$. Because the bulk intervals are well separated and the bulk fields are nearly free, the bulk mutual information is (see Appendix~\ref{app:holography}),
\begin{equation}
    \label{eq:migbulk}
    I_{\text{bulk}} \sim G_{\text{bulk}}^2,
\end{equation}
with $G_{\text{bulk}}$ being the propagator associated with the minimal separation between a point in one interval and a point in the other interval. Looking at panel (C) of Fig.~\ref{fig:wormhole_entanglement}, we see that the distance of closest approach corresponds to the distance between the purple star and the aqua star in panel (B) of Figure~\ref{fig:wormhole_corr}.

In the previous section we estimated $G_{\text{bulk}}$ between the left boundary and the midpoint of the wormhole and found $G_{\text{bulk}} \sim \sqrt{G(\beta/2)}$, see panel (B) of Fig.~\ref{fig:wormhole_corr}. For general $\beta$, we have
\begin{equation}
    I(K) = K I_{\text{bulk}} \sim K \left( \frac{1}{\beta^\Delta} \right)^2.
\end{equation}
Upon taking $\beta \sim N$, we find at last
\begin{equation}
    I(K) \sim \frac{K}{N^{2\Delta}}.
\end{equation}
This means that the mutual information is less than a constant $\delta$ provided 
\begin{equation}
    K < \delta N^{2 \Delta},
\end{equation}
which is our estimate for the distance of this holographic code.

To recap, this result followed from the holographic model using three key facts:
\begin{itemize}
    \item that the number of fermions $K$ is much less than $N$ (to allow us to ignore changes to the QES in panel (C) relative to panels (A) and (B) of Figure~\ref{fig:wormhole_entanglement}, see also Appendix~\ref{app:lr_qes}),
    \item that the bulk fermion fields are nearly free and only weakly correletated with each other (encoded in the $\delta_{ij}$ in \eqref{eq:lr_syk} and used to estimate the bulk mutual information),
    \item and that the wormhole length is very long in the limit of large $\beta$ (also used to give a simple estimate for the bulk mutual information).
\end{itemize}
We discuss the holographic calculations in greater detail in Appendices~\ref{app:holography} and~\ref{app:lr_qes}; in particular, we relax the condition that $K$ be small in Appendix~\ref{app:lr_qes} and obtain a more general formula for the mutual information even when it is no longer small.

\section{Discussion and outlook}
\label{sec:outlook}

We considered the ground spaces of certain non-commuting mean-field quantum many-body systems, the SYK model and low-rank generalizations, as approximate quantum error correcting codes. Using a combination of analytical and numerical methods, we obtained the rates and distances of these codes, with the rate obtained from the ground state entropy and the distance obtained from the two-sided mutual information. For the low-rank codes, the rate and distance were tunable. We also gave an interpretation of the code properties in terms of a dual gravity description and used the gravity model to obtain results consistent with the microscopic calculations. Although we considered a variety of models here, there are many more models that would be worth studying, including different gravitational setups. In the rest of this section, we highlight a few additional results and directions for further study.

\subsection{Encoding and decoding}

One central question is about how to encode and decode the approximate codes we considered here. We do not know an explicit procedure to carry out either process, so we will just give some speculations.

For encoding, one possibility is a recent proposal for a tensor network model of the SYK ground space called NORA (non-local renormalization ansatz)~\cite{nora}. In this model, one has a logarithmic depth circuit which generates a basis for the approximate ground space. The input is an arbitrary state on $k$ ``ground state'' qubits tensored with a fixed state on $N-k$ ``thermal qubits''. The output is an approximate ground state on the full $N$-qubit system. As already pointed out in~\cite{nora}, this picture must be adapted to Majorana fermions to make precise contact with SYK\footnote{Work in progress with V. Bettaque.}, but several qualitative features of the architecture can be compared with our results.

For example, the logarithmic depth of the circuit, combined with the hypothesis that each layer ``renormalizes'' a scaling operator by a fixed factor related to the field's scaling dimension (familiar from MERA circuits~\cite{vidal_mera_scaling}), naturally yields a left-right correlation of the form $N^{-c_1}$ for some constant $c_1$. Similarly, the growth of logical operators through the network also naturally produces distances that scale like $N^{c_2}$ for some other constant $c_2$. Nevertheless, it remains to be seen if the NORA network can provide a detailed accounting of the physics of SYK.

For decoding, we do not know how to proceed. It is the case that acting with fermion operators on the ground space increases the energy by a significant amount on average (i.e. not $1/N$ suppressed). However, the variance is larger than the average, so it is not obvious how to determine which fermion operators acted using only a single round of measurements on one copy of the state. The Hamiltonian has a typical mean-field form, meaning it can be written as
\begin{equation}
    H = \sum_a h_a
\end{equation}
such that the terms $h_a$ commute with one another at leading order, $[h_a,h_b]=\mathcal{O}(1/N)$ for each $a$ and $b$. This can be contrasted with the typical situation in a spin chain, say, in which a given term typically strongly fails to commute with its neighboring terms but exactly commutes with all other terms. Here each term $h_a$ consists of all coupling terms that involve the field $\chi_a$. Each term in $H$ of the form $\chi_{a_1} \cdots \chi_{a_q}$ is split evenly between $h_{a_1}$ through $h_{a_q}$. Since the individual $h_a$s almost commute, we almost have a notion of a commuting ``energy per particle'' operator. Nevertheless, it is not clear if there is any simple decoding strategy. For example, we could try to measure individual $h_a$s to detect errors, but trying to measure many of them at once naively runs into a buildup of effects from the individually weak non-commutativity.

It is interesting to relate the mutual-information-based notion of distance we considered to other relevant measures of code performance, such as the recovery fidelity. We make such a connection in Appendix~\ref{app:approxrecovery}. It would also be interesting to directly study other distance-like notions in SYK, including the recent proposal of ``subsystem variance''~\cite{yi2023complexity} which appeared as this paper was being prepared for submission. 

\subsection{Variations in the code properties}

Another question is how much the code properties vary from one instance of the ensemble to another. In SYK, standard estimates indicate that the fluctuations in various correlations are suppressed relative to their average values by $1/N$ factors. For example, in the $q=4$ SYK model, the correlations are diagonal up to $1/N$ corrections,
\begin{equation}
    \langle \chi_{iL} \chi_{jR} \rangle = i G(\beta/2) \delta_{ij} + \cdots,
\end{equation}
where $\cdots$ is a random variable of size $N^{-3/2} = N^{-(q-1)/2}$. Moreover, the overall size of the fluctuations is also proportional to the Green function, hence the $N^{-3/2}$ is actually a relative fluctuation. This is important since we want to consider regimes of $\beta$ for which the correlator itself is small. 

These facts indicate that the mutual information between one fermion on the left and the same fermion on the right, which is directly controlled by the left-right correlation written just above, will be well approximated by its average value. For the mutual information between a fermion on the left and the entire right, which we have studied in detail in this paper, it is harder to say anything precise. From the bulk point view, this mutual information is controlled by a correlation between the middle of the wormhole and the left boundary. This correlation is presumably also subject to small variations over the ensemble. Hence, we conjecture that the mutual information defining the code distance will be well approximated by its average, at least provided $K \ll N^{2 \Delta}$.

One way to approach this problem is to study the ensemble fluctuations of the various entropic quantities we consider. For example, given a fixed realization of the couplings, the purity is
\begin{equation}
    e^{-S_2(K)} = \frac{\mathrm{tr}(M_K \rho \otimes \rho)}{\mathrm{tr}(\rho)^2}
\end{equation}
where $\rho = e^{-\beta H}$ and the choice of fermions is encoded in the choice of twist operator $M_K$. The normalization, $\mathrm{tr}(\rho)$, is self-averaging (see discussion in Appendix~\ref{app:extrap}), so the main problem is to compute
$\mathrm{tr}( M_K \rho \otimes \rho)$. Much of the technical work of this paper is directed at computing the ensemble average of this and related quantities, e.g. representing $\mathbb{E}_J[ \mathrm{tr}(M \rho \otimes \rho)] $ as a path integral and evaluating it by saddle point. 

To address fluctuations, we can study the second moment,
\begin{equation}
    \mathbb{E}_J[ \mathrm{tr}(M_K \rho \otimes \rho)  \mathrm{tr}(M_K \rho \otimes \rho) ] .
\end{equation}
This object can also be represented as a path integral over a total of four copies of the system (one for each factor of $\rho$), with the copies coupled in pairs by the twist fields. The path integral manifestation of $\mathrm{tr}(\rho)$ being self-averaging is the fact that the dominant saddle point which computes $\mathbb{E}_J[\mathrm{tr}(\rho) ] $, call it $G_1$, and the dominant saddle point which computes  $\mathbb{E}_J[\mathrm{tr}(\rho) \mathrm{tr}(\rho) ] $, call it $G_2$, are related by 
\begin{equation}
    G_2 = \begin{bmatrix}
        G_1 & 0 \\ 0 & G_1
    \end{bmatrix},
    \label{eq:G2factorize}
\end{equation}
that is, the two copy saddle $G_2$ is two uncorrelated (zeros on the off-diagonals) copies of the single copy saddle $G_1$. We expect an analogous situation for the purity calculation, namely that the saddle point which computes $\mathbb{E}_J[ \mathrm{tr}(M_K \rho \otimes \rho)  \mathrm{tr}(M_K \rho \otimes \rho) ] $ will consist of two uncorrelated copies of the saddle which computes $ \mathbb{E}_J[ \mathrm{tr}(M \rho \otimes \rho)   ] $. If this is true, then we will be well on the way to bounding fluctuations. There are of course many sources of $1/N$ corrections that need to be analyzed, making this an interesting topic for further study. 

In fact, these expectations are borne out in the $N \rightarrow \infty$ limit by direct numerical calculation of the second moment as we report in Appendix~\ref{app:numericalmethods}. There, we map the second moment onto a `flagpole' diagram with $k = 4$ replicas, and boundary conditions determined by the twist operators $M$. At large $N$, these diagrams can be computed numerically by iteratively solving the Schwinger-Dyson equations, similar to the methods we used in Sec. \ref{sec:syk_mi}. Our results explicitly show that the saddle-point fields $G$ factorize into two uncorrelated subsystems exactly as claimed in Eq. \eqref{eq:G2factorize} at all but the lowest temperatures (see Fig. \ref{fig:typicalitykAkB}). We therefore conclude that fluctuations about the mean are negligible and the second Renyi entropy is self-averaging for $N \rightarrow \infty$.

\subsection{Eternal wormholes and low-energy adiabatically accessible states}

It is also interesting to consider whether the no low-energy trivial states (NLTS) \cite{freedman2013quantum,eldar2017local,anshu2023nlts} condition holds for the SYK model and its low-rank generalizations. We have not succeeded in answering that question here, but we have answered a related question in which the notion of trivial states is modified from a circuit definition to an adiabatic definition. We will discuss our results and compare and contrast them with NLTS in this subsection.

To properly explain all this, let us first consider the NLTS property as normally formulated and the important notion of a trivial state. This is a state on $N$ qubits that can be produced by a constant depth circuit. Informally, a Hamiltonian on $N$ qubits has the NLTS property if there is an energy per qubit $\epsilon_*$ such that all states with energy less than $E_{\text{gs}} + \epsilon_* N$ are non-trivial, i.e. no state with energy less than $E_{\text{gs}} + \epsilon_* N$ can be produced from a constant depth circuit. Naturally, we need to be careful about quantifying all this and considering a family of Hamiltonians indexed by $N$. 

What is known about the NLTS property for familiar systems? One very general point is that no quantum many-body system in finite-dimensional Euclidean space can have the NLTS property. However, the recently discovered good quantum LDPC codes do have the NLTS property~\cite{anshu2023nlts}. This is no contradiction because these codes are defined on expander-graph-like objects in which the ``surface area'' of a region scales as its ``volume''. SYK and other mean-field type models also have highly connected degrees of freedom and cannot be viewed as living in a finite-dimensional Euclidean space, so NLTS is not immediately ruled out for such models.

In the recently defined fermionic version of NLTS~\cite{herasymenko2023fermionic}, a state is trivial if it can be obtained from a constant depth fermionic circuit supplemented by free Gaussian operations acting on at most $N$ ancilla. So all the energy eigenstates of the $q=2$ SYK model are trivial because they are all Gaussian. In light of our main results about the code properties of SYK and the results of~\cite{anshu2023nlts}, it is natural to wonder if the NLTS property holds for some $\epsilon_*$ for the SYK model with general $q>2$.

As stated at the beginning of this subsection, we have not been able to answer this question for SYK. What we have been able to show is that SYK possesses ``low-energy adiabatically accessible states'' (LAS). A target state $|\psi\rangle$ is adiabatically accessible if there exists a family of gapped Hamiltonians $H(\mu)$, indexed by a tunable parameter $\mu$, such that:
\begin{itemize}
    \item $H(\mu)$ obeys all relevant constraints, e.g. few-body terms, bounded interaction strengths, and, in some cases, geometrical locality, 
    \item for all $\mu$, the ground state of $H(\mu)$ is unique and the gap to the first excited remains bounded away from zero in the thermodynamic limit,
    \item the ground state of $H(\mu_0)$ is a product state,
    \item and the ground state of $H(\mu_1)$ is the target state $|\psi\rangle$.
\end{itemize}
For SYK we find that for any fixed $\epsilon>0$ as $N \to \infty$, there is a state with average energy $E_{gs} + \epsilon N$ which is adiabatically accessible. The minimum gap in the corresponding Hamiltonian $H(\mu)$ will decrease as $\epsilon$ is decreased, but it remains non-zero for fixed $\epsilon$ as $N \to \infty$. Hence, SYK fails to have the ``no low-energy adiabatically accessible states'' (NLAS) property.

Specifically, we can construct a family of states $|\mu\rangle$ defined on two copies ($L$ and $R$) of the system such that:
\begin{itemize}
    \item $|\mu\rangle$ is the unique ground state of a LR coupled Hamiltonian $H_{\text{MQ}}(\mu)$ which has a non-vanishing gap to the first excited state (the gap depends on $\mu$, but does not vanish in the thermodynamic limit),
    \item at large $\mu = \mu_0$, $|\mu \rangle$ is a Gaussian state equivalent to fermionic ``Bell pairs'' connecting L and R (the infinite temperature TFD),
    \item at small $\mu = \mu_1$, $|\mu\rangle$ is close to the low temperature TFD state at an inverse temperature $\beta(\mu)$, and in particular, from the perspective of the Hamiltonian of a single copy, its energy per particle relative to the ground state can be made arbitrarily small.
\end{itemize}

Hence, we can prepare states of arbitrarily low energy density by an adiabatic evolution starting from an infinite temperature TFD state. The ``depth'' of course grows as the energy density decreases, but for any fixed energy density as $N \to \infty$, we only need a finite amount of adiabatic evolution. Our analysis makes heavy use of the Maldacena-Qi eternal wormhole model~\cite{maldacena2018eternal} (the $H_{\text{MQ}}$ above) and is discussed in detail in App.~\ref{app:nlts}. 

Now, how does NLAS relate to NLTS? In many contexts, especially in finite dimensional Euclidean lattices where one has Lieb-Robinson bounds, adiabatic accessibility is regarded as physically equivalent to triviality. They are not strictly equivalent in general, for example, converting an adiabatic path to a finite depth circuit using quasi-adiabatic continuity and Trotter formulas incurs some error due to truncation of tails in the interactions. Nevertheless, they are often used as interchangeable notions, such as when discussing the classification of topological phases of matter. In fact, in the context of finite-dimensional Euclidean systems, we think it can be shown that the existence of LAS implies the existence of LTS since one only needs to prepare a state of low energy density. As already noted, in the Euclidean context there are other more general arguments against NLTS, but it is nevertheless interesting to explore the conditions under which LAS implies LTS.

In the context of SYK, or other mean-field models, the connection between triviality and adiabatic continuity is far less clear. Let us explore the situation taking the $q=4$ SYK model as an example. Consider using a Trotter formula to digitize an adiabatic evolution. Supposing, as in our construction, that the Hamiltonian $H(\mu)$ consists of 2- and 4-body terms, then because the model has $O(N^4)$ terms, any product formula for a small time step will also have $O(N^4)$ gates. Since a finite depth circuit in which each acts on a bounded number of degrees of freedom can have at most $O(N)$ gates, a product formula approximation is immediately outside the finite-depth class. However, this objection strikes us as too fast because while there are many gates, each gate is close to the identity since the strength of each of the $O(N^4)$ 4-body terms is of order $1/N^{3/2}$.

Indeed, in many physics contexts, such as scrambling and entanglement growth, the amount of time one evolves with an SYK Hamiltonian is physically similar to the depth of a random circuit in which each layer of gates couples randomly chosen pairs of qubits (no geometric locality). For example, both models take a time/depth of order $\ln N$ to fully scramble in the sense of out-of-time-order correlators and to produce fully entangled subsystems starting from a product state.

Hence, depending on the context, SYK time can be similar to circuit depth. And while product formulas do not by themselves give a finite depth circuit in the case of a finite amount of SYK time evolution, there are hints that perhaps one can do better. We do not know if this possible, but we think it is an important open problem in the field.

\subsection{Supersymmetric SYK}

Here we discuss the extension of our results to the case of supersymmetric (SUSY) SYK. Ref.~\cite{susy_syk} introduced two models, a $\mathcal{N}=1$ SUSY model in which the system possesses a real supercharge $Q_{\mathcal{N}=1}$ and the Hamiltonian is $H = \{ Q, Q \}$ and a $\mathcal{N}=2$ SUSY model in which the system possesses complex supercharge $Q_{\mathcal{N}=2}$ (equivalently, two real supercharges) and the Hamiltonian is $H = \{ Q_{\mathcal{N}=2} , Q_{\mathcal{N}=2}^\dagger \}$. In the $\mathcal{N}=1$ model, the ground space degeneracy is still approximate, although the splitting between levels is substantially smaller than in SYK. In the $\mathcal{N}=2$ model, the ground state degeneracy is exact and the ground states have exactly zero energy. In the vernacular, SUSY is spontaneously broken by non-perturbative effects in the $\mathcal{N}=1$ model and unbroken in the $\mathcal{N}=2$ model.

Here we will focus on the interesting case of the $\mathcal{N}=2$ model as it brings the most new features to the discussion. Henceforth, we denote the complex supercharge simply as $Q$. The model is defined on $N$ complex fermions $\psi_a$ (as opposed to the real fermions of SYK), and these obey the algebra
\begin{equation}
    \{ \psi_a , \psi_b^\dagger \} = \delta_{ab}.
\end{equation}
The supercharge is a random variable and has the form
\begin{equation}
    Q = i \sum_{a<b<c} C_{abc} \psi_a \psi_b \psi_c
\end{equation}
where $C_{abc}$ are complex Gaussian random variables with mean zero and variance $\mathbb{E}(C_{abc}C_{abc}^*) = \frac{2J}{N^2}$. The Hamiltonian is
\begin{equation}
    H = \{ Q, Q^\dagger \}
\end{equation}
where
\begin{equation}
    Q^\dagger = i \sum_{a <b<c} C_{abc}^* \psi_a^\dagger \psi_b^\dagger \psi_c^\dagger.
\end{equation}

One can explicitly compute the anti-commutator defining $H$ to find that the Hamiltonian still consists of 4-body interactions made up of complex fermions. We also note that the model as a $U(1)$ symmetry, an ``R-symmetry'' in the language of SUSY, so we can consider energy eigenstates to also have a definite R-charge. The fermions conventionally have charge $1/3$ under this $U(1)$, so the supercharge has charge $1$. We also note that the model can be generalized to include $\hat{q}$ fermions in the supercharge; we are considering the $\hat{q}=3$ case but any odd integer greater than $1$ yields a model with similar properties.

Like the SYK model we focused on in the main body of the paper, this SUSY model has a degenerate ground space, a dual gravitational description at low energy, and fermions which are scaling operators albeit with a different scaling dimension of $\Delta=1/6$ (compared to $\Delta=1/4$ for non-SUSY SYK). Moreover, the microscopic calculations using twist fields and fermion determinants will work in a similar way (Section~\ref{sec:syk_mi}) and the gravitational picture is also similar (Section~\ref{sec:gravity_mi}) although the bulk now also contains a gauge field corresponding to the R symmetry. Hence, we expect all the calculations reported in the main body to work similarly for SUSY SYK provided we set $\Delta=1/6$ and stick to temperatures of order $T \sim 1/N$.

So far, the discussion has been quite parallel between non-SUSY and SUSY models. Nevertheless, the SUSY model has two new features that make it, in some ways, an even simpler example of an approximate code:
\begin{itemize}
    \item the ground state energy is zero (the smallest possible value since the Hamiltonian is non-negative in the SUSY case) and the ground state degeneracy is exact instead of approximate,
    \item and there is a gap of order $1/N$ separating the exact the ground space from the set of excited states~\cite{stanford_ferm_loc}.
\end{itemize}
With these properties in hand, many simplifications occur, for example, the choice of code space is more natural---just take the entire ground space. Moreover, direct calculation of correlators within the ground space are possible~\cite{lin2023looking} and one finds that the extrapolations discussed above to $\beta \sim N$ correctly give the order of magnitude of these correlations.

It is therefore quite interesting to make a detailed study of the SUSY SYK code. For example, the fact that the ground space is annihilated by both $Q$ and $Q^\dagger$ may simplify the description of the code space. We already have unpublished exact diagonalization results on this model and hope to return to a more detailed study soon.

%%%%%%%

\subsection{Comparison to ETH codes}

It is also instructive to compare our work here to the notion of eigenstate thermalization hypothesis (ETH) codes~\cite{brandao_eth_qecc,bao2019eigenstate}, since both constructions consider specially chosen energy eigenstates to construct the approximate codes. For example, Theorem 1 in~\cite{brandao_eth_qecc} considers a situation in which energy eigenstates in an $n$-qubit spin chain are randomly chosen from an energy window of size $n \pm \sqrt{n}$. ETH proposes that the off-diagonal matrix elements of ``simple'' operators between these states are small~\cite{PhysRevE.50.888,PhysRevA.43.2046}, and from this one can deduce various code properties. Theorem 1 of~\cite{brandao_eth_qecc} gives a fully rigorous version of this construction for a translation invariant spin chain in the case in which the number of randomly chosen eigenstates is of order poly$(n)$. There are many other interesting regimes one can consider and the ETH can, at least heuristically, give approximate codes with excellent properties.

In contrast, in our work we are considering the entire set of energy eigenstates within a certain energy window, leading to a much larger code space and comparatively weaker distance results. To give one perspective on the comparison, suppose we carried out a version of our construction at general inverse temperature $\beta = 1/T$ in a translation invariant spin chain. We take all the energy eigenstates in some energy band, construct the microcanonical thermofield double, and study its mutual information. Just from the correlations of local fields, we would expect a mutual information which is proportional to the sub-system size on the left and to some power of the left-right correlation, structurally similar to our results for SYK. This correlation is finite at any finite $\beta$ (not scaling with system size). In our case, we can maintain the very large number of states in the code space while decreasing this correlation by going to very low temperature, but in the spin chain case we would not typically expect an extensive approximately degenerate ground space.

It is also interesting to ask if we can study situations with a smaller code space, to get closer to the ETH codes discussed in~\cite{brandao_eth_qecc}. In the way we have approached the problem here, this might be achieved by making the temperature much smaller than $1/N$, for example, exponentially small in $N$. In the $\mathcal{N}=2$ SUSY case, this actually doesn't change anything because the ground space is exactly degenerate and the matrix elements in the ground space already reproduce thermal correlations with $\beta \sim N$. However, in the non-SUSY case, lowering the temperature beyond $1/N$ does lead to a further slow decrease of the entropy and correlations.

We will present one simple speculation for how this might go, but we emphasize that it remains an open problem to give a controlled analysis. The idea is to appeal to the so-called quantum Schwarzian regime, obtained when $\beta \gg N$. In this regime, the entropy receives an important 1-loop correction~\cite{maldacena2016remarks} and takes the form
\begin{equation}
    S(\beta) = s_0 N - \frac{3}{2} \ln \beta J.
\end{equation}
The fermion thermal correlator is also modified to~\cite{bagrets_syk_liouville}
\begin{equation}
    G(\beta/2) \sim (\beta J)^{-3/2}.
\end{equation}
If we make the uncontrolled guess that the mutual information continues to be proportional to $G(\beta/2)$, then we find an interesting result. 

Consider a $\beta$ such that the code space size is 
\begin{equation}
    e^{S(\beta)} = e^{s_0 N( 1- f)},
\end{equation}
i.e. some exponentially small fraction of the approximately degenerate ground space. Plugging the corresponding $\beta$ into the thermal correlator gives an exponentially small correlation,
\begin{equation}
    G(\beta/2) \sim e^{- f s_0 N}.
\end{equation}
Including the factor of $K$, this now indicates that the mutual information, $K e^{- f s_0 N}$, is small for any $K$ up to some fraction of $N$. This looks like a linear distance, although we must also be concerned that ETH might break down or become modified when $K$ is large. Indeed, the mutual information will become large at sufficiently large $K$, since it must reproduce the full entropy when $K=N$. In the holographic setting, for example, the analog of the QES might shift to a qualitatively different location, although none of these concepts is clearly defined in this deeply quantum regime. It is therefore interesting to investigate this regime in more detail.

As one check of this picture, suppose we push $\beta$ to the extreme regime where $f=1$ and the code space contains only a small number of very low-lying states. The mutual information is predicted to be $K e^{- s_0 N}$, which is precisely what one would expect from a direction calculation of, say, a pair of energy eigenstates with simple matrix elements of order $e^{- s_0 N/2}$. For example, in the state $|\text{TFD}\rangle = |E_0\rangle_L |0\rangle_R + |E_1\rangle_L |1\rangle_R$, the left-right correlation between a single qubit operator $O_L$ on the left and the operator $O_R = |0\rangle \langle 1| + |1\rangle \langle 0 |$ on the right will be of order $\langle O_L O_R \rangle_{\text{TFD}} \sim e^{- s_0 N/2}$, and since the mutual information bounds correlators as $\text{MI} \geq \langle O_L O_R \rangle_{\text{TFD}}^2$, we expect an MI of order $e^{-s_0 N}$.

\subsection*{Acknowledgements}

We thank Tiangang Zhou, Isaac Kim, Shaokai Jian, Valerie Bettaque, Alexey Milekhin, and Stefano Antonini for useful discussions and very helpful feedback on the manuscript. We acknowledge support from the AFOSR under grant number FA9550-19-1-0360 and the Department of Energy under grant number DESC0019380. 

\bibliographystyle{unsrturl}
\bibliography{References}

\appendix

\section{The KL conditions, mutual information, and exact recoverability}
\label{app:wormhole-kl}

In this appendix we record simple proofs that the KL conditions~\cite{kl} are equivalent to the mutual information diagnostic and the trace distance diagnostic for an exact code.

\subsection*{KL $\to$ MI}

Consider a code $C \subset V$ which obeys the KL condition for all error operators $E_a$ up to weight $t$, meaning
\begin{equation}
    \langle \phi_i | E_a^\dagger E_b | \phi_j \rangle = \delta_{ij} c_{ab}
\end{equation}
for a basis $\{|\phi_i\rangle \}$ of $C$. Let $|\Phi\rangle_{LR}$ denote two copies of $V$ prepared in a purification of the code projector,
\begin{equation}
    |\Phi \rangle = \frac{1}{\sqrt{\dim C}} \sum_i |\phi_i \rangle_L |\phi_i \rangle_R.
\end{equation}
Finally, let $A$ be a subset of $L$ containing $2t$ or fewer qubits. 

We would like to calculate $I(A:R)$. We do this by computing the full density matrix $\rho_{AR}$. We expand $\rho_{AR}$ in a complete basis as
\begin{equation}
    \rho_{AR} = \sum_{P_A, ij} c(P_A, ij) P_A \otimes |\phi_i \rangle \langle \phi_j |_R
\end{equation}
where $P_A$ denotes Pauli strings that identity on the complement of $A \subset L$ and the $c$s are a set of expansion coefficients. These coefficients are determined by
\begin{equation}
    c(P_A, ij) = \langle \Phi | ( P_A \otimes |\phi_j \rangle \langle \phi_i |_R) | \Phi \rangle.
\end{equation}
From the explicit form of $|\Phi\rangle$, we obtain
\begin{equation}
    c(P_A,ij) = \langle \phi_i | P_A | \phi_j \rangle = \delta_{ij} c(P_A)
\end{equation}
where the second equality follows from the KL condition since $P_A$ as weight $\leq 2t$. 

Plugging this formula into the expression for the density matrix, we learn that 
\begin{equation}
    \rho_{AR} \propto \rho_A \otimes \left( \sum_i |\phi_i \rangle \langle \phi_i |_R \right)
\end{equation}
is factorized between $A$ and $R$. This implies the mutual information $I(A:R)$ is zero for all $A$ of size $\leq 2t$.

\subsection*{MI $\to$ KL}

Now consider the same setup but suppose the mutual information between $A$ and $R$ is zero for all $A$ of size $\leq d-1$. The mutual information bounds connected correlations between operators in $A$ and $R$, so since the mutual information is zero we have
\begin{equation}
    0 = \langle P_A \otimes O_R \rangle - \langle P_A \rangle \langle O_R\rangle
\end{equation}
for any operators $P_A$ and $O_R$.

Consider $O_R = |\phi_i\rangle \langle \phi_j |$ for $i\neq j$. Since $\langle O_R\rangle =0$ for such an operator, it follows that
\begin{equation}
  0 = \langle P_A \otimes O_R \rangle =   \langle i | P_A | j\rangle. 
\end{equation}

Now consider $O_R = |\phi_i \rangle \langle \phi_i |$. Since $\langle O_R\rangle$ is independent of $i$ (it just equals $1/\dim C$), we learn that 
\begin{equation}
    \langle P_A \otimes O_R \rangle = \langle \phi_i | P_A | \phi_i \rangle = c(P_A),
\end{equation}
where the last quantity is independent of $i$. 

Putting these two things together, we learn that
\begin{equation}
    \langle \phi_i | P_A | \phi_j \rangle = \delta_{ij} c(P_A),
\end{equation}
which is the KL condition.

\subsection*{KL $\to$ Trace Distance}

We can also show that the KL conditions are equivalent to vanishing trace distance
\begin{equation}
    || \left(\mathbb{I} \otimes \mathcal{D} \circ \mathcal{N}\right) \psi_{RQ} - \psi_{R Q} ||_1 = 0
\end{equation}
for a noise channel $\mathcal{N}$ and decoding channel $\mathcal{D}$, following \cite{nielsen2010quantum}. Suppose that the noise channel $\mathcal{N}$ is represented by the operation elements (Kraus operators) $\{E_i\}$, which satisfy the KL conditions
\begin{equation}
    P \adj{E}_a E_b P = c_{ab} P
\end{equation}
where $P$ is the projector onto the code subspace $C$. Because the matrix $c_{ab}$ is Hermitian by assumption, we can diagonalize the KL conditions using the unitary matrix $u_{ak}$:
\begin{equation}
    P \adj{F}_k F_l P = \delta_{kl} \lambda_{kk} P
\end{equation}
where $F_k = \sum_i u_{ak} E_a$ and $\lambda = \adj{u} c u$ is diagonal. Using a polar decomposition, we find
\begin{equation}
    F_k P = U_k \sqrt{P \adj{F}_k F_k P} = \sqrt{\lambda_{kk}} \ U_k P
\end{equation}
for some unitary operator $U_k$. We take the operators $R_k = P \adj{U}_k = P \adj{F}_k / \sqrt{\lambda_{kk}}$ to be the operation elements of the decoding channel $\mathcal{D}$. The combined noise and decoding channels therefore act as
\begin{equation}
    \mathcal{D} \circ \mathcal{N} (\rho) = \sum_{kl} P \adj{U}_k F_l \rho \adj{F}_l U_k P 
\end{equation}
Then, for states $\psi_Q = P \psi_Q P$ in the code subspace $C$, we have
\begin{equation}
    P \adj{U}_k F_l P \sqrt{\psi_Q} = \frac{1}{\sqrt{\lambda_{kk}}} P \adj{F}_k F_l P \sqrt{\psi_Q} = \delta_{kl} \sqrt{\lambda_{kk}} P \sqrt{\psi_Q}
\end{equation}
and therefore
\begin{equation}
    \mathcal{D} \circ \mathcal{N} (\psi_Q) = \sum_{kl} \delta_{kl} \lambda_{kk} \ \psi_Q = \psi_Q
\end{equation}
which means that the trace distance vanishes exactly, as required.

\subsection*{Trace Distance $\to$ KL}

Conversely, suppose that the trace distance vanishes; because it is a distance measure, this implies that
\begin{equation}
    \left(\mathbb{I} \otimes \mathcal{D} \circ \mathcal{N}\right) \psi_{RQ} = \psi_{R Q}
\end{equation}
which also implies that $\mathcal{D} \circ \mathcal{N} (P \psi_{Q} P) = P \psi_Q P$
where $P$ is the projector onto the code subspace.
Take $\{E_a\}$ and $\{R_i\}$ to be the operation elements (Kraus operators) of the noise channel $\mathcal{N}$ and recovery channel $\mathcal{D}$, respectively. Then we have
\begin{equation}
    \sum_{aj} R_j E_a P \psi_Q P \adj{E}_a \adj{R}_j = c P \psi_Q P
\end{equation}
Due to the unitary freedom of the operator-sum representation, this implies that there exist complex numbers $d_{ki}$ such that
\begin{equation}
    R_k E_a P = d_{ka} P
\end{equation}
Then, summing over $k$ and using the fact that $\mathcal{D}$ is trace-preserving, we conclude that
\begin{equation}
    \sum_k P \adj{E}_a \adj{R}_k R_k E_b P = \sum_k d^*_{ka} d_{kb} P
\end{equation}
and
\begin{equation}
    P \adj{E}_a E_b P = c_{ab} P
\end{equation}
where $c_{ab} = \sum_k d^*_{ka} d_{kb}$ is a Hermitian matrix. These are the KL conditions.

\section{Mutual information and approximate recoverability}
\label{app:approxrecovery}

In this appendix, we show how the smallness of the mutual information $I(A:R)$ is rigorously connected to standard definitions of recoverability for approximate error-correcting codes. In particular, a standard definition for approximate quantum error correction in the literature \cite{yi2023complexity,hayden2020approximate} is to measure recoverability via the trace norm: for any noisy quantum channel $\mathcal{N}$, we say that quantum information encoded into a code state $\rho_Q$ is $\delta$-approximately recoverable if there exists a decoding map $\mathcal{D}$ such that
\begin{equation}
    || \left(\mathbb{I} \otimes \mathcal{D} \circ \mathcal{N}\right) \psi_{RQ} - \psi_{R Q} ||_1 < \delta \label{eq:appcodedef}
\end{equation}
for some small $\delta$, where $\psi_{RQ}$ is a purification of the code state $\rho_Q = \mathrm{tr}over{R}{\psi_{RQ}}$. Throughout, we shall assume that the noise channel has the form $\mathcal{N} = \mathcal{N}_A \otimes \id_B$, acting only on a subset $A$ of the qubits, where $\magn{A}$ is the distance of the code. Here we will prove that the smallness of the mutual information
\begin{equation}
    I(A:R) < \epsilon
\end{equation}
implies the smallness of the trace norm Eq. \eqref{eq:appcodedef}, where $\delta = c' \epsilon^{1/4}$ and $c'$ is an $\mathcal{O}(1)$ constant. We shall also prove the converse, namely that a nonzero mutual information implies a nonzero trace norm for some choice of noise channel $\mathcal{N}$.

\begin{figure}
    \centering
    \includegraphics[width=\textwidth]{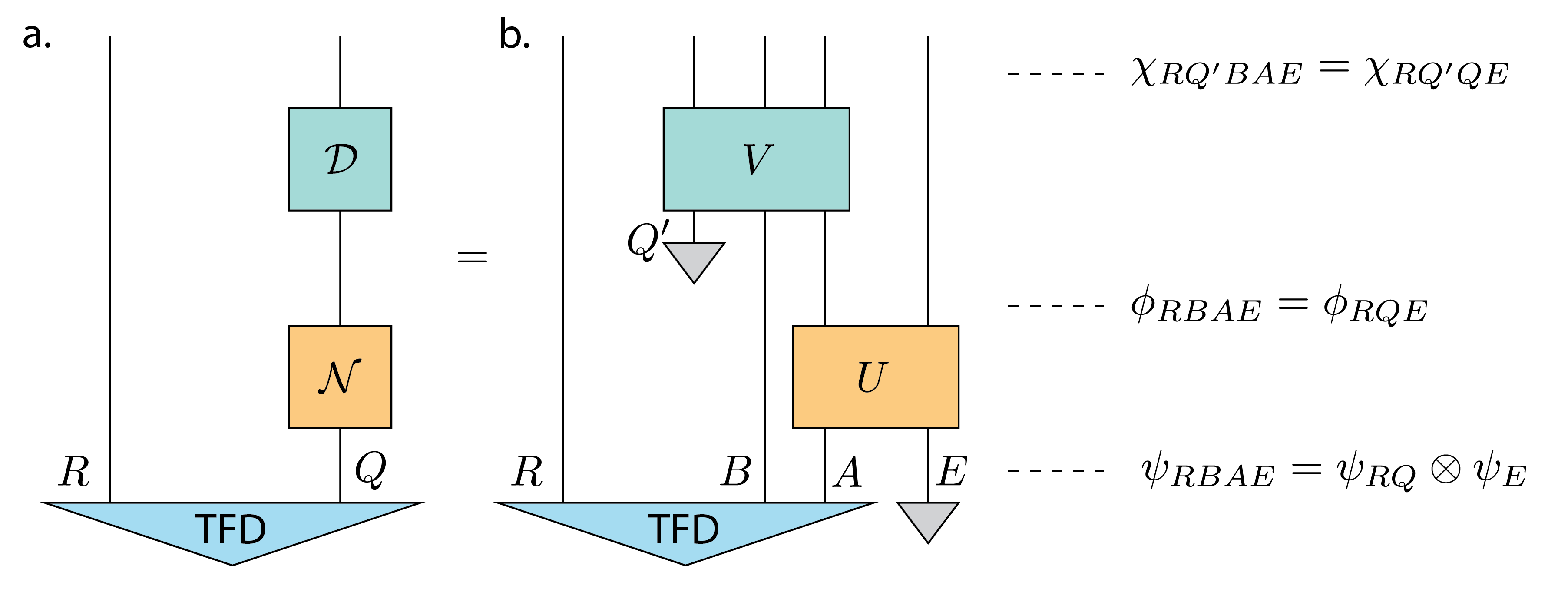}
    \caption{Two equivalent circuit representations of the noisy channel and decoding process for our approximate error-correcting code. In (a) we show the system $Q$ and its purification $R$, where the system is subject to the noisy channel $\mathcal{N}$ followed by the decoding channel $\mathcal{D}$ (both of which may be non-unitary). In (b) we expand the system into two parts $Q = A \cup B$ and introduce an environment $E$ to represent the noisy channel $\mathcal{N}$ as a unitary operation $U$, where the environment is subsequently traced out. Similarly, we introduce an auxiliary system $Q'$ to represent the decoding channel $\mathcal{D}$ as a unitary operation $V$. We exclusively use different symbols $\psi,\phi,\chi$ to represent states at different points in the circuit, where $\psi_{RBAE}$ is the pure state at the bottom of the circuit, $\phi_{RBAE}$ is the pure state after applying the unitary $U$, and $\chi_{RQ'BAE}$ is the pure state after applying the unitary $V$.}
    \label{fig:aqecccircuit}
\end{figure}

To follow the details of the proof, we will refer to the circuit representation of the encoding and decoding process shown in Fig. \ref{fig:aqecccircuit}. In Fig. \ref{fig:aqecccircuit}.a. we illustrate a simplified version of the circuit that only shows the system $Q$ and its purification $R$, and where the noisy channel $\mathcal{N}$ and the decoding channel $\mathcal{D}$ only act on the system $Q$. In Fig. \ref{fig:aqecccircuit}.b. we split the system into two parts $Q = A \cup B$ and assume that the noisy channel only acts on the subsystem $A$: $\mathcal{N} = \mathcal{N}_A \otimes \mathbb{I}_B$. We also introduce an `environment' $E$ that allows us to represent the noisy channel $\mathcal{N}_A$ as a unitary operator $U$ acting on $AE$ followed by tracing out the environment. Similarly, we introduce an auxiliary system $Q'$ to represent the decoding channel $\mathcal{D}$ as a unitary operation $V$. At each stage of the circuit, we exclusively use different symbols $\psi,\phi,\chi$ to keep track of which state we are talking about, where $\psi_{RBAE}$ is the pure state at the bottom of the circuit, $\phi_{RBAE}$ is the pure state after applying the unitary $U$, and $\chi_{RQ'BAE}$ is the pure state after applying the unitary $V$.

We begin by defining our terms and setting notation. The von Neumann entropy for a state $\rho$ is $S(\rho) = - \mathrm{tr}(\rho \log \rho)$. We will sometimes use the shorthand $S_X \equiv S(\rho_X)$, where $\rho_X$ is the reduced state on subsystem $X$. The mutual information between subsystems $X,Y$ for a state $\psi$ is $I(X:Y)_{\psi} = S(\psi_X) + S(\psi_Y) - S(\psi_{XY})$. The fidelity between two states $\rho,\sigma$ is $f(\rho,\sigma) = \mathrm{tr}(\sqrt{\rho^{1/2} \sigma \rho^{1/2}})$, and the trace distance between these two states is $1/2 ||\rho - \sigma||_1$, where $||A||_1 = \mathrm{tr}(\sqrt{A^{\dagger} A})$ is the trace norm. These two quantities are related by the Fuchs-van de Graaf inequalities:
\begin{equation}
    1-f(\rho,\sigma) \leq \frac{1}{2} ||\rho-\sigma||_1 \leq \sqrt{1 - f^2(\rho,\sigma)}
\end{equation}
which we will use several times in the proof below. Finally, we define the coherent quantum information between two subsystems in terms of the conditional entropy $I_c(X;Y)_{\psi} = - S(X|Y)_{\psi} = -S(\psi_{XY}) + S(\psi_Y)$.

The proof proceeds in four steps:
\begin{enumerate}
    \item $I(R:A)_{\psi} < \epsilon \implies I(R:E)_{\phi} < \epsilon$ for all noisy channels $\mathcal{N} = \mathcal{N}_A \otimes \mathbb{I}_B$ acting on the subsystem $A$. In other words, after information leaks from $A$ to the environment $E$, we cannot have a larger mutual information between $E$ and $R$ than we had previously between $A$ and $R$.
    \item $I(R:E)_{\phi} < \epsilon \implies || \phi_{RE} - \phi_R \otimes \phi_E||_1 < \tilde{\epsilon} = c \epsilon^{1/2}$ with $c$ an $O(1)$ constant. In other words, the smallness of mutual information between $R,E$ implies that the state $\phi_{RE}$ is approximately separable, or {\it decoupled} \cite{schumacher2002approximate}.
    \item $|| \phi_{RE} - \phi_R \otimes \phi_E||_1 < \tilde{\epsilon} \implies f(\left(\mathbb{I} \otimes \mathcal{D} \circ \mathcal{N} \right) \psi_{RQ}, \psi_{RQ}) > 1-\frac{1}{2}\tilde{\epsilon}$. In other words, the approximate separability of $\phi_{RE}$ implies that the initial state $\psi_{RQ}$ is approximately recoverable. Or, more succinctly, {\it approximate decoupling implies approximate recoverability} \cite{preskill1998lecture}.
    \item $f(\left(\mathbb{I} \otimes \mathcal{D} \circ \mathcal{N} \right) \psi_{RQ}, \psi_{RQ}) > 1 - \frac{1}{2}\tilde{\epsilon} \implies ||\left(\mathbb{I} \otimes \mathcal{D} \circ \mathcal{N} \right) \psi_{RQ} - \psi_{RQ}||_1 < \delta = \tilde{c} \tilde{\epsilon}^{1/2} = c' \epsilon^{1/4}$, where $\tilde{c},c'$ are $O(1)$ constants, which follows directly from the Fuchs-van de Graaf inequalities.
\end{enumerate}
It remains to prove steps 1,2, and 3.

Step 1 follows from the monotonicity of the mutual information, i.e. the fact that the mutual information never increases under quantum channels (CPTP maps) \cite{nielsen2010quantum}. We have:
\begin{eqnarray}
    \epsilon > I(R:A)_{\psi} &= I(R:AE)_{\psi} \nonumber \\
    &\geq I(R:AE)_{\phi} \nonumber \\
    &\geq I(R:E)_{\phi}
\end{eqnarray}
The first line follows from the fact that $\psi_{RAE} = \psi_{RA} \otimes \psi_E$, the second line from the fact that mutual information doesn't increase under the quantum channel $U$, and the third line from the fact that the mutual information doesn't increase under the partial trace over $A$, which is also a quantum channel.

Step 2 is a result of Schumacher and Westmoreland \cite{schumacher2002approximate}, and is equivalent to the quantum Pinsker's inequality. First, recall that $I(R:E)_{\phi} = S(\phi_{RE} || \phi_R \otimes \phi_E)$ where $S(\rho || \sigma) = \mathrm{tr}(\rho \log \rho) - \mathrm{tr}(\rho \log \sigma)$ is the relative entropy. Then the quantum Pinsker's inequality states that
\begin{equation}
    S(\rho || \sigma) \geq \frac{1}{2 \ln 2} ||\rho - \sigma ||_1^2
\end{equation}
which establishes the desired result.

Step 3 is really the crux of the argument, and is a result presented in Preskill's quantum information notes \cite{preskill1998lecture}. We review it here for completeness, using the notation presented in Fig. \ref{fig:aqecccircuit}. The purification of $\phi_{RE}$ is $\phi_{RQE}$ by construction. We similarly construct a purification $\tilde{\phi}_{RQE}$ of the separable state $\phi_R \otimes \phi_E$, which always has the form
\begin{equation}
    \tilde{\phi}_{RQE} = W_Q\left( \tilde{\phi}_{RQ_1} \otimes \tilde{\phi}_{Q_2E} \right)
\end{equation}
where $W_Q$ is an arbitrary unitary operation acting on $Q$ and $Q = Q_1 \cup Q_2$ is a decomposition of the system Hilbert space. By Uhlmann's theorem \cite{nielsen2010quantum}, we are free to choose the unitary operation $W_Q$ such that the fidelity is simply an overlap between pure states:
\begin{equation}
    f(\phi_{RE}, \phi_R \otimes \phi_E) = \left| \bracket{\phi_{RQE}}{\tilde{\phi}_{RQE}} \right| = f(\phi_{RQE},\tilde{\phi}_{RQE})
\end{equation}
for some particular choice of $W_Q$. Then, by the Fuchs-van de Graaf inequalities, we have
\begin{equation}
    f(\phi_{RQE},\tilde{\phi}_{RQE}) = f(\phi_{RE}, \phi_R \otimes \phi_E) \geq 1 - \frac{1}{2}||\phi_{RE} - \phi_R \otimes \phi_E ||_1 > 1 - \frac{1}{2}\tilde{\epsilon}.
\end{equation}
Next, the fidelity is monotonic under the quantum operations $\mathrm{tr}over{E}{\cdot}$ and $(\mathbb{I} \otimes \mathcal{D})(\cdot)$, so we have:
\begin{equation}
    f((\mathbb{I} \otimes \mathcal{D})\phi_{RQ},(\mathbb{I} \otimes \mathcal{D})\tilde{\phi}_{RQ}) > 1 - \frac{1}{2}\tilde{\epsilon}
\end{equation}
By definition we have that $(\mathbb{I} \otimes \mathcal{D}) \phi_{RQ} = (\mathbb{I} \otimes \mathcal{D} \circ \mathcal{N}) \psi_{RQ}$. And finally, we may choose the decoding map $\mathcal{D}$ such that $(\mathbb{I} \otimes \mathcal{D})\tilde{\phi}_{RQ} = \psi_{RQ}$, which follows from the principle that {\it exact decoupling implies exact recoverability.} To show this, recall that $\tilde{\phi}_{RQE} = W_Q\left( \tilde{\phi}_{RQ_1} \otimes \tilde{\phi}_{Q_2E} \right)$ is the purification of $\phi_R \otimes \phi_E$. This allows us to construct an isometric decoder $\mathcal{D} = S \adj{W}_Q$ for some choice of $S$, which transfers the purification of $R$ onto $Q_1 \subset Q$. Because all purifications of $R$ differ by an isometry, we may choose $S$ to output the desired state $\psi_{RQ}$. We therefore have
\begin{equation}
    f((\mathbb{I} \otimes \mathcal{D} \circ \mathcal{N})\psi_{RQ},\psi_{RQ}) > 1 - \frac{1}{2}\tilde{\epsilon}
\end{equation}
as desired.

We can also establish the converse, that nonzero mutual information implies nonzero trace norm for some choice of noise channel $\mathcal{N}$. Suppose that $I(R:A)_{\psi} > \epsilon > 0$. Then there exists a noise channel $\mathcal{N} = \mathcal{N}_A \otimes \id_B$ such that $I(R:E)_{\phi} > \epsilon > 0$. For example, choose the unitary gate $U$ to be the SWAP operator that simply exchanges $A$ with $E$, and we have $I(R:A)_{\psi} = I(R:E)_{\phi} > \epsilon > 0$. Next, the coherent quantum information between $R,Q$ is
\begin{equation}
    I_c(R;Q)_{\phi} = S(\phi_Q) - S(\phi_{RQ}) = S(\phi_{RE}) - S(\phi_E) < S(\phi_R) - \epsilon
\end{equation}
where in the second equality we have used $S(\phi_{RQ}) = S(\phi_E)$ and $S(\phi_Q) = S(\phi_{RE})$ since $\phi_{RQE}$ is pure, and in the final inequality we have used the definition of the mutual information $I(R:E)_{\phi}$. Next, by the data processing inequality \cite{nielsen2010quantum}, we know that the coherent information of the final state $\chi$ is no larger than the coherent information in the earlier state $\psi$:
\begin{align}
    I_c(R;Q)_{\chi} &= S(\chi_Q) - S(\chi_{RQ}) = S(\chi_{RQ'E}) - S(\chi_{Q'E}) \nonumber \\
    &\leq I_c(R;Q)_{\phi} < S(\phi_R) - \epsilon = S(\chi_R) - \epsilon
\end{align}
where in the first equality we have used $S(\chi_Q) = S(\chi_{RQ'E})$ and $S(\chi_{RQ}) = S(\chi_{Q'E})$ since $\chi_{RQ'QE}$ is pure, and in the last equality we have used the fact that $S(\chi_R) = S(\phi_R)$. By collecting terms, we therefore conclude:
\begin{equation}
    I(R:Q'E)_{\chi} > \epsilon
\end{equation}
Then, due to the monotonicity of the mutual information, we have
\begin{equation}
    I(RQ:Q'E)_{\chi} = 2 S(\chi_{RQ}) \geq I(R:Q'E)_{\chi} > \epsilon
\end{equation}
where the first equality follows from the fact that $\chi_{RQ'QE}$ is pure, so that $S_{RQ'QE} = 0$ and $S_{RQ} = S_{Q'E}$.
Hence the final state $\chi_{RQ}$ has residual entanglement with the environment $E$ and so cannot have perfect fidelity with the pure state $\psi_{RQ}$.

To prove this last fact, let us consider the fidelity between any pure state $\psi = \ket{\psi} \bra{\psi}$ and any mixed state $\chi$ with entropy $S(\chi) > \epsilon$. The fidelity is
\begin{equation}
    f(\chi,\psi) = \bra{\psi} \chi \ket{\psi}.
\end{equation}
We consider an optimization problem over density matrices $\chi$, where we wish to maximize the fidelity $f(\chi,\psi)$ subject to the constraints $S(\chi) = S > \epsilon$ and $\mathrm{tr}[\chi] = 1$. To solve this optimization problem, consider the following objective function
\begin{equation}
    I = \mathrm{tr}[\ket{\psi} \bra{\psi} \chi] - \gamma \left( -\mathrm{tr}[\chi \log \chi] - S \right) - \alpha \left( \mathrm{tr}[\chi] - 1 \right)
\end{equation}
where $\gamma, \alpha$ are Lagrange multipliers for the two constraints in the problem. To find the optimum, we look for vanishing gradients. In particular $\partial I / \partial \chi = 0$ gives:
\begin{equation}
    \ket{\psi} \ket{\psi} + \gamma \log \chi + \gamma - \alpha = 0
\end{equation}
whose solution is
\begin{equation}
    \chi = \frac{1}{Z} e^{c \ket{\psi} \bra{\psi}} = \frac{1}{Z} \left( \id + (e^c - 1) \ket{\psi} \bra{\psi} \right)
\end{equation}
for some constant $c$, where we have used the fact that $\ket{\psi} \bra{\psi}^n = \ket{\psi} \bra{\psi}$ for any integer $n \geq 1$. Enforcing the constraint $\mathrm{tr}[\chi] = 1$, we find
\begin{equation}
    Z = d-1 + e^c
\end{equation}
where $d$ is the dimension of the Hilbert space. Parameterizing the fidelity by a small parameter $\delta$, we find
\begin{equation}
    \mathrm{tr}[\ket{\psi} \bra{\psi} \chi] = 1-\delta = \left( (d-1)e^{-c} + 1 \right)^{-1}.
\end{equation}
which provides a relation between the constant $c$ and the parameter $\delta$. And finally, enforcing the constraint $S(\chi) = S$ we find
\begin{equation}
    \delta \log(d-1) + H(\delta) = S > \epsilon
\end{equation}
where $H(\delta)$ is the Shannon entropy. From this relation we see that if $\epsilon > 0$, then $\delta > 0$, and therefore the fidelity is less than unity. Finally, using the Fuchs-van de Graaf inequalities we similarly conclude that the trace distance is larger than zero.

\section{Fermion entanglement}
\label{app:fermion}

In this appendix we review the replica trick for computing Renyi entropies, and the twist field formalism of fermion entanglement, which together give meaning to expressions like $\mathrm{tr}(\rho^k)$ for $\rho$ corresponding to some subset of the fermion modes. We first review the replica trick for computing the $k$th order Renyi entropy $S^{(k)} = \frac{1}{1-k} \mathrm{tr}( \rho^k )$. Consider $k=2$ to see the basic outline of the argument: to compute $\mathrm{tr}(\rho^2)$ we introduce two copies, or `replicas,' of the state $\rho$ and compute the expectation value of the twist, or `SWAP,' operator $M_2$ \cite{ekert2002direct,daley2012measuring}. By definition the twist operator $M_2$ swaps the states belonging to the two replicas:
\begin{equation}
    M_2 \ket{\psi_1} \otimes \ket{\psi_2} = \ket{\psi_2} \otimes \ket{\psi_1}
\end{equation}
for all $\ket{\psi_{1,2}}$.
By direct computation, we therefore find:
\begin{align}
    \mathrm{tr}(\rho \otimes \rho \ M_2) &= \sum_{ijkl} \rho_{ij} \rho_{kl} \mathrm{tr}(M_2 \ket{i}\bra{j} \otimes \ket{k}\bra{l}) \nonumber \\
    &= \sum_{ijkl} \rho_{ij} \rho_{kl} \mathrm{tr}( \ket{k}\bra{j} \otimes \ket{i}\bra{l}) \nonumber \\
    &= \sum_{ijkl} \rho_{ij} \rho_{kl} \delta_{kj} \delta_{il} \nonumber \\
    &= \sum_{ij} \rho_{ij} \rho_{ji} = \mathrm{tr}(\rho^2)
\end{align}
Similarly, to compute $\mathrm{tr}(\rho^k)$ for arbitrary integer $k$, we introduce $k$ replicas and compute the expectation value of the $k$th-order twist operator $M_k$, which by definition cyclically permutes the states belonging to the $k$ replicas:
\begin{equation}
    M_k \ket{\psi_1} \ket{\psi_2} \cdots \ket{\psi_k} = \ket{\psi_k} \ket{\psi_1} \cdots \ket{\psi_{k-1}}
\end{equation}
Then a similar calculation to above shows that
\begin{equation}
    \mathrm{tr}(\rho \otimes \rho \cdots \otimes \rho \ M_k) = \mathrm{tr}(\rho^k).
\end{equation}

It remains to show how to apply this technology to fermionic systems. As in the main text, our starting point is $N$ fermion modes obeying the algebra
\begin{equation}
    \{ \chi_i , \chi_j \} = \delta_{ij},
    \label{eq:majoranaanticomm}
\end{equation}
These operators are Hermitian and satisfy $\chi_i^2 = 1/2$. 

As above, we will define fermion entanglement using the notion of twist, or `SWAP,' operators. First we introduce $k$ copies of the system and thus $kN$ fermion operators $\chi_i^a$. The upper index is a replica index and the lower index runs over fermions in a single replica. For a single fermion mode and $k=2$, the twist operator is
\begin{equation}
    M_2 =  e^{\frac{\pi}{2} \chi^{2}\chi^{1} }.
\end{equation}
One can see this explicitly by computing $M^{\dagger}_2 \chi_2 M_2$. Expanding the exponential, we find:
\begin{equation}
    M_2 = 1+ \frac{\pi}{2} \chi^2 \chi^1 + \frac{1}{2} \left(\frac{\pi}{2}\right)^2 \chi^2 \chi^1 \chi^2 \chi^1 + \cdots .
\end{equation}
By commuting the $\chi^a$'s past one another and making use of the anticommutation relations Eq. \eqref{eq:majoranaanticomm} we find:
\begin{equation}
    M_2 = \frac{1}{\sqrt{2}} + \sqrt{2} \chi^2 \chi^1.
\end{equation}
Then we have:
\begin{equation}
    M^{\dagger}_2 \chi^2 M_2 = \left( \frac{1}{\sqrt{2}} + \sqrt{2} \chi^1 \chi^2 \right) \chi^2 \left( \frac{1}{\sqrt{2}} + \sqrt{2} \chi^2 \chi^1 \right) = \chi^1.
\end{equation}
A similar calculation holds for $\chi^1$, so the operator $M_2$ implements a twist or `SWAP' $\chi_1 \leftrightarrow \chi_2$ as claimed.
In the general case of $k$ copies, the twist operator is
\begin{equation}
    M_{k} = e^{\frac{\pi}{2} \chi^{k}\chi^{k-1} } \cdots  e^{\frac{\pi}{2} \chi^{2}\chi^{1} } .
\end{equation}
which permutes the fermions cyclically.

To illustrate the formalism, consider a simple example of two correlated modes $\chi_L$ and $\chi_R$. These modes have a correlation given by
\begin{equation}
    \langle \chi_L \chi_R \rangle = \frac{i a}{2}.
\end{equation}
This correlation can be encoded in a density matrix,
\begin{equation}
    \rho = \frac{I}{2} -  i a \chi_L \chi_R,
\end{equation}
and by using $\mathrm{tr}(\chi_L \chi_R)=0$ and $\mathrm{tr}( \chi_L \chi_R \chi_L \chi_R) = -\frac{1}{2}$, we verify that
\begin{equation}
    \mathrm{tr}(\rho \chi_L \chi_R) = \frac{i a}{2}.
\end{equation}
We first compute the $\mathrm{tr}(\rho^2)$ for $L$ alone (the result is the same for $R$ alone). The twist field for $\mathrm{tr}(\rho_L^2)$ is
\begin{equation}
    M_2 = \frac{I + 2 \chi^2_L \chi^1_L}{\sqrt{2}},
\end{equation}
and the full expression is
\begin{equation}
    \mathrm{tr}(\rho_L^2) = \mathrm{tr}( \left[ \frac{I}{2} - i a \chi_L^1 \chi_R^1 \right] \left[ \frac{I}{2} - i a \chi_L^2 \chi_R^2 \right] \left[\frac{I + 2 \chi^2_L \chi^1_L}{\sqrt{2}} \right]).
\end{equation}
Any term containing a factor of $\chi_R^1$ or a factor of $\chi_R^2$ immediately vanish thanks to the trace. This leaves just the identity terms in the states. For the same reason, the $\chi_L^2 \chi_L^1$ term in the twist field vanishes. Since the total $2$-copy $LR$ Hilbert space has dimension $4$, the result is
\begin{equation}
    \mathrm{tr}(\rho_L^2) = \frac{1}{\sqrt{2}}.
\end{equation}
The second Renyi entropy is thus
\begin{equation}
    S_2(\rho_L) = \frac{1}{2}\ln 2,
\end{equation}
which is one of the ways in which a Majorana fermion is like half a qubit.

Next, we compute $\mathrm{tr}(\rho_{LR}^2)$. This will be sensitive to the correlation $a$. The twist field is now
\begin{equation}
    M_2 = \frac{I + 2 \chi^2_L \chi^1_L}{\sqrt{2}} \frac{I + 2 \chi^2_R \chi^1_R}{\sqrt{2}},
\end{equation}
and the purity is
\begin{equation}
    \mathrm{tr}(\rho_{LR}^2) = \mathrm{tr}(\left[ \frac{I}{2} - i a \chi_L^1 \chi_R^1 \right] \left[ \frac{I}{2} - i a \chi_L^2 \chi_R^2 \right] \left[\frac{I + 2 \chi^2_L \chi^1_L}{\sqrt{2}} \frac{I + 2 \chi^2_R \chi^1_R}{\sqrt{2}} \right]).
\end{equation}
There are two terms that contribute. The all identity term gives
\begin{equation}
    \mathrm{tr}(\frac{I}{2^2 \sqrt{2}^2}) = \frac{1}{2}.
\end{equation}
The all fermion term gives
\begin{equation}
    \mathrm{tr}( [-i a \chi_L^1 \chi_R^1 ] [- i a \chi_L^2 \chi_R^2 ] \sqrt{2} \chi_L^2 \chi_L^1 \sqrt{2} \chi_R^2 \chi_R^1 ) = \frac{a^2}{2}.
\end{equation}
The purity is thus
\begin{equation}
    \mathrm{tr}(\rho_{LR}^2) = \frac{1+a^2}{2}.
\end{equation}

The left-right Renyi-$2$ mutual information is 
\begin{equation}
    I_2 = S_2(L) + S_2(R) - S_2(LR) = \frac{1}{2}\ln 2 + \frac{1}{2} \ln 2 - \ln \frac{1+a^2}{2} = \ln(1 + a^2).
\end{equation}
When $a=0$, we have a maximally mixed state and the mutual information is zero. When $a=1$, we have a maximally entangled state where $\rho_{LR}$ is pure and the mutual information is $\ln 2$, which is half the value for a pair of qubits in a maximally entangled state.

This sort of state is a good model of the reduced state for pairs $\chi_{iL}$ and $\chi_{iR}$ so long of the number of pairs we are considering is not too large.

\subsubsection*{The $K_L R$ density matrix}

Next we discuss the $K_L R$ density matrix describing correlations between $K_L$ fermions on the left and the entire right. Consider first the case of $K_L=1$. The $1_L R$ density matrix is structurally similar to the $\chi_{1L}$, $\chi_{1R}$ density matrix, except that the role of $\chi_{1R}$ is instead played by an \emph{emergent fermion} $\tilde{\chi}_{1R}$. This emergent fermion is a complicated operator on the right that is supported on all the right fermions, and its role is to carry the weak correlations with $\chi_{1L}$. We may think of it as a dressed version of $\chi_{1R}$, one which takes into account many-body effects. One of the primary reasons for introducing this model of the density matrix is to allow us to convert our results for the $k = 2$ Renyi mutual information, which is the quantity that is numerically accessible, into results concerning the von Neumann mutual information. We will see below that both quantities are governed by the same correlation parameter $\tilde{a}$, which allows us to convert one into the other.

In terms of the emergent fermion, the $1_L R$ density matrix reads
\begin{equation}
    \rho = \frac{I + 2 i \tilde{a} \chi_{1L} \tilde{\chi}_{1R}}{2}  \rho_R,
    \label{eq:1LRdensity}
\end{equation}
where $\rho_R$ is the density matrix of $R$ alone and $\tilde{a} = G_{\mathrm{bulk}}$ is the bulk correlation between $\chi_{1L}$ and $\tilde{\chi}_{1R}$. This expression is schematic since the product form is not strictly correct, but it captures the fact that $\rho_R$ is weakly perturbed by the inclusion of $\chi_{1L}$, and the perturbation is expressed via the correlation with the emergent fermion $\tilde{\chi}_{1R}$. 

The key implication of the above model concerns the entropy of $1_L R$. The correlation given by $\tilde{a}$ causes a correction to the left-right entropy, which would be $\frac{1}{2} \ln 2 + s_0 N$ if $\tilde{a}$ were zero. By direct calculation starting from the $1_L R$ density matrix one finds the $k$th Renyi mutual information:
\begin{equation}
    I^{(k)}(1_L :R ) = \frac{1}{k - 1} \log \left[\frac{1}{2} \left((1 + \tilde{a})^k + (1 - \tilde{a})^k \right) \right]
\end{equation}
and by taking the replica limit $k \rightarrow 1$ we find the von Neumann mutual information:
\begin{equation}
    I(1_L :R ) = \frac{1}{2} \left[ (1+\tilde{a}) \log (1+\tilde{a})+(1-\tilde{a}) \log (1-\tilde{a}) \right].
\end{equation}
Thus we find that the $k$th Renyi mutual information and the von Neumann mutual information are both governed by the same parameter $\tilde{a}$. Practically speaking, this is the main point of introducing the $1_L R$ density matrix, since it allows us to estimate, via the correlation parameter $\tilde{a}$, the von Neumann entropy mutual information from the 2nd Renyi entropy mutual information (which is what we can calculate numerically). We caution the reader that $\tilde{a}$ is different from $a$ above. In fact, as we argued in Sec. \ref{sec:gravity_mi}, $G_{\mathrm{bulk}} = \tilde{a}$ is proportional to $\sqrt{G(\beta/2)} = \sqrt{a}$ at low temperatures where $a$ is small.

Furthermore, the mutual information computed using the $1_L R$ density matrix model is consistent with our numerical results presented in the main text. Assuming small $\tilde{a}$, we obtain
\begin{equation}
    I(1_L :R ) = \frac{\tilde{a}^2}{2} + \cdots.
\end{equation}
When considering a general $K \ll N$ fermions on the left, we find the result
\begin{equation}
    I(K_L : R) = K \times \frac{\tilde{a}^2}{2} + \cdots
\end{equation}
Further, with $\beta = \mathcal{O}(N)$, the Green's function scales like
\begin{equation}
    G(\beta/2) \sim \frac{1}{N^{2\Delta}}.
\end{equation}
We now appeal to our holographic argument in Sec. \ref{sec:gravity_mi} where we argued that $\tilde{a} = G_{\mathrm{bulk}} \propto \sqrt{G(\beta/2)}$ due to the geodesic approximation, where the length of the geodesic is equal to half the length of the wormhole (hence the square root). Using this identification, we thus obtain an estimate for the mutual information between $K_L$ fermions on the left with the entire right:
\begin{equation}
    I(K_L : R) \propto \frac{K}{N^{2\Delta}} + \cdots.
\end{equation}
The total mutual information is thus less than a small parameter $\delta$ provided
\begin{equation}
    K < \delta N^{2 \Delta}.
\end{equation}
This is our distance estimate for the ground space code in the SYK model and the low-rank SYK models.

\subsubsection*{Testing the fermion entanglement model}

\begin{figure}[h!]
    \centering
    \includegraphics[width=0.7\columnwidth]{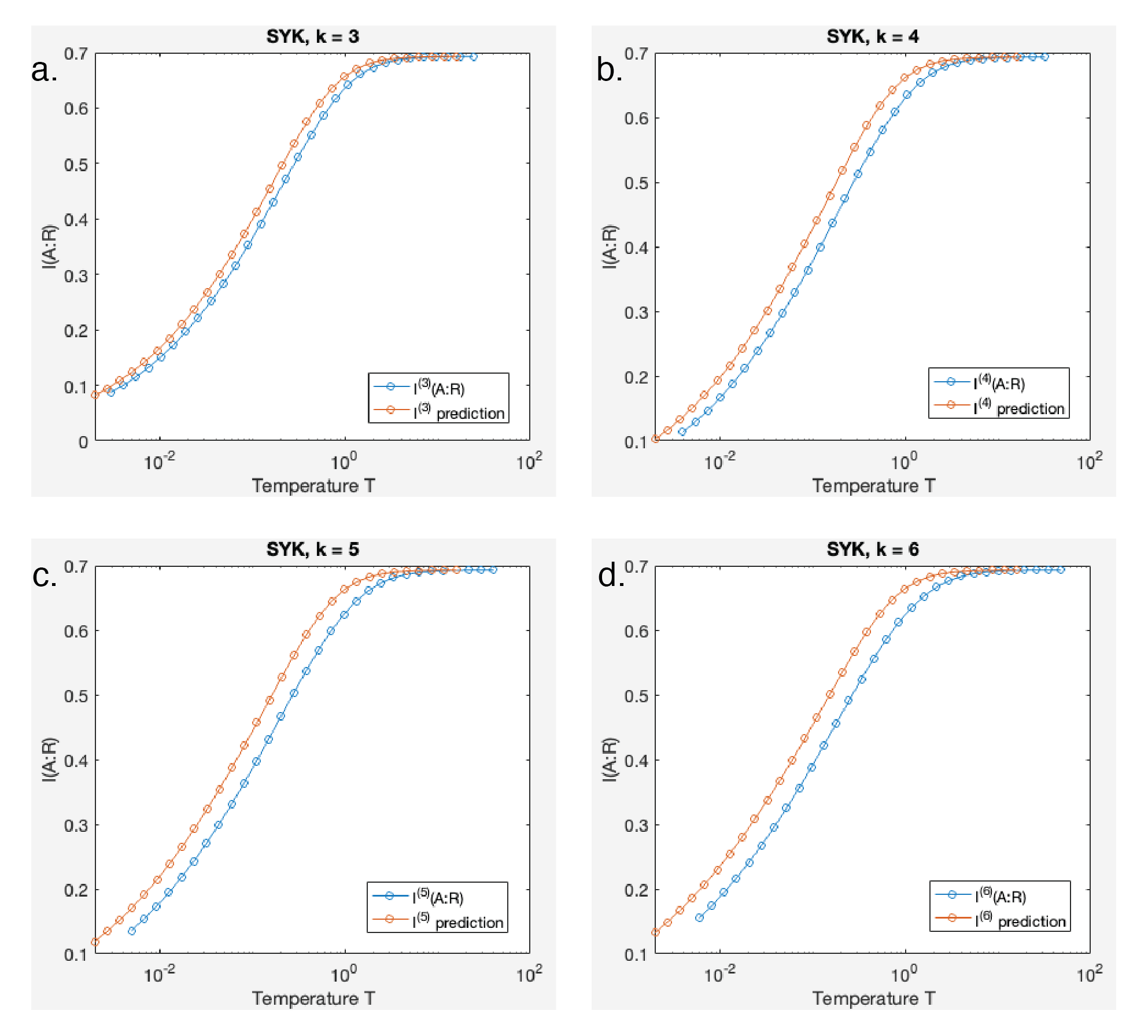}
    \caption{Predicting higher-order Renyi-$k$ mutual information from SYK numerics. The entanglement model presented in Appendix~\ref{app:fermion} predicts a simple form for the mutual information at all orders $k$ that depends on a single parameter $\tilde{a}$. We extract $\tilde{a}$ from $I^{(2)}$ and use it to predict the higher-order Renyi-$k$ mutual informations for $k = 3,4,5,6$ (a-d).}
    \label{fig:miprediction}
\end{figure}

We can directly test this entanglement model by computing fermion determinants in our numerics. The entanglement model predicts that the Renyi-$k$ mutual information for any $k$ is controlled by the single parameter $\tilde{a}$:
\begin{equation}
    I^{(k)}(A:R) = \frac{K}{k - 1} \log \left[\frac{1}{2} \left((1 + \tilde{a})^k + (1 - \tilde{a})^k \right) \right]
\end{equation}
We can test this model numerically by computing the Renyi-$2$ mutual information, extracting the quantity $u$, and using $u$ to predict the higher-order Renyi-$k$ mutual informations. We plot the results of this comparison for the SYK model in Fig. \ref{fig:miprediction}. The agreement between the prediction and the actual mutual information is best for $k = 3$, and we still find decent agreement for higher Renyi-$k$.

\section{Numerical methods}
\label{app:numericalmethods}

Here we provide details of the numerical analysis used to compute the ground-state entropy density $s_0$, the mutual information $I(A:R)$, and the energy gap $E_g$ in the two-sided model.

\subsubsection*{Anti-periodicity and Fourier transforms}

The Green's functions $G(\tau)$ should be antiperiodic on the region $\tau \in [0,\beta)$. To accomplish this in numerics, we extend the range of $\tau$ to $[0,2\beta)$ and make $G(\tau)$ a periodic function of $2\beta$ and an antiperiodic function of $\beta$. We forget about Matsubara frequencies for the moment and naively take the Fourier transform over all frequency components,
\begin{equation}
    G(\tau) = \frac{1}{\beta} \sum_{\omega} G(\omega) e^{- i \omega \tau}.
\end{equation}
To enforce periodicity $G(\tau+2\beta) = G(\tau)$, we must have $\omega = \pi n / \beta$ for $n$ an integer. Further, to enforce antiperiodicity $G(\tau+\beta) = -G(\tau)$, we must have $G(2 \pi n / \beta) = 0$, i.e. the `even' Fourier components all vanish identically. By contrast, the `odd' components $G(\pi (2 n + 1) / \beta)$ are unrestricted. So we end up with only the fermionic Matsubara frequencies anyway, which is what we wanted.

So in the numerics we include twice as many components in the vectors $G(\tau)$ and $G(\omega)$ as necessary, where half of these components are redundant. But this is the most organized way to perform the FFT and IFFT, and provides an independent check on the function $G(\tau)$ being antiperiodic as necessary (just check the even components $G(2 \pi n / \beta)$ and make sure they vanish).

\subsubsection*{Weighted update iteration method}

To find solutions to the Schwinger-Dyson equations, we iteratively search using a weighted update \cite{maldacena2016remarks,maldacena2018eternal}. We illustrate this explicitly for the SYK model, but the steps are similar for all three models considered in this paper. We initially set $\Sigma = 0$ and $G(\omega_f) = (-i \omega_f)^{-1}$. We then take the inverse Fourier transform $G(\tau)$ and use it to calculate an updated $\Sigma'(\tau) = J^2 [G(\tau)]^{q-1}$.
Finally, we take the inverse Fourier transform $\Sigma'(\omega_f)$ and use it to calculate an updated $G'(\omega_f)$ that is weighted by a parameter $x \in [0,1]$:
\begin{equation}
    G'(\omega_f) = (1-x) G(\omega_f) + x \left(-i \omega_f - \Sigma'(\omega_f) \right)^{-1}
\end{equation}
This procedure is iterated repeatedly while monitoring the error $\epsilon = \sum_{\omega_f} \magn{G'(\omega_f) - G(\omega_f)}$. If the error ever grows during a single iteration, we divide $x \rightarrow x / b$ by a constant factor and continue iterating. Usually we begin with $x = 0.5, b = 2$ but have found it useful to pick other values in order to get good convergence.

\subsubsection*{Ground-space entropy}

To calculate the entropy $S(T)$ as a function of temperature $T$, we calculate the energy $E$ at various temperatures and integrate down the heat capacity. We have
\begin{equation}
    C_V = \frac{\partial E}{\partial T} = T \frac{\partial S}{\partial T}.
\end{equation}
Therefore,
\begin{equation}
    S(\infty) - S(T) = \int_{T}^{\infty} \frac{dT'}{T'} C_V = \int_{1/T}^0 d \beta' \beta' \frac{\partial E}{\partial \beta'}
\end{equation}
and so
\begin{equation}
    \label{eq:entropyenergy}
    S(T) = S(\infty) + \int_0^{1/T} \beta' dE
\end{equation}
where $S(\infty) = N/2 \ln 2$. Therefore, given a profile of energies $E(\beta')$ at various temperatures $0 < \beta' \leq 1/T$, we can immediately use \eqref{eq:entropyenergy} to calculate the system's entropy at a desired temperature $T$.

\subsubsection*{Entanglement entropy}

We compute the entanglement entropy by computing fermion determinants. Suppose we prepare two systems $L,R$ in a thermofield double state at inverse temperature $\beta$. That is,
\begin{equation}
    \ket{\Psi}_{LR} \propto \sum_n e^{-\beta E_n / 2} \ket{n}_L \ket{n}_R
\end{equation}
where $\ket{n}$ are the energy eigenstates. We split $L$ into subregions $A,\overline{A}$ and we are interested in computing the $k$th Renyi entropy $S^{(k)}_{\overline{A}} = 1/(1-k) \log \mathrm{tr}[(\rho_{\overline{A}})^k]$ of the region $\overline{A} \in L$. This is equivalent to introducing $k$ copies of the $L,R$ TFD state, and computing the expectation value of the SWAP$_{\overline{A}}$ operator as shown in Fig. \ref{fig:tfddet}. Rearranging the diagram, we find that the $k$th Renyi entropy is equivalent to computing a `flagpole' diagram with $k$ rungs, where the fermions in region $\overline{A}$ have antiperiodic boundary conditions on a single circle of length $k \beta$, whereas fermions in region $A$ have antiperiodic boundary conditions on $k$ independent circles, each of length $\beta$.

The calculation can be cast as a path integral over mean fields $G(\tau),\Sigma(\tau)$ and over fermion fields $\chi_a(\tau)$ with $a = 1,\ldots,N$. The boundary conditions on $A$ impose antiperiodic boundary conditions on $k$ small loops of length $\beta$ for fermions in that region, but otherwise the path integral is unchanged from the usual imaginary-time $G$-$\Sigma$ action on a thermal circle of length $k \beta$. When $K$ is small, we argue that the bulk fields $G(\tau),\Sigma(\tau)$ are approximately unchanged by the presence of the small loops in region $A$. So we simply take the numerical saddle-point solutions to the Schwinger-Dyson equations as the bulk solutions here. Then the only remaining change to the path integral comes from changing the boundary conditions on $K$ fermions from a single loop of length $k \beta$ to $k$ small loops of length $\beta$. For a fermion $\chi(\tau)$ on a single large loop of length $k \beta$, the discretized path integral takes the form
\begin{align}
    &\int \mathcal{D} \chi \exp{\left[ -\int_0^{k \beta} d \tau d \tau' \chi(\tau) \left( \partial_{\tau} \delta_{\tau \tau'} - \Sigma(\tau-\tau') \right) \chi(\tau') \right]} \nonumber \\
    &\approx \prod_{\tau} \int D \chi_{\tau} \exp{\left[ -\sum_{\tau,\tau'} \chi_{\tau} \left( \Delta_{\tau\tau'} - (k \beta / n)^2 \Sigma_{\tau \tau'} \right) \chi_{\tau'} \right]}
\end{align}
where we have discretized the integral into $n$ segments of length $d \tau = k \beta / n$. The fermion fields $\chi_{\tau}$ are now Grassmann vectors of length $n$, and $\Delta_{\tau\tau'},\Sigma_{\tau\tau'}$ are complex $n \times n$ matrices, where $\Sigma_{\tau\tau'} = \Sigma(\tau-\tau')$ is the numerical solution of the Schwinger-Dyson equations on a circle of length $k \beta$, and
\begin{equation}
    \Delta_{\tau\tau'} = \begin{bmatrix}
1 & 0 & 0 & 0 & 0 & \cdots & 1\\
-1 & 1 & 0 & 0 & 0 && 0\\
0 & -1 & 1 & 0 & 0 && 0\\
0 & 0 & -1 & 1 & 0 && 0\\
0 & 0 & 0 & -1 & 1 && 0\\
\vdots &&&&& \ddots \\
0 & 0 & 0 & 0 && -1 & 1\\
\end{bmatrix}
\end{equation}
is a discretization of the partial derivative $\partial_{\tau}$ where the $1$ in the upper-right corner imposes antiperiodic boundary conditions on the large loop of length $k \beta$. The integral is quadratic in the fermions, so we can evaluate it to obtain
\begin{equation}
    \det{\left( \Delta_{\tau\tau'} - d \tau^2 \Sigma_{\tau \tau'} \right)}^{1/2}
\end{equation}
where the square root comes from the Pfaffian $\mathrm{pf}(x) = \sqrt{\det{(x)}}$.

Now consider what changes when we modify the boundary conditions to be antiperiodic on $k$ small loops each of length $\beta$. All that changes in the path integral is the discretization of the partial derivative $\Delta_{\tau\tau'} \rightarrow \Delta^{(k)}_{\tau\tau'}$, which becomes block diagonal and antiperiodic on subspaces of size $n/k$. For example, for $n = 6$ and $k = 2$,
\begin{equation}
    \Delta^{(k)}_{\tau\tau'} = \begin{bmatrix}
1 & 0 & 1 & 0 & 0 & 0\\
-1 & 1 & 0 & 0 & 0 & 0\\
0 & -1 & 1 & 0 & 0 & 0\\
0 & 0 & 0 & 1 & 0 & 1\\
0 & 0 & 0 & -1 & 1 & 0\\
0 & 0 & 0 & 0 & -1 & 1\\
\end{bmatrix}
\end{equation}
Performing the Gaussian integral over the fermions in this case leads to 
\begin{equation}
    \det{\left( \Delta^{(k)}_{\tau\tau'} - d \tau^2 \Sigma_{\tau \tau'} \right)}^{1/2}
\end{equation}

To estimate the $k$th Renyi entropy for finite $K \ll N$, we want to start with the answer for $K = 0$, divide by the contribution from $K$ large-loop fermions, and multiply by the contribution from $K$ small-loop fermions.
That is, we estimate the Renyi entropy by:
\begin{equation}
    \mathrm{tr}[\left( \rho_{\overline{A}} \right)^k] \approx \left. \mathrm{tr}[ \left(\rho_{\overline{A}} \right)^k] \right \rvert_{K = 0} \times \left[ \frac{\det{\left( \Delta_{\tau\tau'}^{(k)} - d \tau^2 \Sigma_{\tau \tau'} \right)}}{\det{\left( \Delta_{\tau\tau'} - d \tau^2 \Sigma_{\tau \tau'} \right)}} \right]^{K/2}
    \label{eq:fermiondetrenyidiff}
\end{equation}

We refer to the determinant part as $\Delta S^{(k)}$, where
\begin{equation}
    \Delta S^{(k)} = \frac{1}{1-k} \log \left[ \frac{\det{\left( \Delta_{\tau \tau'}^{(k)} - d \tau^2 \Sigma_{\tau \tau'} \right)}}{\det{\left( \Delta_{\tau \tau'} - d \tau^2 \Sigma_{\tau \tau'} \right)}} \right]^{1/2}
\end{equation}
so that $S^{(k)}_{\overline{A}} \approx S^{(k)}_L + K \times \Delta S^{(k)}$. We note that this determinant expression can be formally derived from the exact expression for the flagpole diagram in the limit $K/N \ll 1$, see Eq.~\eqref{eq:flagpolealphaaction} below.

\subsubsection*{Numerically Evaluating Flagpole Diagrams}

To check our assertion that the saddle points are not shifted very much for small $K \ll N$, we also directly calculate the Renyi entropy using numerics. In particular, by pushing our numerical Schwinger-Dyson methods a little further we can directly compute the the `flagpole' diagrams for $k = 2$ and arbitrary $\magn{A} = K = \alpha N$. Recall from the main text that the 2nd Renyi entropy $S^{(2)}_{\overline{A}}$ ($k = 2$) is given by the `flagpole' diagram in Fig. \ref{fig:tfddet}. This diagram is equivalent to computing the partition function of the SYK model at temperature $2 \beta$ where the $\magn{A}$ fermions on the left are antiperiodic on two separate loops of length $\beta$, while the $\magn{\overline{A}}$ fermions on the right are antiperiodic on a single loop of length $2\beta$.

By splitting up the fermions in this way, the exact $k=2$ disorder-averaged partition function can be written
\begin{equation}
    \mathbb{E} [Z] = \mathbb{E} \mathrm{tr}(\rho \otimes \rho \ M_{2}^{\overline{A}}) = \int \prod_{ijkl} dJ_{ijkl} \int \prod_i \mathcal{D} \chi_i \ e^{-\mathcal{I}}
\end{equation}
with action
\begin{equation}
    \mathcal{I} = \int_0^{2 \beta} d \tau \left( \frac{1}{2} \sum_{i \in A} \chi_i \partial_{\tau}^{(2)} \chi_i + \frac{1}{2} \sum_{j \in \overline{A}} \chi_j \partial_{\tau} \chi_j - \frac{1}{4!} \sum_{ijkl} J_{ijkl} \ \chi_i \chi_j \chi_k \chi_l \right)
\end{equation}
where the propagator $\partial_{\tau}$ is the usual partial derivative with antiperiodic boundary conditions in $2 \beta$, while the propagator $\partial_{\tau}^{(2)}$ is antiperiodic on two loops of length $\beta$. Here $M_2^{\overline{A}}$ is the SWAP operator on the subregion $\overline{A}$. Performing the disorder average over the couplings $J_{ijkl}$ and discretizing the time variable $\tau$ into steps of size $d \tau = 2 \beta / n$, we find
\begin{equation}
    \mathbb{E}[Z] = \int \prod_{i,\tau}  D \chi_{i,\tau} \ e^{-I}
\end{equation}
with action
\begin{equation}
    I = \sum_{\tau,\tau' = 0}^{2\beta} \left( \frac{1}{2} \sum_{i \in A} \chi_{i,\tau} \Delta_{\tau \tau'}^{(2)} \chi_{i, \tau'} + \frac{1}{2} \sum_{j \in \overline{A}} \chi_{j,\tau} \Delta_{\tau \tau'} \chi_{j, \tau'} - \frac{J^2}{8 N^3} \left( \sum_i \chi_{i,\tau} \chi_{i,\tau'} \right)^4 \right)
\end{equation}
where $\Delta_{\tau \tau'}$, $\Delta^{(2)}_{\tau \tau'}$ are the same $n \times n$ matrices as defined in the previous subsection. Finally, we introduce mean fields
\begin{align}
    G_{\tau\tau'}^A &= \frac{1}{\magn{A}} \sum_{i \in A} \chi_{i,\tau} \chi_{i,\tau'} \nonumber \\
    G_{\tau\tau'}^{\overline{A}} &= \frac{1}{\magn{\overline{A}}} \sum_{j \in \overline{A}} \chi_{j,\tau} \chi_{j,\tau'}
\end{align}
along with Lagrange multipliers $\Sigma_{\tau \tau'}^A, \Sigma_{\tau \tau'}^{\overline{A}}$ and integrate out the fermions to obtain:
\begin{align}
    \frac{I}{N} &= - \alpha \frac{1}{2} \log \det \left( \hat{\Delta}^{(2)} - \left( \frac{2 \beta}{n} \right)^2 \hat{\Sigma}^{A} \right) - (1-\alpha) \frac{1}{2} \log \det \left( \hat{\Delta} - \left( \frac{2 \beta}{n} \right)^2 \hat{\Sigma}^{\overline{A}} \right) \nonumber \\
    &+ \frac{1}{2} \left( \frac{2 \beta}{n} \right)^2 \sum_{\tau,\tau'} \left( \alpha \Sigma_{\tau \tau'}^{A} G_{\tau \tau'}^{A} + (1-\alpha) \Sigma_{\tau \tau'}^{\overline{A}} G_{\tau \tau'}^{\overline{A}} - \frac{J^2}{4} \left( \alpha G^A_{\tau \tau'} + (1-\alpha) G^{\overline{A}}_{\tau \tau'} \right)^4 \right).
    \label{eq:flagpolealphaaction}
\end{align}
At large $N$, the equations of motion for this action are:
\begin{align}
    G^{A}_{\tau \tau'} &= - \left[ \left( \hat{\Delta}^{(2)} - \left( 2 \beta / n \right)^2 \hat{\Sigma}^{A} \right)^{-1} \right]^T_{\tau \tau'} \nonumber \\
    G^{\overline{A}}_{\tau \tau'} &= - \left[ \left( \hat{\Delta} - \left( 2 \beta / n \right)^2 \hat{\Sigma}^{\overline{A}} \right)^{-1} \right]^T_{\tau \tau'} \nonumber \\
    \Sigma^{A}_{\tau \tau'} &= \Sigma^{\overline{A}}_{\tau \tau'} = J^2 \left( \alpha G^A_{\tau \tau'} + (1-\alpha) G^{\overline{A}}_{\tau \tau'} \right)^3
    \label{eq:flagpolealphaeoms}
\end{align}
As mentioned above, one can obtain the fermion determinant expression Eq.~\eqref{eq:fermiondetrenyidiff} for small $\alpha \ll 1$ by expanding the action $I/N$ and saddle-point solutions as power series in $\alpha$; the determinant expression appears as the first-order term in $\alpha$.

We solve the large-$N$ equations of motion numerically by iteration using a similar weighted updated procedure as described above. Finally, we note that the state normalization $\mathbb{E} \mathrm{tr}(\rho \otimes \rho) = e^{-I_{\alpha = 1}}$ is given by the $\alpha = K/N = 1$ action, which must be divided out in order to obtain normalized results. This is equivalent to subtracting off the $\alpha = 1$ action; i.e. to compute Renyi entropies, we have:
\begin{equation}
    S^{(2)}_{\overline{A}} = -\ln \left( \mathbb{E} \mathrm{tr}(\rho \otimes \rho \ M_{2}^{\overline{A}}) \ / \ \mathbb{E} \mathrm{tr}(\rho \otimes \rho) \right) = -\ln \left( e^{-I_{\alpha}} / e^{-I_{\alpha = 1}} \right) = I_{\alpha} - I_{\alpha = 1}.
\end{equation}
We can use the same methods to calculate the entropy $S^{(2)}_A$, but in this case we apply the SWAP operator to the fermions in region $A$, meaning that the region $A$ has a single long loop of length $2 \beta$ and the region $\overline{A}$ has two short loops each of length $\beta$. To avoid confusion, we always refer to $K = \magn{A}$ as the number of fermions in region $A$, regardless of which entropy we are computing.

\begin{figure}[h!]
    \centering
    \includegraphics[width=0.8\textwidth]{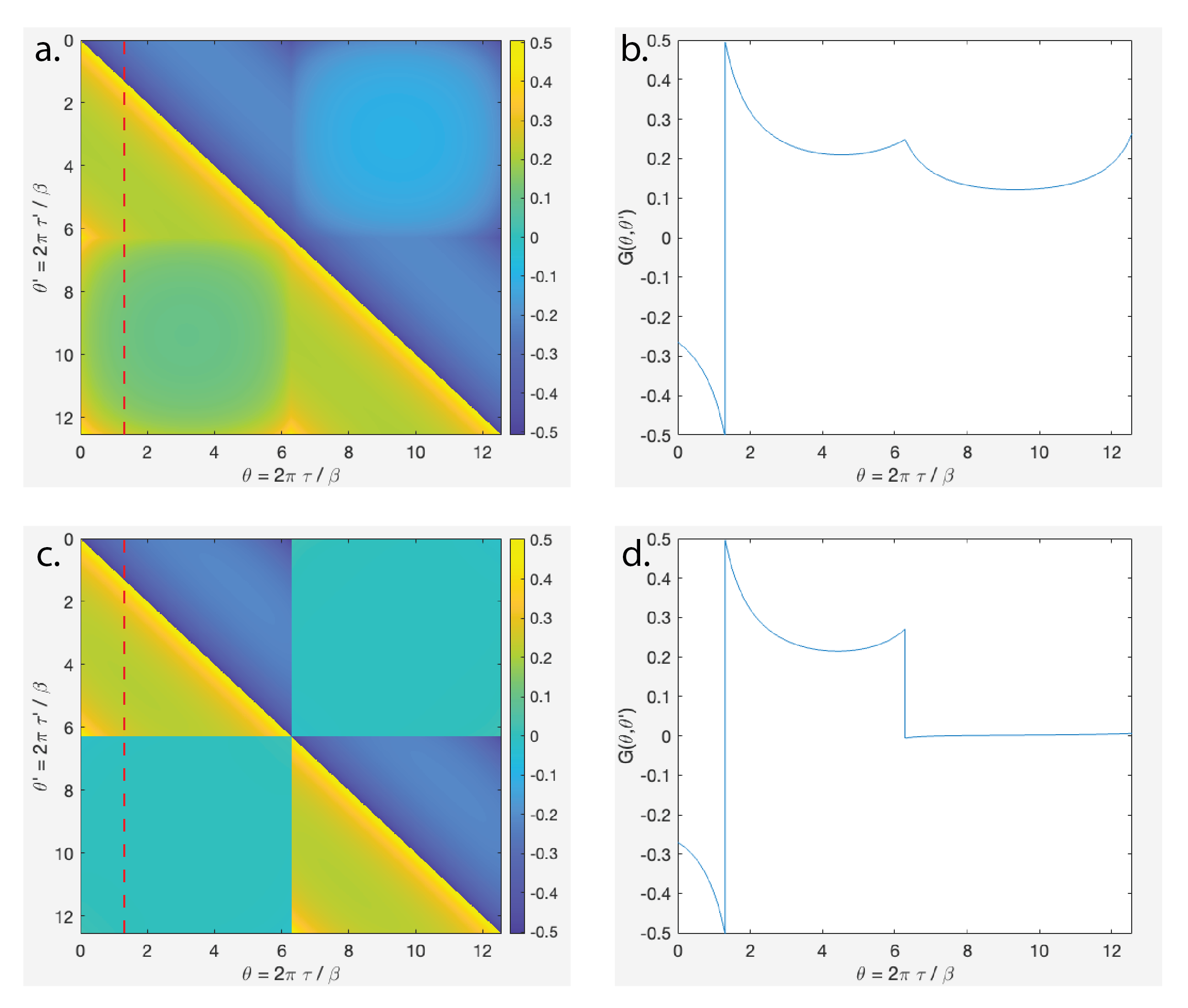}
    \caption{Green's functions $G_{\tau \tau'}^{\overline{A}}$ (top) and $G_{\tau \tau'}^A$ (bottom) obtained from numerically calculating the flagpole diagram for $S^{(2)}_{\overline{A}}$, computed at $J\beta = 18$ and $K/N = 0.5$. In (a) and (c) we plot the full $n \times n$ matrices $G_{\tau \tau'}^{\overline{A}}$ and $G_{\tau \tau'}^A$, respectively, as a function of $\theta = 2 \pi \tau / \beta$ and $\theta' = 2 \pi \tau' / \beta$. In (b) and (d) we take 1D cuts (red dashed line) at fixed $\theta = 1.309$ to show the profile more clearly.}
    \label{fig:gagbexampleplots}
\end{figure}

\begin{figure}[h!]
    \centering
    \includegraphics[width=0.8\textwidth]{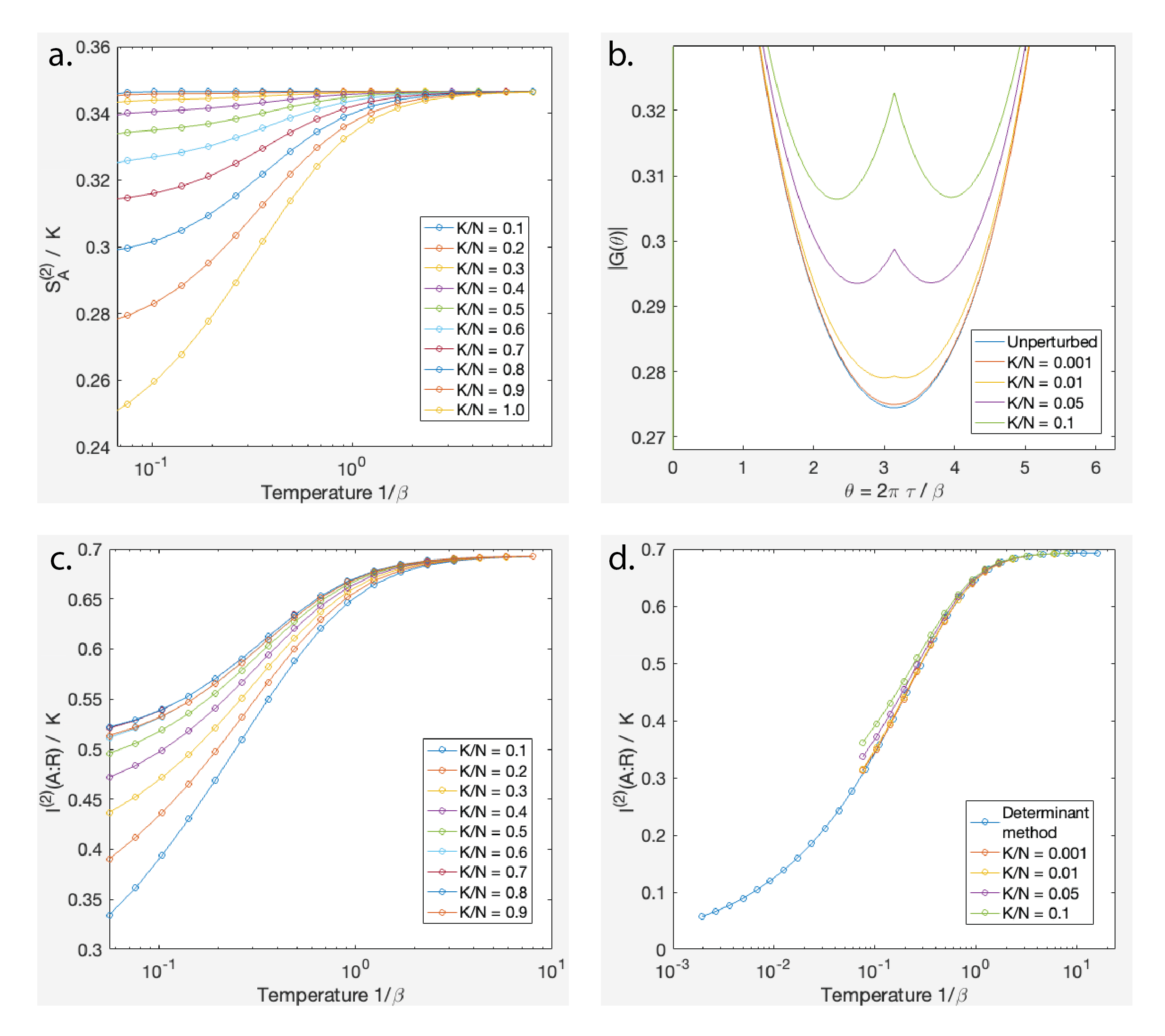}
    \caption{Results of numerically computing flagpole diagrams. The normalized 2nd Renyi entropy $S_A^{(2)} / K$ for a region $A$ (a) decays as a function of temperature except for very small $K = \magn{A}$, in which case it is approximately constant $S_A^{(2)} = \frac{K}{2} \log(2)$, as expected. For $K/N = 1$ we are computing the Renyi entropy of the full $L$ side, and we find good agreement with the entropy density results in Fig. \ref{fig:numericalsyk}. In (b) we show the difference in the Green's function $G(\theta)$ of the perturbed saddle point (with twist fields) compared to the unperturbed saddle point (without twist fields) and find good agreement when $K/N \ll 1$. The resulting normalized mutual information $I^{(2)}(A:R) / K$ decays with temperature as expected (c), and we find excellent agreement between the flagpole numerics and the determinant method (d) when $K/N \ll 1$. Due to memory and time constraints, we are limited to matrix sizes of $n \leq 3840$, which thereby limits the lowest temperatures $J \beta \leq 18$ we are able to access.}
    \label{fig:gagbnumerics}
\end{figure}

The results of these numerical simulations are plotted in Figs. \ref{fig:gagbexampleplots} and \ref{fig:gagbnumerics}. In Fig. \ref{fig:gagbexampleplots} we show the resulting saddle-point Green's functions $G_{\tau \tau'}^{\overline{A}}$ and $G_{\tau \tau'}^A$ for the Renyi entropy $S^{(2)}_{\overline{A}}$ at inverse temperature $J \beta = 18$ and $K/N = 0.5$, plotted as a function of $\theta = 2 \pi \tau / \beta$ and $\theta' = 2 \pi \tau' / \beta$. We explicitly see the $2\beta$-antiperiodicity in $G_{\tau \tau'}^{\overline{A}}$ (Fig. \ref{fig:gagbexampleplots}.a. and \ref{fig:gagbexampleplots}.b.) and the $\beta$-antiperiodicity in $G_{\tau \tau'}^A$ (Fig. \ref{fig:gagbexampleplots}.c. and \ref{fig:gagbexampleplots}.d.). For $K/N = 0$ we find that the resulting Green's functions are translation-invariant (not shown) as expected from our earlier numerical simulations. Further, for $K/N = 0$, we recover the same ground-state entropy $s_0 \approx 0.23$ that we obtained in the main text by integrating down the system's heat capacity. This agreement gives us some confidence that both methods are correct.

Using these saddle-point Green's functions, we can evaluate the Renyi entropies $S^{(2)}_A$, $S^{(2)}_{\overline{A}}$ and mutual information $I^{(2)}(A:R)$ (Fig. \ref{fig:gagbnumerics}) for arbitrary region sizes $K = \magn{A}$, instead of being limited to small sizes as was the case with our determinant methods. We show the resulting Renyi entropy $S^{(2)}_A$ as a function of temperature for a range of region sizes $0 < K/N \leq 1$ in Fig. \ref{fig:gagbnumerics}.a. The entropy decays at low temperatures, except for very small $K/N \ll 1$, in which case the entropy is approximately independent of temperature $S^{(2)}_A \approx \frac{K}{2} \log(2)$ as expected. If we take the region $A$ to span the entire $L$ side we have $K/N = 1$, and we find good agreement between the entropy $S^{(2)}_L$ computed this way and the ground space entropy density $s$ computed in Fig. \ref{fig:numericalsyk}. To see the effect of the twist fields on the saddle point Green's functions, we plot a cut through the perturbed Green's function $G_{\tau \tau'}^{\overline{A}}$ (with twist fields) and compare to the unperturbed Green's function $G(\theta)$ (without twist fields) in Fig. \ref{fig:gagbnumerics}.b. We find good agreement between the two when $K/N \ll 1$, validating our arguments in Appendices~\ref{app:path_int_perturb} and~\ref{app:extrap} below. The Renyi entropies obtained in this way allow us to immediately calculate the normalized mutual information $I^{(2)}(A:R) / K$, which we plot as a function of temperature in Fig. \ref{fig:gagbnumerics}.c. for a variety of system sizes $0 < K/N < 1$. Finally, in Fig. \ref{fig:gagbnumerics}.d. we compare the mutual information obtained using these methods to the mutual information obtained using the fermion determinant methods, finding excellent agreement for small $K/N \ll 1$. Due to time and memory constraints, these flagpole numerics are limited to matrix sizes $n \leq 3840$, which thereby limits the temperatures $\beta \leq 18$ we are able to access before numerical instabilities become a problem in the simulations.

\begin{figure}[h!]
    \centering
    \includegraphics[width=0.4\textwidth]{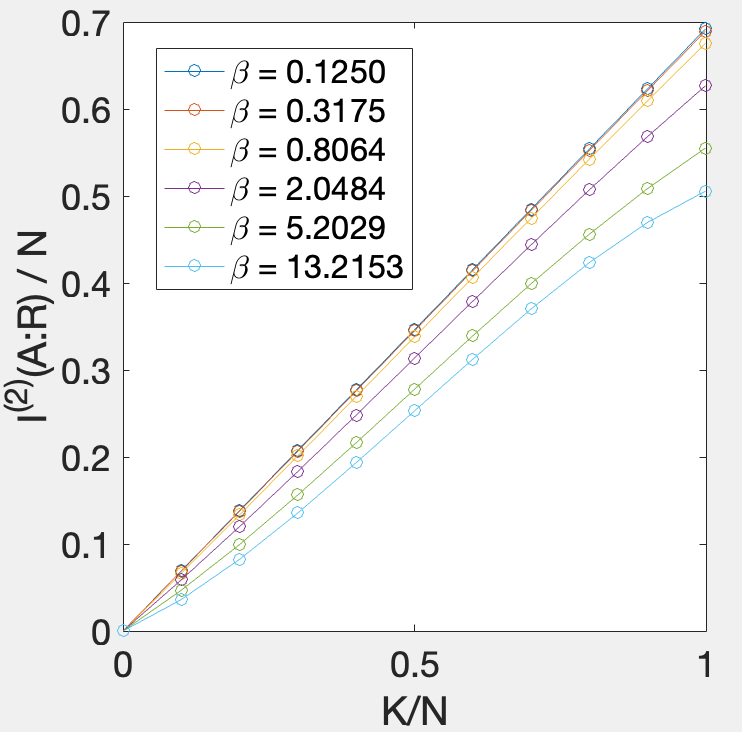}
    \caption{Normalized mutual information $I^{(2)}(A:R) / N$ versus subregion size $K/N$ at various temperatures $\beta$. At high temperatures, the mutual information is a simple linear function of subregion size, while at lower temperatures the functional form is more complicated.}
    \label{fig:miqescompare}
\end{figure}

Next, we compare the SYK numerics with the holographic QES prediction by plotting the mutual information as a function of subregion size $K/N$ for different temperatures $\beta$ in Fig. \ref{fig:miqescompare}. At high temperatures the mutual information is a simple linear function of the subregion size, while at lower temperatures the functional form is more complicated. In particular, at the lowest temperature we can access, we begin to see an initial linear growth regime followed by an upturn around $K/N \sim .1$ and then further growth until saturation. This is roughly what we expect from the QES calculation as discussed in Appendix~\ref{app:lr_qes}, and we would expect these feeatures to further sharpen at larger $\beta$ but we have not been able to access this regime due to numerical instabilities. We also emphasize that the holographic approach only really applies at large $\beta$ so we should not expect precise agreement.

\begin{figure}
    \centering
    \includegraphics[width=0.8\textwidth]{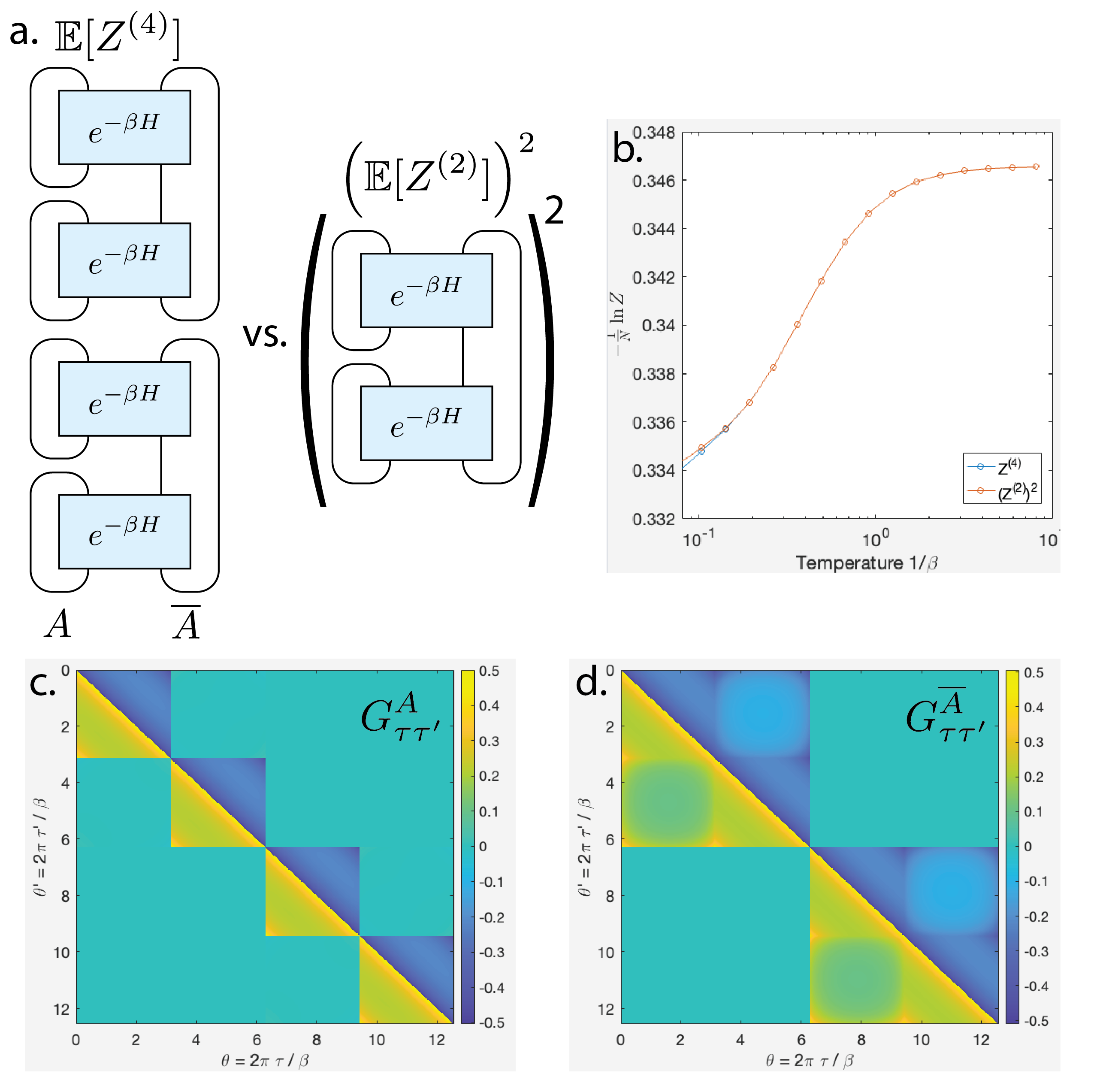}
    \caption{Characterizing the variance in the Renyi entropy $S_{\overline{A}}^{(2)}$ by calculating the flagpole diagrams $\mathbb{E}[Z^{(4)}]$ and $\left(\mathbb{E}[Z^{(2)}]\right)^2$for $K/N = 0.5$. The diagram for $\mathbb{E}[Z^{(4)}]$ involves $k = 4$ replicas with $\beta$-antiperiodic boundary conditions in region $A$ and $2\beta$-antiperiodic boundary conditions in region $\overline{A}$, while the diagram for $\left(\mathbb{E}[Z^{(2)}]\right)^2$ is the square of a $k = 2$ diagram (a). The resulting saddle-point actions $I/N = -1/N \log Z$ for these two diagrams are nearly identical except at the lowest temperatures (b). Further evidence of factorization $\mathbb{E}[Z^{(4)}] \approx \left(\mathbb{E}[Z^{(2)}]\right)^2$ is provided by examining the saddle-point Green's functions $G_{\tau \tau'}^{A}$ and $G_{\tau \tau'}^{\overline{A}}$ for the $\mathbb{E}[Z^{(4)}]$ diagram at temperature $J \beta = 18$ and $K/N = 0.5$ (c,d), which clearly factorize into two disjoint subsystems.}
    \label{fig:typicalitykAkB}
\end{figure}

Finally, we can use similar numerical methods to address questions of typicality. Our calculations of mutual information throughout this paper are ensemble averaged quantities, so one could be concerned that individual instances in the ensemble might have a very different mutual information compared to the ensemble average. To address this question, we can compute the variance
\begin{equation}
    \mathbb{E}[Z^{(4)}] - \left(\mathbb{E}[Z^{(2)}]\right)^2 = \mathbb{E} \mathrm{tr}(\rho \otimes \rho M_2^{\overline{A}})^2 - \left(\mathbb{E} \mathrm{tr}(\rho \otimes \rho M_2^{\overline{A}}) \right)^2
\end{equation}
which measures fluctuations in the 2nd Renyi entropy $S_{\overline{A}}^{(2)}$. The quantity $\mathbb{E} [Z^{(4)}]$ on the left is a `flagpole' diagram that involves $k=4$ replicas with $\beta$-antiperiodic boundary conditions on $A$ and 2$\beta$-antiperiodic boundary conditions on $\overline{A}$ as illustrated in Fig. \ref{fig:typicalitykAkB}.a., whereas the quantity $(\mathbb{E} [Z^{(2)}])^2$ on the right is the square of a `flagpole' diagram that involves only $k=2$ replicas. We can compute both of these quantities using the same methods discussed above. By direct numerical calculation we find that these two quantities are nearly identical for $K/N = 0.5$ except at the lowest accessible temperatures as shown in Fig. \ref{fig:typicalitykAkB}.b., implying a small variance. Further evidence of this fact is provided by examining the Green's function solutions $G_{\tau \tau'}^{A}$ and $G_{\tau \tau'}^{\overline{A}}$ for $Z^{(4)}$, plotted in Figs. \ref{fig:typicalitykAkB}.c,d. for $K/N = 0.5$, which clearly show factorization into two uncorrelated subsystems. This factorization implies that $\mathbb{E}[Z^{(4)}] \approx \left(\mathbb{E}[Z^{(2)}]\right)^2$ and therefore a small variance in the Renyi entropy $S_{\overline{A}}^{(2)}$. The main takeaway here is that our averaged mutual information $I^{(2)}(A:R)$ is a typical value in the ensemble because the variance is small. We suspect that the slight disagreement between $\mathbb{E}[Z^{(4)}]$ and $\left(\mathbb{E}[Z^{(2)}]\right)^2$ at low temperatures is due to numerical instabilities, but we leave a detailed study of this phenomenon for future work.

\section{Path integrals and perturbations}
\label{app:path_int_perturb}

In this appendix, we briefly review the path integral manipulations which justify the determinant formula for fermion entanglement discussed in Appendix~\ref{app:numericalmethods}. As discussed there and in Appendix~\ref{app:fermion}, one includes twist fields in the path integral to compute fermionic Renyi entropies. When the number $K$ of fermions involved is small, these insertions can be viewed as a small perturbation to the action. We want to show that the dominant effect of this inclusion, to lowest order in $K/N$, is to shift the value of the path integral by precisely the perturbation evaluated on the unperturbed saddle point.

The goal is to calculate a path integral or functional integral of the form
\begin{equation}
    F = \int DG e^{ - N I}
\end{equation}
where $I(G)$ is the action and $G$ collectively denotes all the integration variables (for example, the correlation $G$ and self-energy $\Sigma$ in SYK). We suppose that $N$ is large and evaluate the functional integral via saddle point. We further suppose that the action takes the form $I = I_0 + \epsilon \Delta I$ for some small parameter $\epsilon$. This parameter can be $1/N$ but we keep it more general. We will also assume for simplicity that $I_0$ and $I$ have minima that are perturbatively connected as $\epsilon \to 0$.

We begin with the case $\epsilon=0$. Let $G_0$ denote the configuration which minimizes $I_0$. We write the integration variable as
\begin{equation}
    G = G_0 + \delta g
\end{equation}
and expand the action to quadratic order in $\delta g$,
\begin{equation}
    I(G+\delta g) = N I_0(G_0) + N I_0'(G_0) \delta g + \frac{1}{2} N I_0''(G_0) \delta g^2 + \cdots.
\end{equation}
Since $G_0$ minimizes $I_0$ we have $I_0'(G_0)=0$ and the functional integral is
\begin{equation}
    F \approx \int D(\delta g) e^{- N I_0(G_0) - \frac{1}{2} N I_0''(G_0) \delta g^2} = \det[N I_0''(G_0)]^{-1/2} e^{- N I_0(G_0)} \equiv F_0.
\end{equation}

Now we consider $\epsilon \neq 0$. The minimum is shifted to 
\begin{equation}
    G_0 + \Delta G
\end{equation}
where $\Delta G \sim O(\epsilon)$ can be expanded as a power series in $\epsilon$. We now want to expand the action to first order in $\epsilon$ and up to second order in fluctuations, so to start we expand up to second order in $\Delta G$ and $\delta g$,
\begin{equation}
    I(G_0+\Delta G + \delta g) = I(G_0) + I'(G_0) (\Delta G + \delta g) + \frac{1}{2} I''(G_0) (\Delta G + \delta g)^2 + \cdots.
\end{equation}
Suppressing arguments of $G_0$ and remembering that $I_0'(G_0) = 0$, this reduces to
\begin{equation}
    I(G_0+\Delta G + \delta g) = (I_0 + \epsilon \Delta I) + \epsilon \Delta I'(\Delta G + \delta g) + \frac{1}{2} ( I_0'' + \epsilon \Delta I'') (\Delta G^2 + 2 \Delta G \delta g + \delta g^2) + \cdots. 
\end{equation}

We first fix $\Delta G$. $G_0 + \Delta G$ minimizes the action if the coefficient of $\delta g$ vanishes, so we have
\begin{equation}
    \epsilon \Delta I' + (I_0'' + \epsilon \Delta I'') \Delta G = 0
\end{equation}
or
\begin{equation}
    \Delta G = - \epsilon (I_0'' + \epsilon \Delta I'')^{-1} \Delta I' = - \epsilon (I_0'')^{-1} \Delta I'  + O(\epsilon^2).
\end{equation}

We now plug this formula for $\Delta G$ into the expanded action and keep terms up to first order in $\epsilon$. The result is 
\begin{equation}
    I_0 + \epsilon \Delta I + \frac{1}{2} ( I_0'' + \epsilon \Delta I'') \delta g^2.
\end{equation}
The functional integral is thus approximated as
\begin{equation}
    F \approx  \det[N I_0'' + N \epsilon \Delta I'' ]^{-1/2} e^{- N I_0 - N \epsilon \Delta I} .
\end{equation}
Compared to $F_0$, we have 
\begin{equation}
    F = F_0 e^{- N \epsilon \Delta I} \left[ \frac{ \det[N I_0'' ]}{\det[N I_0'' + N \epsilon \Delta I'' ]}\right]^{-1/2}.
\end{equation}
Using standard formulas, the ratio of determinants is
\begin{equation}
    \frac{ \det[N I_0'' ]}{\det[N I_0'' + N \epsilon \Delta I'' ]} = e^{ \mathrm{tr} \ln [ N I_0'' ] - \mathrm{tr} \ln [N I_0'' + N \epsilon \Delta I'' ] } = e^{ - \mathrm{tr}[ (I_0'')^{-1} \epsilon \Delta I''] + O(\epsilon^2)}.
\end{equation}
We therefore obtain
\begin{equation}
    F = F_0 e^{- N \epsilon \Delta I + \frac{1}{2} \epsilon \mathrm{tr}[(I_0'')^{-1} \Delta I''] + O(\epsilon^2)}.
\end{equation}
Both terms in the exponent are $O(\epsilon)$ but only the first is enhanced by $N$, so the dominant change to $F$ at large $N$ is
\begin{equation}
    F_0 \to F_0 e^{- N \epsilon \Delta I(G_0)}.
    \label{eq:path_int_perturb_formula}
\end{equation}
Hence, the dominant effect of including the additional perturbative term $\epsilon \Delta I$ in the action is to shift the action by this perturbation evaluated on the unperturbed saddle point $G_0$.

Because it comes up repeatedly in this paper, we emphasize that one of the key ingredients in \eqref{eq:path_int_perturb_formula} is the fact that $I_0$ is minimized by $G_0$. This means that the value of $I_0$ does not vary to first order in the perturbation; its variation only begins at second order and is thus suppressed relative to the leading term in the exponent of \eqref{eq:path_int_perturb_formula}.

As stated at the beginning of this appendix, these manipulations are how we justify evaluating fermionic Renyi entropies using unperturbed saddles. In particular, the effect of the twist fields is to change boundary conditions, so that some of the fermion determinants contributing to the action are modified. It is these modifications which are captured by $\Delta I$ with $\epsilon$ playing the role of $K/N$.

\section{Extrapolating the mutual information to large $\beta$}
\label{app:extrap}

Having now explained the technical details of our calculations, we must address one final point which is the extrapolation of our results to to $\beta \sim N/J$. This is a potentially dangerous extrapolation since the various techniques discussed in the main text, specifically the QES formalism and the evaluation of path integrals by saddle point, break down at such large $\beta$'s. Indeed, we know that new physics appears for $\beta$ larger than $N/J$, e.g.~\cite{bagrets_syk_liouville,lin2023looking}. On the other hand, we do not expect new physics prior to reaching $\beta \sim N/J$, so such an extrapolation is at least not obviously implausible.

For the remainder of this appendix, we will focus on the microscopic Renyi entropy calculation in the SYK model and describe step by step our justification for the extrapolation. We discuss in more detail the analogous holographic analysis in the following two appendices,~\ref{app:holography} and \ref{app:lr_qes}.

We first discuss the state of interest. As argued in the main text, the maximally mixed state on the code is a microcanonical state obtained from a certain energy window. For the purposes of computing the mutual information, we argued that we could use a related canonical state which is simpler to analyze. For a fixed realization of the couplings, that state is the Gibbs state at inverse temperature $\beta$,
\begin{equation}
    \hat{\rho}(\beta) = \frac{e^{-\beta H}}{Z}
\end{equation}
where the partition function is $Z = \mathrm{tr}(e^{-\beta H})$. Because the partition function is self-averaging in the SYK model (see~\cite{Baldwin_2020} for a recent discussion), $Z \approx \mathbb{E}(Z)$, we can accurately replace $\hat{\rho}$ with 
\begin{equation}
    \hat{\rho} \to \tilde{\rho} = \frac{e^{-\beta H}}{\mathbb{E}(Z)}.
\end{equation}
In other words, we can separately average the numerator and denominator of the thermal state.

To obtain the left-right mutual information, we need three different entropies in this state: (i)  the entropy of all $N$ fermions, which is just the thermal entropy, (ii) the entropy of $K$ fermions, and (iii) the entropy of $N-K$ fermions. We focus the discussion on the 2nd Renyi entropy for concreteness, although we caution readers that the quantity of greatest interest is still the von Neumann mutual information, for which we rely on the density matrix model or the QES approach.

Let us first discuss (i), $S_2(N)$, which is the easiest case. Since we are computing the 2nd Renyi entropy of all the fermions, there is a twist field for each fermion and we have
\begin{equation}
    e^{-S_2(N)} =  \frac{\mathrm{tr}( M_{N} \rho \otimes \rho)}{\mathrm{tr}(\rho)^2} =  \frac{\mathrm{tr}(\rho^2)}{\mathrm{tr}(\rho)^2}
\end{equation}
where $\rho = e^{-\beta H}$ is the un-normalized state. Because $Z = \mathrm{tr}(\rho)$ is self-averaging, we see that the average of $e^{-S_2(N)}$ is approximately
\begin{equation}
    \mathbb{E}\left(e^{-S_2(N)}\right) \approx \frac{Z(2\beta)}{Z(\beta)^2}.
\end{equation}
The known answer in this case is 
\begin{equation}
   Z(\beta) = e^{ - \beta F}
\end{equation}
with
\begin{equation}
    \beta F = \beta N  e_0  - s_0 N -  \frac{c N}{\beta J} + O(1/\beta^2).
\end{equation}
The first term is the ground state energy, the second term is the ground state entropy, and the third term is the contributions from low-temperature excitations. This formula also receives a 1-loop correction which does not depend on $N$ but is important once $\beta \sim N/J$. Indeed, the Schwarzian mode becomes strongly coupled at large $\beta$, leading to a slow logarithmic decrease in the thermal entropy arising from a 1-loop correction (which turns out to be exact~\cite{stanford_ferm_loc}). The 2nd Renyi entropy is thus
\begin{equation}
    e^{-S_2(N)} \approx e^{- s_0 N - \frac{3}{2} \frac{c N}{\beta J} + \cdots}
\end{equation}

Since the $s_0 N$ term is large but fixed, the dimensionless parameter which controls the validity of the saddle point approximation at low temperature is evidently $\frac{N}{\beta J}$. For example, this is the coefficient of the Schwarzian action. We can still rely on a saddle point analysis provided $\frac{N}{\beta J}$ remains large.

Let us now consider the contribution (ii), $S_2(K)$. The twist field expression is 
\begin{equation}
    e^{-S_2(K)} = \frac{\mathrm{tr}( M_{K} \rho \otimes \rho)}{\mathrm{tr}(\rho)^2},
\end{equation}
where again $\rho = e^{-\beta H}$ is the unnormalized thermal state and $M_K$ is the twist field. When evaluating $e^{-S_2(K)}$ in the main text, we used two copies of the same saddle point $G_*$ to evaluate $\mathrm{tr}(M_K \rho \otimes \rho)$. This is justified whenever $K \ll N$ since the effect of the twist fields is small compared to the value of the untwisted action. In particular, the inclusion of the twist fields will not shift the saddle point appreciably. Since $G_*$ already solves the untwisted equations of motion, it follows that the untwisted action only changes by an amount of order $N (K/N)^2$. The dominant contribution therefore comes from evaluating the perturbation to the action on the unperturbed saddle point. This is the content discussed just above in Appendix~\ref{app:path_int_perturb}.

Similar comments apply to contribution (iii), $S_2(N-K)$. In this case the unperturbed $K=0$ saddle corresponds to taking two copies of the TFD state and including a complete set of twist fields for the right system. As discussed in detail in the main text, this saddle is closely related to the $G_*$ saddle discussed above but evaluated as $2\beta$ instead of $\beta$. We again treat the small number, $K$, of additional twist fields on the left as a perturbation and evaluate the ratio of fermion determinants using the unperturbed saddle.

However, as we just discussed in the context of contribution (i), at low temperature the relevant parameter is not $N$ but $\frac{N}{\beta J}$. Hence, it is logically possible that the saddle point could be appreciably shifted if $K \gg \frac{N}{\beta J}$ despite maintaining $K \ll N$. In particular, if we try to take $\beta J \to N$, then it appears that any value of $K$ could potentially shift the saddle point significantly. 

We will now argue that this concern is unfounded. The crucial point is that the parameter which controls the ability of the $K$ twist fields to shift the saddle point is not $K$ but $K/(\beta J)^{2 \Delta}$. Hence, as long as we maintain $K/(\beta J)^{2\Delta} \ll N/(\beta J)$, the determinant analysis is reliable. However, we caution readers that we cannot fully prove this last point, even at a physics level of rigor.

Thus, we will distinguish two types of scalings depending on whether $K$ is bigger or small than $N/(\beta J)$. These are:
\begin{itemize}
    \item Scaling 1 --- $\beta J \sim N^a$, $N \to \infty$, $a < a_*$: we maintain $K \ll N/(\beta J)$ and are able to conclude that the distance is lower bounded by $\min( N^{1-a}, N^{2 \Delta a})$ with the best bound obtained as $a \to a_* = \frac{1}{1+2 \Delta}$,
    \item Scaling 2 --- $\beta J \sim N^a$, $N \to \infty$, $a_* < a < 1$: now $K$ can be larger than $N/(\beta J)$, and if our analysis continues to hold, then the distance is lower bounded by $N^{2\Delta a} > N^{2 \Delta a_*}$.
\end{itemize}
We argue that Scaling 1, which includes the case of constant but large $\beta J$, is under control. From this we establish that the distance is at least greater than any constant and in fact scales at least as a power of $N$ as $N \to \infty$. Scaling 2 is harder to control, but we nevertheless argue that our calculation of the mutual information is still reliable. If this is so, then the distance scales at least as $N^{2\Delta a}$. We then achieve the extrapolation to $\beta J \sim N$ via the limit $a \to 1$.

\subsection*{Scaling 1: $\beta J  \sim N^a$, $N \to \infty$, $a < a_*$}

Here $\beta J$ grows with $N$, but as a relatively slow fractional power of $N$. As a consequence $\frac{N}{\beta J}$ still diverges with $N$, indicating that we are still safe to evaluate path integrals by saddle point.
Turning to the mutual information, recall that both $e^{-S_2(K_L)}$ and $e^{-S_2(K_L \cup R)}$ have a factor of $2^{-K/2}$. For $S_2(K_L)$, this factor is just representing the maximally mixed effective state of the $K$ fermions on the left. For $S_2(K_L \cup R)$, this factor is corrected by a much smaller contribution which ultimately controls the mutual information since the two $(K/2) \ln 2$ terms from $S_2(K_L)$ and $S_2(K_L \cup R)$ cancel.
Scaling 1 is distinguished by demanding that $K \ll \frac{N}{\beta J}$, so that the twist field contributions to $S_2(K_L)$ and $S_2(K_L \cup R)$ are individually small compared to the unperturbed actions.

To get the best lower bound on the distance, we'd therefore like to choose the parameter $a$ such that $K$ can be as large as possible subject to the condition that our calculation in the main text is still reliable, 
\begin{equation}
    K \ll \frac{N}{\beta J} \sim N^{1-a},
\end{equation}
and the mutual information calculated this way is small,
\begin{equation}
    K \ll (\beta J)^{2\Delta} \sim N^{2 \Delta a}.
\end{equation}
These two conditions on $K$ degenerate when $a$ approaches
\begin{equation}
    a_* = \frac{1}{1+ 2\Delta}.
\end{equation}

Thus, for any $a < a_*$, the code has asymptotic rate $s_0$ and a distance which is lower bounded by $\min(N^{1-a},N^{2\Delta a})$. The limit of this lower bound as $a \to a_*$ is
\begin{equation}
    N^{1-a_*} = N^{\frac{2\Delta}{1 + 2\Delta}}.
\end{equation}
For SYK$_q$, $a_* = q/(q+2)$ and the distance exponent is $2/(q+2)$; this is $1/3$ for $q=4$ SYK. For low-rank SYK, we can tune the rank to approach $a_* \to 1/2$ and distance exponent $\to 1/2$.

\subsection*{Scaling 2: $\beta J  \sim N^a$, $N \to \infty$, $a_* < a < 1$}

Next we address the more challenging case where $K$ can be greater than $\frac{N}{\beta J}$, while still requiring that $K \ll N$. By assumption, we still have that $\frac{N}{\beta J}$ grows weakly with $N$, so the saddle point evaluation of the path integral should still be justified. However, the full contribution from the $K$ twist fields can naively cause a significant shift in the saddle point value of $G$ relative to its $K=0$ value unless $K \ll N^{1-a}$ (see Scaling 1 discussion). But by hypothesis, $a > a_*$, so the distance bound is worse than that obtained from Scaling 1. The question is then whether we can control the calculation for large values of $K$ to get a better lower bound. 

We argue as follows. Consider $e^{- S_2(K)}$ as a function of $\beta$. This quantity is simply equal to $2^{-K/2}$ independent of $\beta$, provided that $K$ and $\beta$ are fixed as $N \to \infty$. This means that the contribution of the twist fields is independent of the saddle point value of $G$ near the unperturbed thermal saddle point (see the next subsection for a derivation of this fact). But if a term in the action is independent of $G$, then it will never cause a shift in the saddle point because its contribution to the equations of motion is zero. 

Similar considerations apply to $e^{- S_2(N-K)}$. This quantity also contains a factor of $2^{-K/2}$ as well as another contribution of the form $e^{- \# K / (\beta J)^{2\Delta}}$. The latter contribution is part of the perturbed action evaluated on the unperturbed saddle and is the mutual information we found in the main text and in Appendix~\ref{app:numericalmethods}. It is only this second contribution which depends on $\beta$ and hence on the saddle point value of $G$. We should therefore be concerned about shifts in the saddle point value of $G$ away from the unperturbed thermal saddle point only if the $K/(\beta J)^{2\Delta}$ contribution is comparable to $N/(\beta J)$. 

The condition to have a small shift of the saddle point is thus 
\begin{equation}
    \frac{K}{(\beta J)^{2\Delta}}\ll \frac{N}{\beta J}.
\end{equation}
If this condition is met, then the saddle point is not significantly shifted and the mutual information will be given by our result from the main text,
\begin{equation}
    \text{MI} = \# \frac{K}{(\beta J)^{2\Delta}}.
\end{equation}
The distance is defined as the largest $K$, call it $K_\epsilon$, for which this quantity is still $< \epsilon$. Hence, shifts of the saddle point can be neglected for determining the distance because 
\begin{equation}
    \frac{K}{(\beta J)^{2\Delta}} < \epsilon < \frac{N}{\beta J}.
\end{equation}
The first inequality is the condition $K < K_\epsilon$ and the second inequality is $ \epsilon < \frac{N}{\beta J} \sim N^{1-a}$. This second inequality will be true for any $a <1$. With $a$ free to approach unity, we find a lower bound on the distance of
\begin{equation}
    N^{2 \Delta a},
\end{equation}
with exponent approaching $2\Delta$ as $ a \to 1$. The result in the main text can be understood as the limit $a \to 1$. 

In Appendix~\ref{app:lr_qes} we offer a holographic perspective on the same question and reach the same conclusion. In particular, it is the size of $K/(\beta J)^{2\Delta}$ compared to $N/(\beta J)$ which determines whether the QES shifts significantly from its $K=0$ value.

\subsection*{Lowest-order contributions from twist fields}

Above we asserted that the lowest-order contribution $2^{-K/2}$ to the path integral for $S_{A}^{(2)}$ coming from the twist fields is independent of the saddle-point fields $G,\Sigma$ when $K \ll N$. We already see evidence of this fact from our numerical simulations (see Fig.~\ref{fig:gagbnumerics}.a.), since the Renyi entropy $S_A^{(2)} \approx \frac{K}{2} \log(2)$ is independent of temperature. We also asserted that the lowest-order contribution to $S_{\overline{A}}^{(2)}$ is also $2^{-K/2}$ and that the next order contribution is $\exp [K (\#) / (\beta J)^{2\Delta} ]$. Here we demonstrate these statements explicitly.

First, consider the Renyi entropy $S_{\overline{A}}^{(2)}$ (a similar analysis holds for $S_{A}^{(2)}$) and take $K \ll N$. This is given by
\begin{equation}
    S^{(2)}_{\overline{A}} = -\ln \left( \mathbb{E} \mathrm{tr}(\rho \otimes \rho \ M_{2}^{\overline{A}}) \ / \ \mathbb{E} \mathrm{tr}(\rho \otimes \rho) \right) = -\ln \left( e^{-I_{\alpha}} / e^{-I_{\alpha = 1}} \right) = I_{\alpha} - I_{\alpha = 1}.
\end{equation}
where $I_{\alpha}$ is the action Eq.~\eqref{eq:flagpolealphaaction} evaluated on its associated large-$N$ equations of motion Eq.~\eqref{eq:flagpolealphaeoms}. Expanding all terms as a power series in $\alpha = K/N \ll 1$, we find that the lowest-order terms are
\begin{equation}
    S_{\overline{A}}^{(2)} / N = S_{L}^{(2)} / N - \alpha \frac{1}{2} \log \det \left( \hat{\Delta}^{(2)} - d\tau^2 \hat{\Sigma} \right) + \alpha \frac{1}{2} \log \det \left( \hat{\Delta} - d\tau^2 \hat{\Sigma} \right) + \mathcal{O}(\alpha^2)
\end{equation}
where $S^{(2)}_L = I_{\alpha = 0} - I_{\alpha = 1}$, and $\hat{\Sigma} = \hat{\Sigma}^{A} = \hat{\Sigma}^{\overline{A}}$ is the fermion self-energy evaluated on the $\alpha = 0$ saddle-point. This expression is equivalent to the fermion determinant expression in Eq.~\eqref{eq:fermiondetrenyidiff}. To find the leading order contribution coming from the twist fields, we must therefore consider a ratio of determinants
\begin{equation}
    \Delta S^{(2)} = -\frac{1}{2} \log \left[ \frac{\det \left( \hat{\Delta}^{(2)} - d\tau^2 \hat{\Sigma} \right) }{\det \left( \hat{\Delta} - d\tau^2 \hat{\Sigma} \right)} \right] = -\frac{1}{2} \log \det \hat{M}
\end{equation}
where
\begin{equation}
    \hat{M} = \left( \hat{\Delta}^{(2)} - d\tau^2 \hat{\Sigma} \right) \left( \hat{\Delta} - d\tau^2 \hat{\Sigma} \right)^{-1}
\end{equation}

\begin{figure}[h!]
    \centering
    \includegraphics[width=0.7\textwidth]{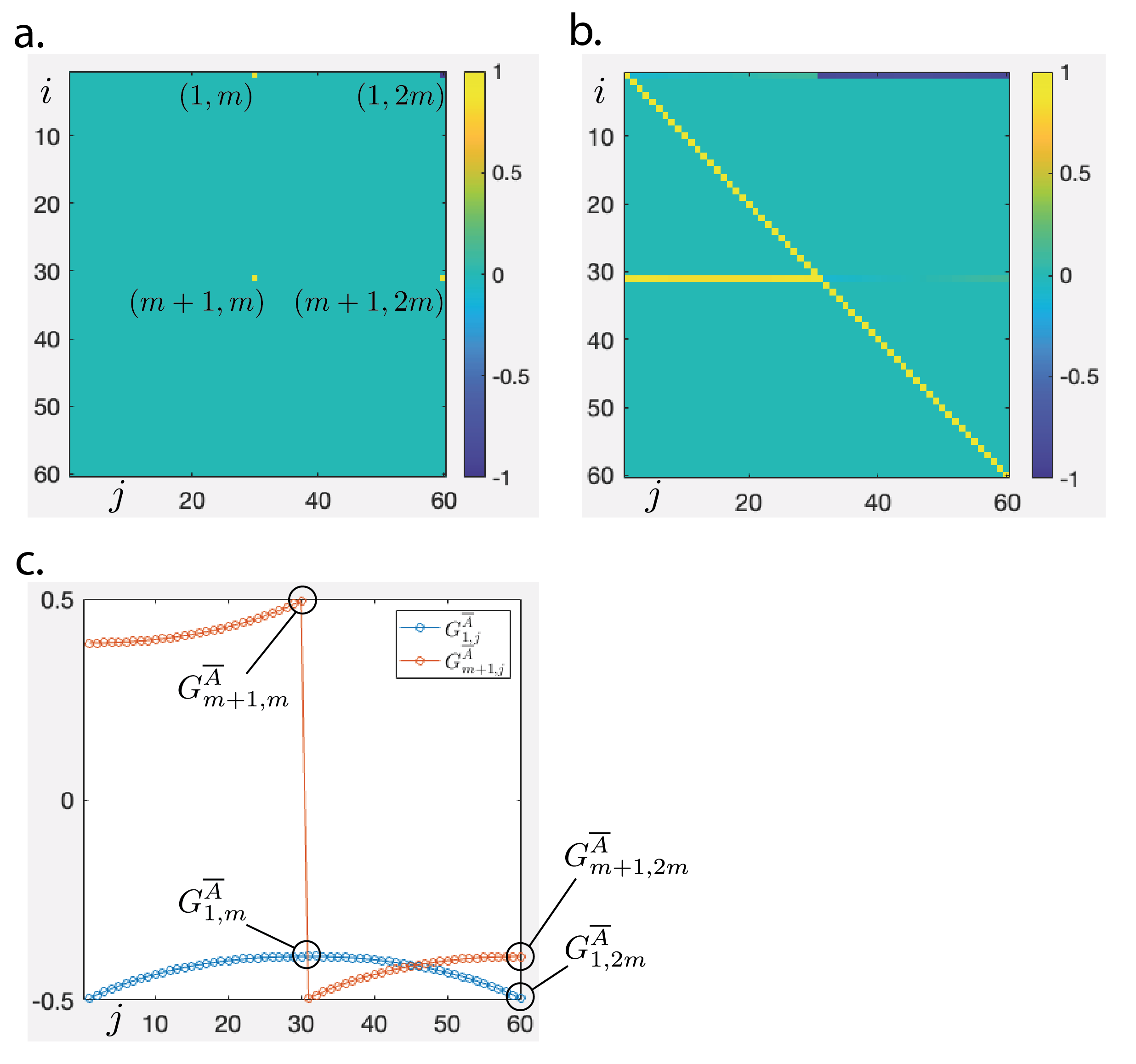}
    \caption{Evaluating the fermion determinant to obtain the lowest-order contribution from the twist fields. The matrix $\hat{d}$ (a, plotted for $n = 2m = 60$) has only four nonzero entries, which simplifies the remainder of the calculation. The matrix $\hat{M}$ (b) has a simple structure, with ones along the diagonal and two nonzero strips along rows $i = 1$ and $i = m+1$. The resulting determinant $\det \hat{M}$ only includes four entries (c) from the Green's function $G^{\overline{A}}_{ij}$.}
    \label{fig:detmatrices}
\end{figure}

To proceed, it is easiest to discretize the functions $\hat{\Delta}, \hat{\Delta}^{(2)}, \hat{\Sigma}$ into $n \times n = 2m \times 2m$ matrices (with $n = 2m$ even) and consider the limit $d \tau \rightarrow 0, n \rightarrow \infty$ at the end. As illustrated in Fig. \ref{fig:detmatrices}.a., the matrices $\hat{\Delta},\hat{\Delta}^{(2)}$ differ at only four discrete points, so we write
\begin{equation}
    \hat{\Delta}^{(2)} = \hat{\Delta} + \hat{d}
\end{equation}
where
\begin{equation}
    d_{ij} = \delta_{(i,j)(1,m)} + \delta_{(i,j)(m+1,m)} + \delta_{(i,j)(m+1,2m)} - \delta_{(i,j)(1,2m)}
\end{equation}
and the matrix indices run from $i,j = 1,\ldots,2m$. Using the saddle-point equations of motion Eq.~\eqref{eq:flagpolealphaeoms}, we may therefore write the matrix $\hat{M}$ as
\begin{equation}
    \hat{M} = \mathbb{I} + \hat{d} (\hat{\Delta} - d\tau^2 \hat{\Sigma})^{-1} = \mathbb{I} - \hat{d} \left(\hat{G}^{\overline{A}} \right)^T.
\end{equation}

Due to the simple form of the matrix $\hat{d}$, the matrix $\hat{M}$ has an extremely simple structure, as illustrated in Fig. \ref{fig:detmatrices}.b.: it has ones along the diagonal, coming from the identity $\mathbb{I}$, plus two strips of nonzero elements along row $i = 1$ and row $i = m+1$, which are picked out by the nonzero elements of the matrix $\hat{d}$. We can easily compute the determinant of this matrix using an expansion in minors to find
\begin{equation}
    \det \hat{M} = M_{1,1} M_{m+1,m+1} - M_{1,m+1} M_{m+1,1}.
\end{equation}
All that remains is to find and evaluate these four elements. Multiplying together $- \hat{d} \left(\hat{G}^{\overline{A}} \right)^T$ we find:
\begin{align}
    M_{1,1} &= 1 + G^{\overline{A}}_{1,2m} - G^{\overline{A}}_{1,m} \nonumber \\
    M_{m+1,m+1} &= 1 - G^{\overline{A}}_{m+1,m} - G^{\overline{A}}_{m+1,2m} \nonumber \\
    M_{1,m+1} &= G^{\overline{A}}_{m+1,2m} - G^{\overline{A}}_{m+1,m} \nonumber \\
    M_{m+1,1} &= -G^{\overline{A}}_{1,m} - G^{\overline{A}}_{1,2m}
\end{align}
which only involve four discrete elements of the Green's function $G^{\overline{A}}_{ij}$. Taking the limits $d \tau \rightarrow 0, n \rightarrow \infty$, and using the fact that $G^{\overline{A}}_{ij} \rightarrow G^{\overline{A}}(\tau,\tau') \approx G(\tau - \tau')$ for small $K \ll N$, where $G(\tau'-\tau)$ is the unperturbed Green's function on a thermal circle of length $2 \beta$, we have two UV elements
\begin{align}
    G^{\overline{A}}_{1,2m} &\rightarrow G(\epsilon - 2\beta) = -\frac{1}{2} \nonumber \\
    G^{\overline{A}}_{m+1,m} &\rightarrow G(\epsilon) = \frac{1}{2}
\end{align}
and two IR elements
\begin{align}
    G^{\overline{A}}_{1,m} &\rightarrow G(\epsilon-\beta) = - \frac{b}{(\beta J)^{2 \Delta}} \nonumber \\
    G^{\overline{A}}_{m+1,2m} &\rightarrow G(\epsilon-\beta) = - \frac{b}{(\beta J)^{2 \Delta}}
\end{align}
where $b$ is an $\mathcal{O}(1)$ constant, $\epsilon \rightarrow 0$ is an infinitesimal element, and $\Delta$ is the SYK scaling exponent ($\neq \hat{\Delta}$). We can also find these values directly from numerics as shown in Fig. \ref{fig:detmatrices}.c. Putting this all together, we finally find
\begin{equation}
    \det \hat{M} = \frac{1}{2} \left( 1 + \frac{2 b}{(\beta J)^{2 \Delta}} \right)^2
\end{equation}
and therefore
\begin{align}
    e^{-S_{\overline{A}}^{(2)}} &= e^{-S_L^{(2)}} e^{\frac{K}{2} \log \det \hat{M}} \nonumber \\
    &= e^{-S_L^{(2)}} 2^{-K/2} e^{K \log (1 + \frac{2b}{(\beta J)^{2 \Delta}})} \nonumber \\
    &\approx e^{-S_L^{(2)}} 2^{-K/2} e^{ K \frac{(\#)}{(\beta J)^{2 \Delta}}}
\end{align}
where in the last line we have expanded the logarithm assuming $\beta J \gg 1$. We therefore find that the lowest-order contribution from the twist fields is $2^{-K/2}$, which is independent of the Green's function $G$, followed by the smaller contribution $e^{ K \frac{(\#)}{(\beta J)^{2 \Delta}}}$, which does depend on the Green's function, as claimed. Similar arguments applied to the Renyi entropy $S_A^{(2)}$ yield
\begin{equation}
    e^{-S_A^{(2)}} = 2^{-K/2}
\end{equation}
to lowest order. Putting this all together, we find that the mutual information scales like
\begin{equation}
    I(A:R) \approx \frac{K (\#)}{(\beta J)^{2 \Delta}}
\end{equation}
as claimed in the main text.

\subsection*{Comparison to exact diagonalization}

Here we discuss a comparison to small size exact diagonalization calculations. We consider the $q=4$ SYK model on $N$ fermions ($N/2$ qubits via Jordan-Wigner) and set 
\begin{equation}
    \rho_C = \frac{e^{-\beta H}}{\mathrm{tr}(e^{-\beta H})}\bigg|_{\beta = N/J}.
\end{equation}
We then directly evaluate the 2nd Renyi mutual information for $K=2$, equivalent to one qubit. We do this by defining the analog of the reduced density matrix for subsets of fermions as a sum over all operators built from this subset, with coefficients determined by matching expectation values with the full state. Then the 2nd Renyi is proportional to the logarithm of the sum of the squares of these coefficients. This is also what the twist field definition of entanglement yields.

\begin{figure}
    \centering
    \includegraphics[width=.6\textwidth]{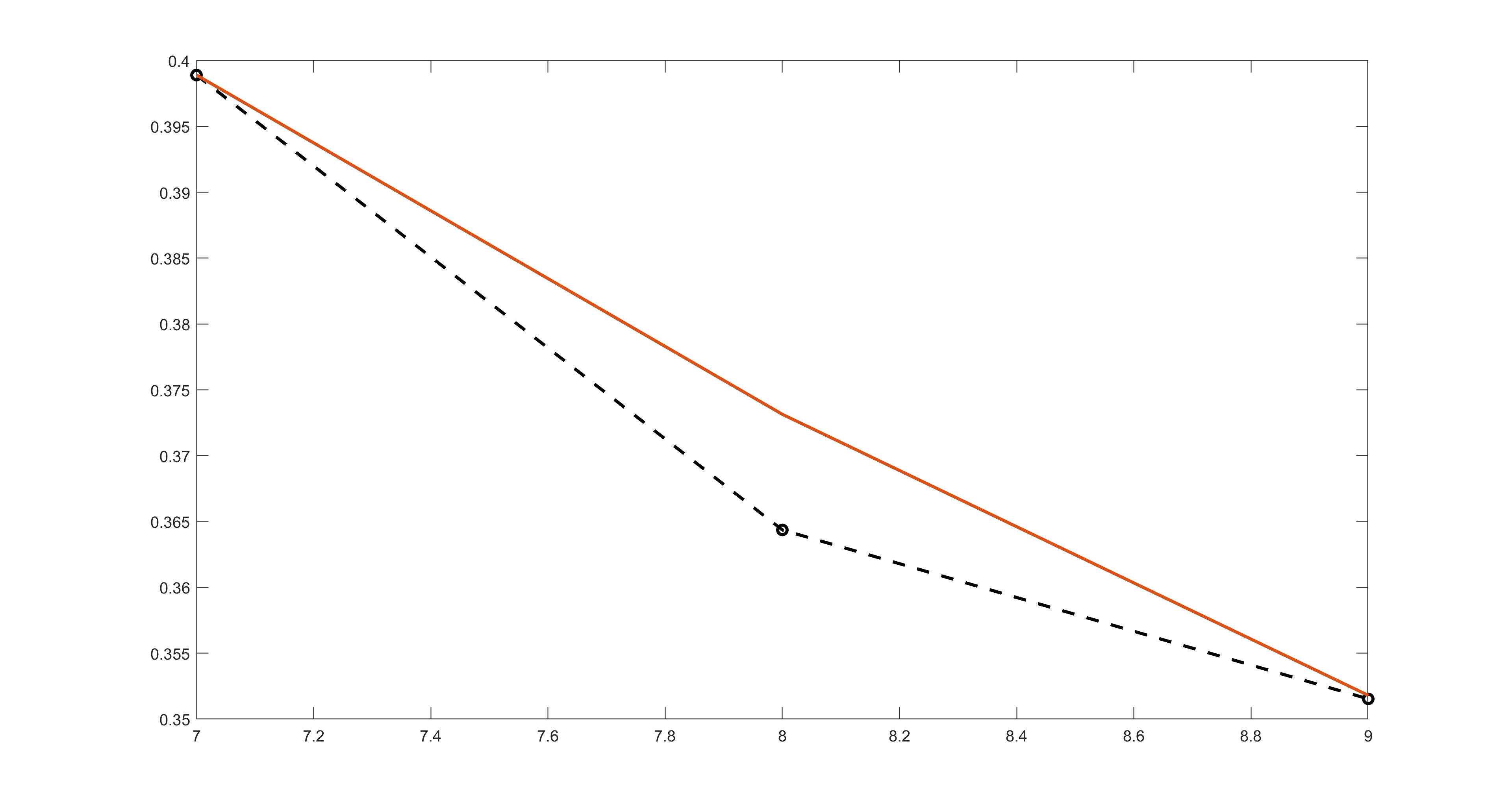}
    \caption{The $K=2$ mutual information as a function of $N/2$ with $\beta=N/J$. The black dots are the data, with each point representing $4$ samples. The red curve is $1/N^{1/2}$, normalized to the first point. It is hard to conclude anything very definite from these small sizes, but we do see a slow decrease of the mutual information with $N$ which is roughly consistent with our prediction.}
    \label{fig:ferm_mi_ed}
\end{figure}

Since we iterate over all fermion operators, the code is slow to run in its present form. It would be interesting to improve it and carry out a more systematic numerical investigation. For now, we content ourselves with a small scale consistency check of our main results, looking at the $K=2$ mutual information as a function of $N$ at $\beta=N/J$ for $N=14, 16, 18$. The data is shown in Fig.~\ref{fig:ferm_mi_ed}.

\section{Holographic calculation of the mutual information}
\label{app:holography}

In this appendix, we give a detailed holographic calculation of $I(K)$ by computing the following combination of generalized entropies:
\begin{equation}
I(K_{L} : R) = S_{\text{gen}}(K_{L}) + S_{\text{gen}}(R) - S_{\text{gen}}(K_{L} \cup R)
\end{equation}
The standard context in which in one computes entropies using the generalized entropy and quantum extremal surface prescription is for spatial subregions. That setting does not make sense for SYK / 2d gravity where there are no spatial subregions on the boundary. In the main text, we discussed a purely 2d model of $S_{\text{gen}}$ which extends the usual QES formalism to address subsets of fermions~\cite{qglab_2,syk_code_cl,antonini_holo_meas_2d}. 

We will return to the 2d model in the following Appendix~\ref{app:lr_qes} where we explicitly compute the shift in the QES as a function of $K$. This appendix represents a small detour in our logic in which we consider another completion of the 2d gravity theory in terms of a higher dimensional black hole setup. The resulting low energy physics, namely many fermions coupled to JT gravity, is the same, but the shift in perspective has been valuable for us when thinking about subtleties in the 2d QES prescription. All the results we obtain are consistent with the microscopic SYK calculations and the 2d QES calculations where they overlap. The analysis discussed here may be of independent interest beyond the main results of this paper.

In this appendix, we provide a fuller analysis, including an alternative perspective on the 2d QES prescription via a dimensional reduction. Specifically, we map a choice of subset of fermions to a choice of spatial partition in a higher dimensional setup and argue that the 2d QES prescription used in the main text is obtained after dimensional reduction. We address some key subtleties, including the fact that - from the higher-dimensional perspective - the Ryu-Takayanagi surface of a boundary subregion does not have rotational symmetry around the transversal sphere. So, the usual dimensional reduction to obtain JT gravity technically speaking does not work in this case.

The idea is to start with the magnetic Reissner-Nordström-AdS solution in 3+1 dimensions. It is a solution of the equations of motion arising from the Einstein-Hilbert action for a metric $h$ coupled to a $U(1)$ gauge field $A$:
\begin{equation}
I = \int d^{4}x \sqrt{-h} \left( \frac{R}{16\pi G_{N}} - \frac{F^{2}}{4g^{2}} \right)
\end{equation}
The solution is given by:
\begin{equation}
    ds^{2} = -\left( 1 - \frac{2M G_{N}}{r} + \frac{\pi q^{2} G_{N}}{g^{2} r^{2}} + \frac{r^{2}}{L^{2}} \right) dt^{2} + \left( 1 - \frac{2M G_{N}}{r} + \frac{\pi q^{2} G_{N}}{g^{2} r^{2}} + \frac{r^{2}}{L^{2}} \right)^{-1} dr^{2} + r^{2} (d\theta^{2} + \sin^{2}{\theta}d\phi^{2})
\end{equation}
\begin{equation}
    A = \frac{q}{2} \cos{\theta} d\phi.
\end{equation}

The position of the horizon of the black hole is obtained as a solution to 
\begin{equation}
    f(r) =  \left( 1 - \frac{2M G_{N}}{r} + \frac{\pi q^{2} G_{N}}{g^{2} r^{2}} + \frac{r^{2}}{L^{2}} \right) = 0 .
\end{equation}
We are interested in the near extremal limit  of this black hole, corresponding to nearly zero temperature. In this limit, the black hole develops a long throat described by a nearly AdS$_2$ geometry (times the transverse sphere). We can thus think of the geometry as having an AdS$_4$ region at the asymptotic boundary which crosses over into a nearly AdS$_2$ region at some radial scale $r_c$ set by $q$.

Just to give some concrete formulas, let us consider cases where the black hole is much larger than the AdS radius. We thus drop the $1$ term in $f$ above. In the extremal limit, both $f$ and $f'$ vanish at the horizon. Setting $a = 2 G_N M$ and $b = \pi q^2 G_N / g^2$ and working in units where $L=1$, we must solve
\begin{equation}
    - a / r + b / r^2 + r^2 = 0
\end{equation}
and
\begin{equation}
    a/r^2 - 2 b/r^3 + 2r = 0.
\end{equation}
The solution is 
\begin{equation}
    r_h = (b/3)^{1/4} \sim q^{1/2} G_N^{1/4} / g^{1/2}
\end{equation}
with
\begin{equation}
    a = 4 r_h^3 = 4 (b/3)^{3/4}.
\end{equation}

The entropy of the black hole in the extremal limit is 
\begin{equation}
    S_{T=0} = \frac{4 \pi r_h^2}{4 G_N} = \frac{\pi^{3/2} }{3^{1/2} G_N^{1/2} g} q.
\end{equation}
It is useful to recast this in terms of the magnetic length, $\ell_B$, and the Planck length, $\ell_P$. These are defined by
\begin{equation}
    \ell_B^2 \sim r_h^2/q \sim  G_N^{1/2}/g.
\end{equation}
and
\begin{equation}
    \ell_P = G_N^{1/2}.
\end{equation}
In terms of these lengths, the black hole entropy can be written 
\begin{equation}
    S \sim q \frac{\ell_B^2}{\ell_P^2}.
\end{equation}
We assume that the magnetic length is much larger than the Planck length, $\ell_B \gg \ell_P$, in which case the black hole entropy is much larger than $q$.

This discussion has all concerned one exterior region, say the left. However, we can introduce a second exterior region on the right, with the maximal analytic extension of the above metric describing a wormhole geometry corresponding, via AdS/CFT, to a thermofield double state of two coupled CFTs in the presence of opposite background magnetic fluxes.

So, we now have our AdS$_2$ region and wormhole. The next step is to introduce fermions. This is done by allowing a Dirac field to propagate on the black hole spacetime. The quantum Hall effect in 3 spatial dimensions predicts that the fermions are localized to the magnetic field lines, and indeed, a calculation similar to \cite{maldacena2020traversable} confirms that there are massless fields moving along the field lines. Thus, each field line gives us a dimensionally reduced system in 2 dimensions.

The number of field lines is obtained by integrating the magnetic flux over the transverse sphere and dividing by the flux quantum, which is $2\pi$ in the present units. The field strength is
\begin{equation}
    F = dA = - \frac{q}{2} \sin \theta d\theta d \phi,
\end{equation}
which gives a total magnetic flux of 
\begin{equation}
    \int_{S^2} F = 2 \pi q,
\end{equation}
independent of the radius of the $S^2$. The number of field lines---and the number of 2d fermion modes---is thus $q$.

\begin{figure}
    \centering
    \includegraphics[width=.8\textwidth]{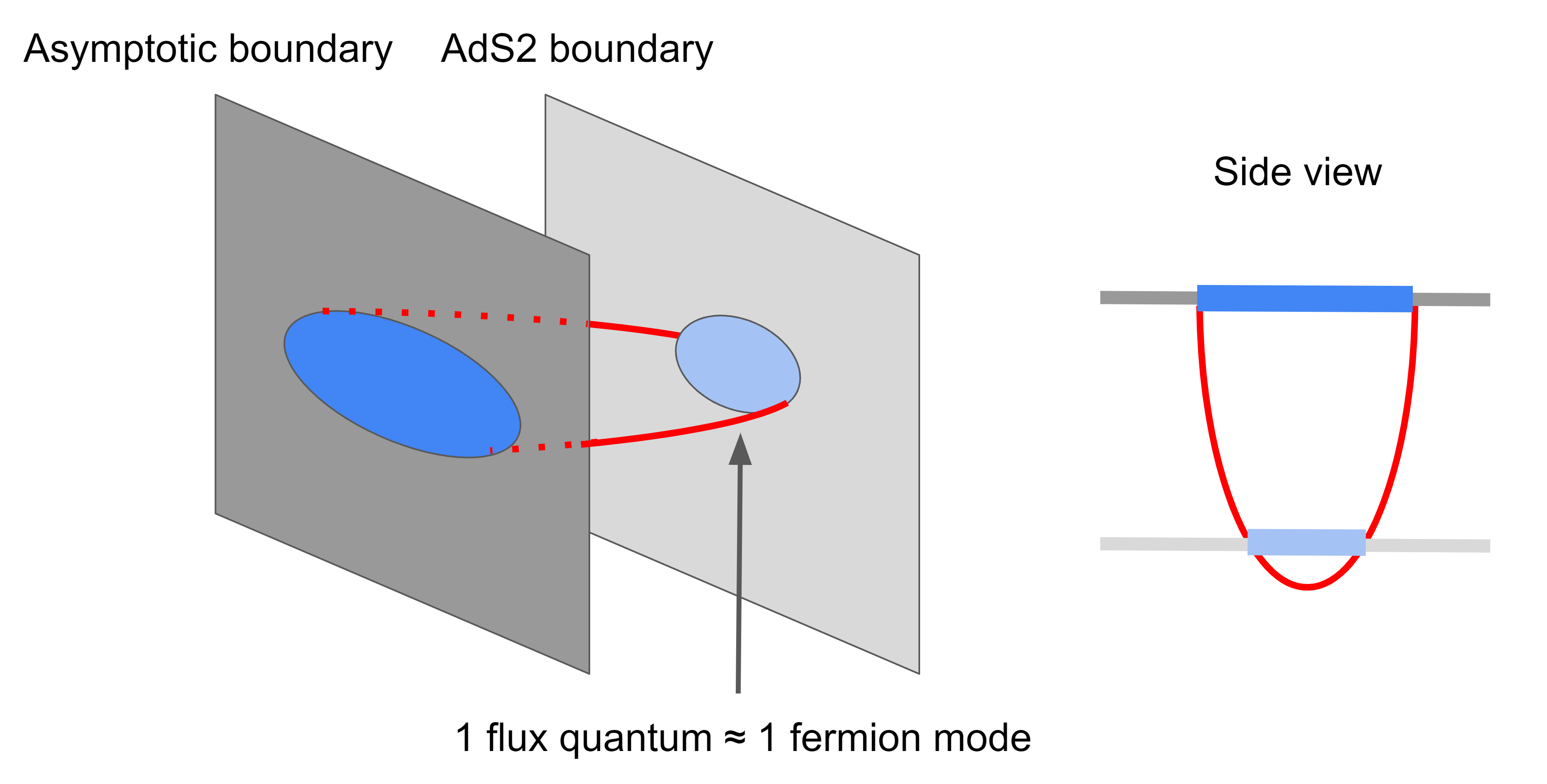}
    \caption{Schematic showing how regions at the asymptotic boundary can be chosen so as to enclose one flux quantum and thereby gain access to the degrees of freedom of one of the fermion modes at the cutoff of the nearly-AdS$_2$ region.}
    \label{fig:mag_regions}
\end{figure}

Next we establish a correspondence between spatial subregions and subsets of fermions. The key intuition is that each emergent 2d fermion mode is associated with a patch of the transverse sphere enclosing one flux quantum. Therefore, to access one fermion mode, we need access to the algebra of fermion observables in such a patch at the boundary, $r_c$, of the nearly AdS$_2$ region. This can be achieved using a spatial subregion of the asymptotic boundary chosen so that its Ryu-Takayanagi (RT) surface intersects the nearly AdS$_2$ cutoff at $r_c$ in a curve that encloses one flux quantum, as depicted in Figure~\ref{fig:mag_regions}.  

Note that, as always, we should in principle use the quantum extremal surface (QES) prescription, which replaces the RT surface by the QES. Here the QES prescription for a given boundary spatial subregion instructs us to construct bulk surfaces $\gamma$ homologous to the boundary region and minimize the generalized entropy,
\begin{equation}
    S_{\text{gen}}(\gamma) = \frac{A_\gamma}{4 G_N} + S_{\text{bulk}}(\gamma).
\end{equation}
However, if the RT surface includes only a small number of fermion modes and we are working in the regime $\ell_B \gg \ell_P$, then the QES is only slightly different from the naive RT surface, i.e. the geometrical piece of the generalized entropy dominates.

\begin{figure}
    \centering
    \includegraphics[width=.6\textwidth]{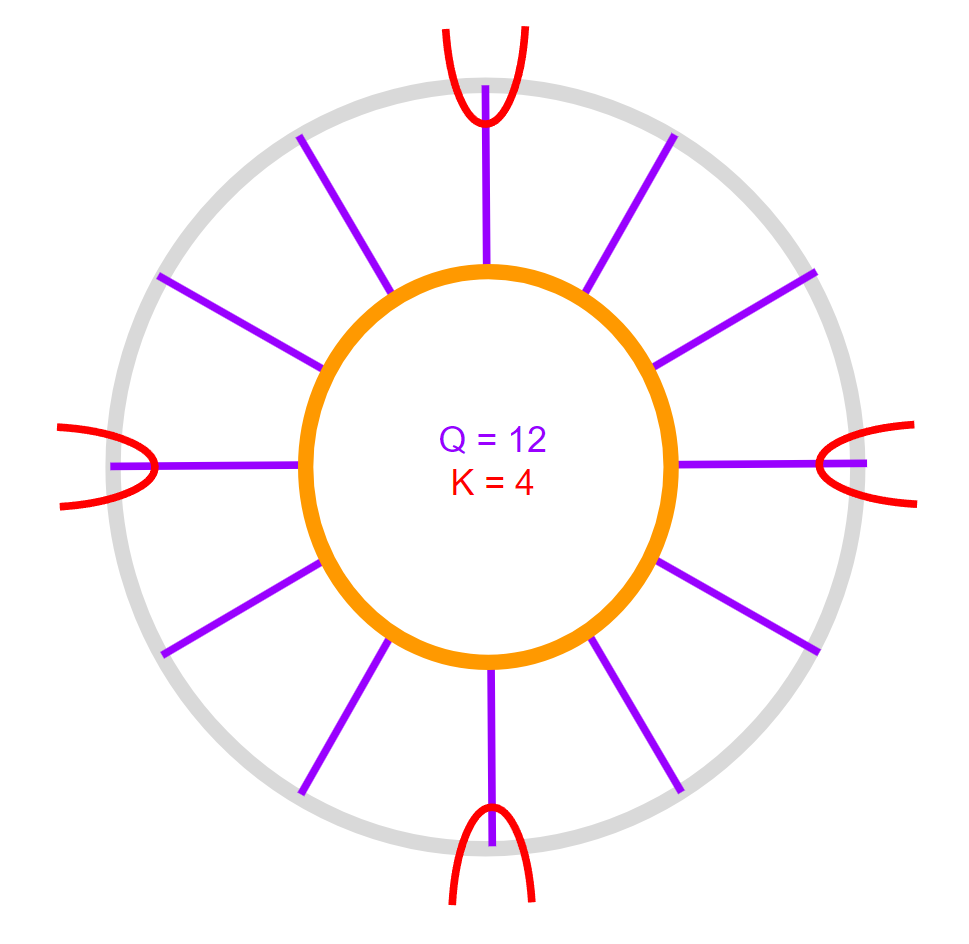}
    \caption{This is a sketch of the left exterior where the circles represent one of the two transverse spatial directions of the S$^2$. The grey outer circle represents the cutoff of the nearly AdS$_2$ region, where the geometry smooth joins onto the asymptotic AdS$_4$ region. The orange circle represents the black hole horizon. The purple lines indicate magnetic field lines. The red arcs denote the lower tips of RT surfaces as in Figure~\ref{fig:mag_regions}. By choosing a sufficiently dilute set of boundary regions, the full RT surface is the union of the red arcs and, in particular, does not dip further in towards the horizon.}
    \label{fig:dim_red_left}
\end{figure}

The above procedure gives access to one 2d fermion mode. To access $K$ modes, we need $K$ such regions. However, there are some important constraints. Clearly not all choices of spatial subregion enclosing roughly the same magnetic flux can have the same entropy, e.g. one big region versus many small regions will certainly have significantly different RT surfaces. However, at least in the regime where $K \ll q$ (with $q$ analogous to $N$ in the SYK model), we can isolate $K$ modes by choosing $K$ well separated boundary subregions as sketched in Fig.~\ref{fig:dim_red_left}. For such well separated regions, the RT surface of the union of boundary subregions is the union of the RT surfaces of each region separately. And this will be true independent of how precisely the regions are arranged.

The conclusion is that we have access to many choices of $K$ modes by choosing different well separated boundary subregions. This situation is analogous to the many choices of $K$ fermions in SYK. A choice which ``forgets'' the locality in the transverse space as much as possible is to distribute the spatial subregions randomly across the sphere in a distribution of approximately uniform density, such as one which maximizes the minimum distance between spatial subregions.

This mapping is certainly sensible in the regime $K \ll q$ of most interest here. If we take the limit of large $q$, it also makes sense for $K$ up to a small constant fraction of $q$. For even larger $K$s, we must begin to worry about individual RT surface components merging and about the RT surface qualitatively changing shape. For example, if we choose the entire boundary as the subregion, then the RT surface will reside at the bifurcation surface. This is relevant for the part of the mutual information involving the entire right boundary.

Even with this choice, the (higher-dimensional) QES for the union of $K$ modes on the left and the entire right will not exactly include the bifurcation surface of the black hole as one of its connected components. Rather, that connected component will have ripples due to forces localized at the angular directions of the $K$ left fermions. Such ripples are a potential danger for the 2-dimensional picture which we seek, because no single value of the transvere area of the sphere, which becomes the dilaton field in JT gravity, can describe such ripples. Fortunately, it is possible to argue that such ripples are small (and $1/N$ suppressed):
\begin{itemize}
    \item In the regime where $K \ll q$, the shift in the QES away from the naive RT result at the bifurcation surface is small in absolute magnitude, although it can have ripples of comparable size to the overall shift.
    \item In the regime where $K \propto q$, the QES can shift substantially away from the naive RT result (and into the left exterior). However, the ripples must still be small. This is because the characteristic scale of any rippling would be set by the density of left modes, $K/r_h^2 = \frac{K}{q} \frac{1}{\ell_B^2}$, leading to a length scale proportional to the magnetic length, $\ell_B \sqrt{q/K}$. Hence, if we work in a regime where the magnetic length is small (but still much larger than the Planck length), we can suppress transverse gradients in the QES and ensure an approximately uniform radial displacement.
\end{itemize}

\begin{figure}
    \centering
    \includegraphics[width=.8\textwidth]{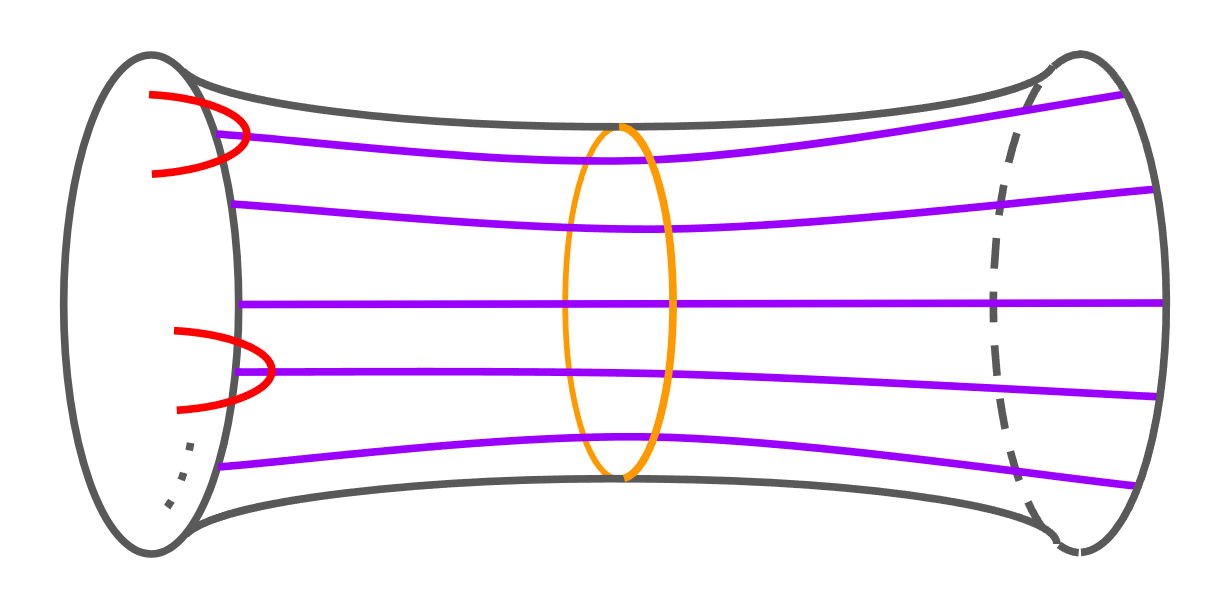}
    \caption{This illustrates the wormhole geometry and the possible higher-dimensional RT surfaces. The RT surface of the entire right side corresponds to the region to the right of the orange curve. This should be compared to panel (A) of Figure~\ref{fig:wormhole_entanglement}. In the regime where $K \ll N$, the RT surface of the $K$ left regions corresponds to the region to the left inside the red arcs. This should be compared to panel (B) of Figure~\ref{fig:wormhole_entanglement}. The RT surface of the combination of $K$ regions on the left and entire right is the union of the previous regions, at least when $K \ll N$. This should be compared to panel (C) of Figure~\ref{fig:wormhole_entanglement}. When $K$ is larger, comparable to $N^{2\Delta}$ or greater, then the RT surface must be generalized to a full QES and this QES can substantially shift from its small $K$ location. When a large fraction of the left boundary is included in the left subregion, the RT surfaces for just the left subregion and the combination of the left subregion and the entire right can even qualitatively change, morphing to include almost all the left exterior apart from small cutout regions associated with the complement of the left boundary subregion.}
    \label{fig:dim_red_left_right}
\end{figure}

Putting all these ingredients together, the following consistent 2d picture for the relevant QESs emerges (higher dimensional picture in Figure~\ref{fig:dim_red_left_right}, dimensionally reduced picture in Figure \ref{fig:wormhole_entanglement}): the QES for $K_{L}$ is some small interval near the left-boundary; the QES for $R$ is located at the bifurcation surface; and the QES for $K_{L} \cup R$ is defined by two intervals as in panel (C) of Figure \ref{fig:wormhole_entanglement}: a small interval $l$ near the left-boundary, and a larger interval $r$ from somewhere near the bifurcation surface to the right boundary. There are $K$ fermions contributing to the bulk entropy of both $l$ and $r$, and the remaining $N-K$ fermions contribute to the bulk entropy of $r$ only.

Given this 2d QES prescription, the bulk mutual information is then the main contribution to the full mutual information in the regime where $K$ is small compared to $N$ and $N^{2\Delta}$. In order to compute the mutual information in the bulk, we can use either the techniques developed in \cite{agon2016quantum, cardy2013some} and \cite{syk_code_cl}, or the ones developed in \cite{casini2009entanglement}.

\subsection*{Applying the Cardy/Agon-Faulkner method}

Here we review method of Cardy and Agon-Faulkner for computing the mutual information in the large separation limit and then apply it to our case.

Let $\mathcal{M}$ be the manifold on which the bulk quantum field theory (QFT) lives. Let's start with the R\'{e}nyi entropy of a subregion $X$:
\begin{equation}
    S_{X}^{(n)} = \frac{1}{1-n} \log \mathrm{Tr}_{X} \rho_{X}^{n}
\end{equation}
which can also be written in terms of a path-integral over a conifold defined by taking $n$ copies of the QFT and sewing them together appropriately along the subregion $X$:
\begin{equation}
    S_{X}^{(n)} = \frac{1}{1-n} \log{\left( \frac{Z(\mathcal{C}_{X}^{(n)})}{Z^{n}} \right)}
\end{equation}
where $Z(\mathcal{C}_{X}^{(n)})$ is the partition function on the conifold, and $Z$ is the partition function on the original space. We are interested in the mutual information:
\begin{equation}
    I(A,B) \equiv \lim_{n \rightarrow 1} I^{(n)}{(A,B)}
\end{equation}
with $I^{(n)}{(A,B)}$ the R\'{e}nyi mutual information, defined by:
\begin{equation}
    I^{(n)}{(A,B)} \equiv S_{A}^{(n)} + S_{B}^{(n)} - S_{AB}^{(n)}
\end{equation}
or, in terms of path integrals,
\begin{equation}
    I^{(n)}{(A,B)} = \frac{1}{1-n} \log{\left( \frac{Z(\mathcal{C}_{AB}^{(n)})Z^{n}} {Z(\mathcal{C}_{A}^{(n)}) Z(\mathcal{C}_{B}^{(n)})} \right)}.
\end{equation}

To make progress from this rather general formula, we invoke the idea that, when the two subregions are very far away from each other, we can think about the sewing operation of the replica sheets as some semi-local operators that couple the sheets together:
\begin{equation}
    \frac{Z(\mathcal{C}_{AB}^{(n)})}{Z^{n}} = \langle \Sigma_{A}^{(n)} \Sigma_{B}^{(n)} \rangle_{\mathcal{M}^{n}}
\end{equation}
where $\mathcal{M}^{n}$ is n copies of the uncoupled manifold, and the semi-local operator $\Sigma_{A}^{(n)}$ takes the form of an expansion:
\begin{equation}
    \Sigma_{A}^{(n)} = \frac{Z(\mathcal{C}_{A}^{(n)})}{Z^{n}} \sum_{\{ k_{j} \}} C^{A}_{\{ k_{j} \}} \prod_{j=0}^{n-1}  \Phi^{(j)}_{k_j} (r_{A})
\end{equation}
for some complete set of operators $\Phi_{k_j}^{(j)}{(r_{A})}$ of the $j$th copy of the QFT, some location $r_{A}$ chosen conventionally in the subregion $A$ (for example the midpoint or center of the subregion), and some coefficients $C^{A}_{ \{ k_j \} }$. Without loss of generality, those operators can be chosen to have zero 1-point function, by a appropriate subtraction. The coefficients can in principle be found by considering the expectation value of a product of operators in the conifold geometry:
\begin{eqnarray}
\langle \prod_{j'=0}^{n-1} \Phi_{k'_{j'}}^{(j')} (r) \rangle_{\mathcal{C}_{A}^{(n)}} &=& \langle \prod_{j'=0}^{n-1} \Phi_{k'_{j'}}^{(j')} (r) \sum_{ \{ k_j \} } C^{A}_{\{ k_j \} } \prod_{j=0}^{n-1} \Phi_{k_j}^{(j)} (r_{A}) \rangle_{\mathcal{M}^{n}} \nonumber \\
&=& \sum_{\{ k_j \} } C^{A}_{\{ k_j \} } \prod_{j} \langle \Phi_{k_{j}'}(r) \Phi_{k_{j}}(r_{A}) \rangle_{\mathcal{M}}
\end{eqnarray}
We note that the above does not assume conformal symmetry, so it's valid for a general bulk QFT.

Following Faulkner–Lewkowycz–Maldacena (FLM) \cite{faulkner2013quantum}, the leading-order contribution to the mutual information comes from a pair of exchanges of the lightest operator between the two subregions. So, we end up with
\begin{equation}
    I_{\text{QFT}} \sim C G^{2}
\end{equation}
where $G$ is the correlation of the exchanged bulk operators evaluated at the minimal separation between $A$ and $B$ and $C$ is a constant. Our task then is to apply this formula to our situation, as discussed in the main text.

Here we address one subtlety, which is that we can apply this formula both from the higher-dimensional and 2d points of view. The 2d perspective is relatively simple. We apply the formula to each 2d mode separately. This is justified because we assume that direct interactions between different modes are suppressed, e.g. by $1/N^{\#}$ in SYK. This is essentially the discussion in the main text. What remains is to show that the higher-dimensional perspective is consistent with this estimate.

For the higher-dimensional point of view, we are studying correlations of a bulk fermion in a strong magnetic field. We will assume the fermion is massless, but it could have a bare mass as well which would then result in massive 2d modes. The lowest Landau level contains massless effectively 2d modes which move radially along the field lines. By contrast, correlations amongst different 2d modes, which are separated on the transverse sphere, decay exponentially with a length scale set by the magnetic length. Higher Landau levels have a similar decay, as well as an effectively large mass for the 2d motion. Because of the effective exponential decay, while different 2d modes may be locally correlated, that correlation falls off rapidly with distance along the transverse sphere. In particular, in the dilute limit of small $K$, the different modes associated with the $K$ boundary subregions on the left are all well separated. Hence, we again arrive at a picture in which the correlations of different 2d modes are approximately decoupled. We also need to assume that no other bulk fields contribute significantly to the mutual information, which will be true if their masses are sufficiently great in units of the AdS$_2$ length.

\section{Calculating the shift of the $S(K_L \cup R)$ QES}
\label{app:lr_qes}

In this appendix, we revisit the question of shifting saddle points, this time in the context of the gravitational picture. We make use of the QES formula to evaluate the entropies entering into the two-side mutual information. The analysis reported here is complementary to the arguments in Appendix~\ref{app:extrap}.

In fact, it is worth emphasizing just how parallel the arguments are. Within the QES formalism, the dilaton contribution to the generalized entropy is $\frac{\phi(\sigma)}{4 G_N}$ with $G_N \sim 1/N$ and $\phi(\sigma)$ given by \eqref{eq:dil_profile} which is proportional to $1/\beta$ at large $\sigma$. Hence, the dilaton contribution goes like $N/\beta$ in the large $\sigma$ regime. Once $N/\beta$ is no longer large, the minimization procedure to extract the QES is no longer justified. A related issue is the possible shift of the $K_L \cup R$ QES from that of the $R$ alone due to the inclusion of the $K$ left fermions. These issues may seem somewhat different from the question of whether we can use an unperturbed saddle point to compute the mutual information, but they are in fact closely related. Indeed, standard derivations of the Ryu-Takayanagi formula (as well as the QES formula) start by inserting twist fields to compute the $\alpha$-th Renyi entropy. These twist fields on the boundary are dual to special branes in the bulk. These branes have a large tension when $\alpha \neq 1$, of order the number of degrees of freedom, and they contribute a term proportional to their area times their tension to the action. One then computes the gravitational path integral by saddle point and takes the $\alpha \to 1$ limit to obtain the von Neumann entropy. 

This process is quite parallel to how we are evaluating Renyi entropies for the microscopic SYK analysis. As with SYK, there are two major issues which could arise: (a) the saddle point approximation itself could break down or (b) the saddle point could shift appreciably. In the holographic context, these translate to: (a') the minimization procedure inherent in the QES formula could break down or (b') the QES of $K_L \cup R$ could shift relative to that of $R$ alone. 

Similar to the two scalings considered in Appendix~\ref{app:extrap}, we will consider a limit in which $\beta \sim N^{a}$ and take this to mean that we are always safe to minimize over the location of the QES. This addresses point (a'). To address (b'), we must estimate the potential shift in the $K_L \cup R$ QES relative to the $R$ QES. We first give an intrinsically 2d analysis and then revisit the analysis from the dimensional reduction perspective. 

Within the 2d framework, we make a few technical assumptions in the  analysis:
\begin{itemize}
    \item The entanglement entropy of a single bulk interval does not depend on the endpoint of the interval. 
    \item The only component of the QES which shifts for general $K$ is the part which limits onto the bifurcation surface as $K \to 0$. 
    \item The bulk mutual information is estimated using the geodesic approximation with $m \to \Delta$.
\end{itemize}
The third of these is purely for simplicity and will not affect the result in the limit of large separations. The first and second assumptions are more important and we discuss them further below from the higher dimensional perspective. It is also worth noting that, unfortunately, analytical formulas for the bulk entanglement entropies of 1- and 2-interval regions are not known for general masses, so we cannot be as concrete as we might wish. Nevertheless, the crucial physics of the long-range correlations is under control.

With these assumptions, it follows that the ``force'' pulling the QES into the left exterior region comes from the bulk mutual information, since the bulk entropy for each interval separately is independent of endpoint. Indeed, without the bulk mutual information, the dilaton would be the only part of the generalized entropy that varies non-trivially and the dilaton alone is minimized at the bifurcation surface. So the bulk mutual information wants to pull the QES into the left exterior, and this tendency is resisted by the dilaton contribution.

We need to find the minimum of
\begin{equation}
    \frac{\phi}{4 G_N} - I_{\text{bullk}}.
\end{equation}
The dilaton profile is known, see \eqref{eq:dil_profile}, which we copy here for convenience,
\begin{equation}
    \phi = \phi_0 + \phi_r \frac{2\pi}{\beta} \frac{1}{\tanh \frac{2\pi \sigma}{\beta}}.
\end{equation}
The other contribution is the bulk mutual information. Suppose the QES shifts from $\sigma=\infty$ (bifurcation surface) to some $\sigma = \sigma_*$ (left exterior). The proper distance from the AdS$_2$ boundary to the QES is
\begin{equation}
    d(\epsilon,\sigma_*) = \int_{2\pi \epsilon/\beta}^{2\pi \sigma_* /\beta} du \frac{1}{\sinh u} = \ln \frac{\tanh \frac{\pi \sigma_*}{\beta}}{\tanh \frac{\pi \epsilon}{\beta}}.
\end{equation}
At large separation, the mutual information can be approximated as proportional to $G_{\text{bulk}}^2$ (see Appendix~\ref{app:holography}), so we obtain
\begin{equation}
    I_{\text{bulk}} \sim K G_{\text{bulk}}^2(\epsilon,\sigma_*) \sim K e^{- 2 m d(\epsilon,\sigma_*)}.
\end{equation}

It is convenient to define the variable $w = e^{-2\pi \sigma_*/\beta}$, in terms of which the dilaton contribution is
\begin{equation}
    \frac{\phi_r}{4 G_N} \frac{2\pi}{\beta} \frac{1+w^2}{1 - w^2}
\end{equation}
and the mutual information is
\begin{equation}
    I_{\text{bulk}} = K (\pi \epsilon/\beta)^{2\Delta} \left( \frac{1 + w}{1 - w}\right)^{2\Delta}.
\end{equation}
Note that we have freely dropped constant terms that do not depend on $w$. Absorbing all numerical factors into two constants, $a$ and $b$, and using $\epsilon \sim 1/J$ to match to SYK conventions, we must minimize
\begin{equation}
    \mathcal{S}(w) = a \frac{N}{\beta J} \frac{1+w^2}{1-w^2} - b \frac{K}{(\beta J)^{2\Delta}} \left( \frac{1 + w}{1 - w}\right)^{2\Delta}
    \label{eq:qes_min_fun}
\end{equation}
with respect to $w$.

When $K=0$, the minimum is $w=0$ corresponding to $\sigma_*=\infty$ and the bifurcation surface. This is the QES of $R$ alone. When the shift from the $R$ QES is small, we can find it by expanding in $w$. The Taylor expansion of $\mathcal{S}$ near $w=0$ is
\begin{equation}
    \mathcal{S}(0) - 4 \Delta b \frac{K}{(\beta J)^{2\Delta}} w + 2 a \frac{N}{\beta J} w^2 + \cdots
\end{equation}
where $\cdots$ denotes terms of order $w^3$, $K w^2$, and higher. The new minimum is at
\begin{equation}
    w_{\min} = \frac{b \Delta}{a} \frac{K}{N} (\beta J)^{1- 2\Delta}.
\end{equation}
For $\beta J \sim N$, this shift is of order
\begin{equation}
    w_{\min} \sim \frac{K}{N^{2\Delta}},
\end{equation}
which is small provided $K \ll N^{2\Delta}$. This is in complete accord with our arguments in Appendix~\ref{app:extrap}.

One virtue of the gravitational setup is that the QES calculation for larger values of $K$ is also analytically tractable. In particular, we can consider the situation when $K > N (\beta J)^{2\Delta -1}$. In this case, the QES can shift dramatically from its location at $K=0$. As an illustration, consider the case where $\sigma_* \ll \beta$ and $w \approx 1$. Introduce $u = 1-w$ so that $\mathcal{S}$ becomes
\begin{equation}
    \mathcal{S} \approx a \frac{N}{\beta J} \frac{1}{u} - b \frac{K}{(\beta J)^{2\Delta}} \left( \frac{2}{u} \right)^{2\Delta}.
\end{equation}
This is now minimimized at 
\begin{equation}
    u_{\min} \sim \left( \frac{N}{K} \right)^{1/(1-2\Delta)} \frac{1}{\beta J},
\end{equation}
and it is consistent to treat $u$ as small provided $K/(\beta J)^{2\Delta} \gg N/(\beta J)$. The cutoff surface representing the left boundary corresponds to $u$ equal to
\begin{equation}
    u_{\text{cut}} \sim \frac{1}{\beta J},
\end{equation}
so the QES remains far from the cutoff as long as we also have $K/N \ll 1$.

Setting $\beta J \sim N$, we can estimate the full boundary mutual information (not just the bulk mutual information) using the value of $\mathcal{S}$, since the other terms in the mutual information are subleading. The result is 
\begin{equation}
    \mathcal{S} \sim \left( \frac{K}{N^{2 \Delta}}\right)^{1/(1-2\Delta)},
\end{equation}
which increases as $K$ increases beyond $N^{2\Delta}$.

Once $K$ is of order $N$, $u_{\min}$ approaches $u_{\text{cut}}$ and we must consider the possibility of a qualitatively different QES. Indeed, we know that when $K$ is large enough, for example, as $K/N \to 1$, it must be the case that the mutual information approaches $2 s_0 N$ as the $K$ fermions will approach maximal entanglement with the reference. In the above formula, one finds that setting $K = O(N)$ gives $\mathcal{S} = O(N)$ and hence the $K$ fermions are now correlated with a significant fraction of the ``logical'' fermions on the right.

In general, we must minimize \eqref{eq:qes_min_fun} over $w \in [0,1-u_{\text{cut}}]$. As just discussed, we should also consider alternative QES configurations when $K/N$ is close to one, but we neglect those here with the understanding that the result below while break down when $K/N \lesssim 1$. Given the minimizer $w_{\min}$, the mutual information is $\mathcal{S}(0) - \mathcal{S}(w_{\min})$, and the above analysis indicates that it has two distinct regimes:
\begin{itemize}
    \item $K \ll N(\beta J)^{2\Delta -1}$ --- MI $\sim K/(\beta J)^{2\Delta}$,
    \item $K \gg N(\beta J)^{2\Delta -1}$ --- MI $\sim N (K/N)^{1/(1-2\Delta)}$.
\end{itemize}
We can confirm that these two formulas are of the same order when $K \sim N (\beta J)^{2\Delta -1}$, which is the crossover regime.

In terms of its dependence on $\alpha = K/N$, we thus expect the mutual information to start out linear in $\alpha$ (first regime just above) at small $\alpha$ and then to curve upward at larger $\alpha$ (second regime just above) before finally saturating at its maximum as $\alpha \to 1$ (not obtained from QES calculation). This behavior can be compared to the numerical results for general $\alpha$ in Figure~\ref{fig:miqescompare} where at the lowest temperature ($\beta \sim 13.2$, light blue curve) we begin to see to the development of an initial linear growth, upturn, and final saturation. We emphasize that the holographic formulation only really applies at large $\beta$, so it is not surprising that the agreement is at best qualitative. It is also worth noting that, in both the SYK and holographic models, it is possible for the mutual information to jump because of the minimization inherent in its calculation at large $N$. There do not appear to be jumps in the SYK numerics at the $\beta$s we accessed; we cannot rule out other solutions but correctly obtaining the full entropy as $\alpha \to 1$ is a clue that we have the right saddle. If the SYK calculations can be pushed to lower temperature, it would be interesting to extend the QES calculation to general $K/N$ and to compare in detail with the SYK results.

\subsection*{Dimensional reduction perspective}

Here we briefly analyze the situation from the higher-dimensional point of view. We consider three entropies in turn: (i) $S(R)$, (ii) $S(K_L)$, and (iii) $S(K_L \cup R)$. For each we use the higher-dimensional QES prescription in which the QES is a surface $\Sigma$ which minimizes the generalized entropy,
\begin{equation}
    S_{\text{gen}}(\Sigma) = \frac{\text{Area}(\Sigma)}{4 G_N} + S_{\text{bulk}}(\Sigma).
\end{equation}
For a given boundary region, we denote the ``geometric'' QES (which just minimizes the area term) by $\Sigma_0$ and the full QES by $\Sigma$. Throughout the analysis in this subsection, we consider $K \ll N$ with the $K$ higher-dimensional boundary regions chosen to be approximately uniformly spaced. 

(i): For $S(R)$, the true QES, $\Sigma_R$, will be equal to the geometric QES which is the bifurcation surface of the black hole.

(ii): For $S(K_L)$, we argued in Appendix~\ref{app:holography} that the geometric QES should be approximately a union of $K$ copies, suitably translated, of the $K=1$ QES. Call this surface $\Sigma_{K,0}$. As was discussed, there will be weak correlations between the bulk field degrees of freedom inside each component of the full QES. Moreover, $S_{\text{bulk}}$ can depend on the precise shape of $\Sigma$ via more than just its total area. Hence, the the true QES, $\Sigma_K$, may be slightly deformed away from $\Sigma_{K,0}$ due to these contributions to $S_{\text{bulk}}$. However, we expect any such deformation to be small so that the entropy is still linear in $K$ at small $K$. 

(iii): For $S(K_L \cup R)$, the geometric QES, $\Sigma_{K\cup R,0}$, is the union of the geometric QES from (i) and (ii). The bulk entropy is now even more complicated, since it includes all the effects from (ii) as well as new long-range correlations between the fields inside $\Sigma_{K,0}$ with those inside $\Sigma_{R}$. The adjective long-range is appropriate because, as was argued in Appendix~\ref{app:holography}, the components of $\Sigma_{K\cup R,0}$ associated with (i) and (ii) are well separated at large $\beta$. Hence, we expect the true $\Sigma_{K \cup R}$ to be the union of $\Sigma_K$ and $\Sigma_R$ plus additional small deformations due to the long-range correlations across the wormhole. This long-range part of $S_{\text{bulk}}$, which is what we called $I_{\text{bulk}}$, can lead to small shifts of both $\Sigma_R$ and $\Sigma_K$ components (the second technical assumption above amounts to the neglect of shifts in $\Sigma_K$). However, given such a small shift, we can appeal to the fact that $\Sigma_R$ and $\Sigma_K$ are separately minimal without $I_{\text{bulk}}$ to conclude that the leading shift in the generalized entropy of $K \cup R$ is just given by $I_{\text{bulk}}$ evaluated with $\Sigma_{K \cup R} = \Sigma_K \cup \Sigma_R$. (This is similar to the logic in Appendix~\ref{app:path_int_perturb} where we reviewed how a perturbation to a minimum due to a perturbation of the objective function only changes the unperturbed objective function at second order in the perturbation.)

From this analysis, we conclude that 
\begin{equation}
    \Sigma_{K \cup R} = (\Sigma_K + \delta \Sigma_K ) \cup (\Sigma_R + \delta \Sigma_R)
\end{equation}
and
\begin{equation}
    S_{\text{gen}} = S_{\text{gen}}(\Sigma_K + \delta \Sigma_K) + S_{\text{gen}}(\Sigma_R + \delta \Sigma_R) - I_{\text{bulk}}(\Sigma_{K \cup R}).
\end{equation}
Since $\Sigma_K$ and $\Sigma_R$ are already minimizers of $S_{\text{gen}}$, we have 
\begin{equation}
    S_{\text{gen}}(\Sigma_K + \delta \Sigma_K) = S_{\text{gen}}(\Sigma_K) + O(\delta \Sigma_K^2)
\end{equation}
and similarly for $S_{\text{gen}}(\Sigma_R + \delta \Sigma_R)$. Moreover, $I_{\text{bulk}}$ is already linear order in small quantities, so we have
\begin{equation}
    I_{\text{bulk}}(\Sigma_{K \cup R}) = I_{\text{bulk}}(\Sigma_{K} \cup \Sigma_{R}) + O(\delta \Sigma_{K \text{or} R}^2).
\end{equation}
Putting everything together, the left-right mutual information is 
\begin{equation}
    \text{MI} = S_{\text{gen}}(\Sigma_R) + S_{\text{gen}}(\Sigma_K) - S_{\text{gen}}(\Sigma_{K \cup R}) \approx I_{\text{bulk}}( \Sigma_K \cup \Sigma_R).
\end{equation}
Note that this is true regardless of precisely how $\Sigma_K$ shifts away from $\Sigma_{K,0}$.

\section{Eternal wormholes and NLAS}
\label{app:nlts}

In this appendix, we discuss how, after first taking the large $N$ limit, states with arbitrarily low energy density can be adiabatically prepared. Previous work has also argued that one can cool into these low energy states in a time independent of system size~\cite{maldacena2020syk}.

We first describe the construction for the case of SYK$_2$, since it is analytically tractable. Then we present numerical evidence consistent with \cite{maldacena2018eternal} that the corresponding SYK$_4$ setup also has a gapped family of auxiliary Hamiltonians whose ground states interpolate between high and low temperature. 

Consider the Hamiltonian
\begin{equation}
    H_2= \sum_{ab} i J_{ab} \chi_a \chi_b.
\end{equation}
Now introduce a second copy made of $\psi_a$ operators. We think of the $\chi$s as the left and the $\psi$s as the right. The Maldacena-Qi Hamiltonian is
\begin{equation}
    H=  \sum_{ab} i J_{ab} \chi_a \chi_b +  \sum_{ab} (-i) J_{ab} \psi_a \psi_b + i \mu \sum_a \chi_a \psi_a.
\end{equation}
Our goal is show that this $H$ has a gap for all non-zero $\mu$ and that its ground state has tunable energy as measured by $H_2$ (for the $\chi_a$ or $\psi_a$ systems).

Consider first a single pair of fermions on either side,  
\begin{equation}
    H = i J \chi_1 \chi_2 - i J \psi_1 \psi_2 + i \mu (\chi_1 \psi_1 + \chi_2 \psi_2).
\end{equation}
One can define the complex fermion operators,
\begin{equation}
    a=(\chi_1+i\chi_2)/\sqrt{2}
\end{equation}
\begin{equation}
    b = (\psi_1 + i\psi_2)/\sqrt{2},
\end{equation}
and then compute all the bilinears,
\begin{equation}
    a^\dagger a = 1/2 + i \chi_1 \chi_2 = \frac{1 + 2 i \chi_1 \chi_2}{2},
\end{equation}
and similarly for $b$, and,
\begin{equation}
    \chi_1 \psi_1 = (a+a^\dagger)(b+b^\dagger)/(\sqrt{2})^2
\end{equation}
\begin{equation}
    \chi_2 \psi_2 = (a-a^\dagger)(b-b^\dagger)/(\sqrt{2}i)^2
\end{equation}
\begin{equation}
    \chi_1 \psi_1 + \chi_2 \psi_2 = a b^\dagger + a^\dagger b.
\end{equation}

Using these formulae, the Hamiltonian can be rewritten as
\begin{equation}
    H = J a^\dagger a - J b^\dagger b + i \mu (a^\dagger b + a b^\dagger).
\end{equation}
In the $a$, $b$ space, we have a coupling matrix
\begin{equation}
    \begin{bmatrix}
        J & i \mu \\ -i \mu & - J
    \end{bmatrix},
\end{equation}
which has eigenvalues  
\begin{equation}
    \pm \sqrt{\mu^2 + J^2}.
\end{equation}
We see that $\mu$ opens a gap in the spectrum of a single pair of modes, i.e. there is no state at zero energy.

The corresponding eigenmodes are
\begin{equation}
   c_+ =  a \cos \theta  +b  \sin \theta 
\end{equation}
and
\begin{equation}
    c_- = - a \sin \theta  + b \cos \theta 
\end{equation}
where
\begin{equation}
    \tan \theta = \frac{\sqrt{J^2 + \mu^2} - J}{\mu}.
\end{equation}
The ground state is given by $c_-^\dagger$ acting on the state annhilated by $c_\pm$.

Returning to the case of a general coupling matrix $J_{ab}$ with eigenvalues $\epsilon_a$, the inclusion of $\mu$ changes the single particle spectrum to 
\begin{equation}
    \pm \sqrt{\epsilon_a^2 + \mu^2},
\end{equation}
which is gapped. The many-body ground state is obtained by filling all the negative energy states, and the resulting many-body gap is also proportional to $\mu$.

If we look at the occupations in the ground state from a one-sided perspective, any state with $|\epsilon_a| \gg \mu$ is either filled or empty with high probability depending on the $\pm$ sign. By contrast, states with $\epsilon_a \sim \mu$ are smeared out, being neither filled nor empty from the one-sided perspective. Now, this smearing is different in detail than the one provided by the Fermi-Dirac distibution. This means the ground state of $H$ is not literally the thermofield double. However, we can still make the one-sided energy density as low as desired by tuning $\mu$ to be small. This effectively puts more and more $\epsilon_a$ states into their one-sided ground state, thus smoothly lowering the one-sided energy.

With this concrete example in place, we now consider the SYK$_4$ Maldacena-Qi Hamiltonian. We establish the same two key properties, that the gap is maintained for any $\mu>0$ in the large $N$ limit, and that the one-sided energy density can be tuned as low as desired.

\subsubsection*{Numerical results}
\label{sec:nltsnumerics}

Here we numerically check the energy gap by solving the Schwinger-Dyson equations for the two-sided SYK model with Maldacena-Qi coupling, following \cite{maldacena2018eternal}. The effective $G$-$\Sigma$ action is
\begin{align}
    -S_E/N = &\log \mathrm{Pf} \left( \partial_{\tau} \delta_{ab} - \Sigma_{ab} \right) - \frac{1}{2} \int d \tau_1 d \tau_2 \sum_{ab} \left[ \Sigma_{ab}(\tau_1,\tau_2) G_{ab}(\tau_1,\tau_2) - \sum_q s_{ab} \frac{J_q^2}{2 q^2} \left[ 2 G_{ab}(\tau_1,\tau_2) \right]^q \right] \nonumber \\
    &+ \frac{i \mu}{2} \int d \tau_1 \left[ -G_{LR}(\tau_1,\tau_1) + G_{RL}(\tau_1,\tau_1) \right]
\end{align}
where $a,b = L,R$ label the two sides and $s_{LL} = s_{RR} = 1$, $s_{LR} = s_{RL} = (-1)^{q/2}$. We have included a sum over $q$ to allow for multiple coupling types, each of which is controlled by an independent coupling strength $J_q$. From this action, we may derive equations of motion for the fields $\Sigma_{ab},G_{ab}$. For a model with $q = 2$ and $q = 4$ interactions only, these take the form
\begin{alignat}{2}
    & G_{LL}(\omega_f) = \frac{- i \omega_f - \Sigma_{RR}(\omega_f)}{D(\omega_f)} \quad \quad && G_{LR}(\omega_f) = \frac{\Sigma_{LR}(\omega_f)}{D(\omega_f)} \nonumber \\
    & G_{RR}(\omega_f) = \frac{- i \omega_f - \Sigma_{LL}(\omega_f)}{D(\omega_f)} \quad \quad && G_{RL}(\omega_f) = \frac{\Sigma_{RL}(\omega_f)}{D(\omega_f)} \nonumber \\
    & D(\omega_f) := \left( - i \omega_f - \Sigma_{LL} \right) \left( - i \omega_f - \Sigma_{RR} \right) - \Sigma_{LR} \Sigma_{RL} \\
    & \nonumber \\
    & \Sigma_{LL}(\tau) = \frac{J_4^2}{4} \left[ 2 G_{LL}(\tau) \right]^3 + 2 J_2^2 G_{LL}(\tau) \quad \quad && \Sigma_{LR}(\tau) = \frac{J_4^2}{4} \left[ 2 G_{LR}(\tau) \right]^3 - 2 J_2^2 G_{LR}(\tau) - i \mu \delta (\tau) \nonumber \\
    & \Sigma_{RR}(\tau) = \frac{J_4^2}{4} \left[ 2 G_{RR}(\tau) \right]^3 + 2 J_2^2 G_{RR}(\tau) \quad \quad && \Sigma_{RL}(\tau) = \frac{J_4^2}{4} \left[ 2 G_{RL}(\tau) \right]^3 - 2 J_2^2 G_{RL}(\tau) + i \mu \delta (\tau) \nonumber
\end{alignat}

\begin{figure}
    \centering
    \includegraphics[width=0.8\textwidth]{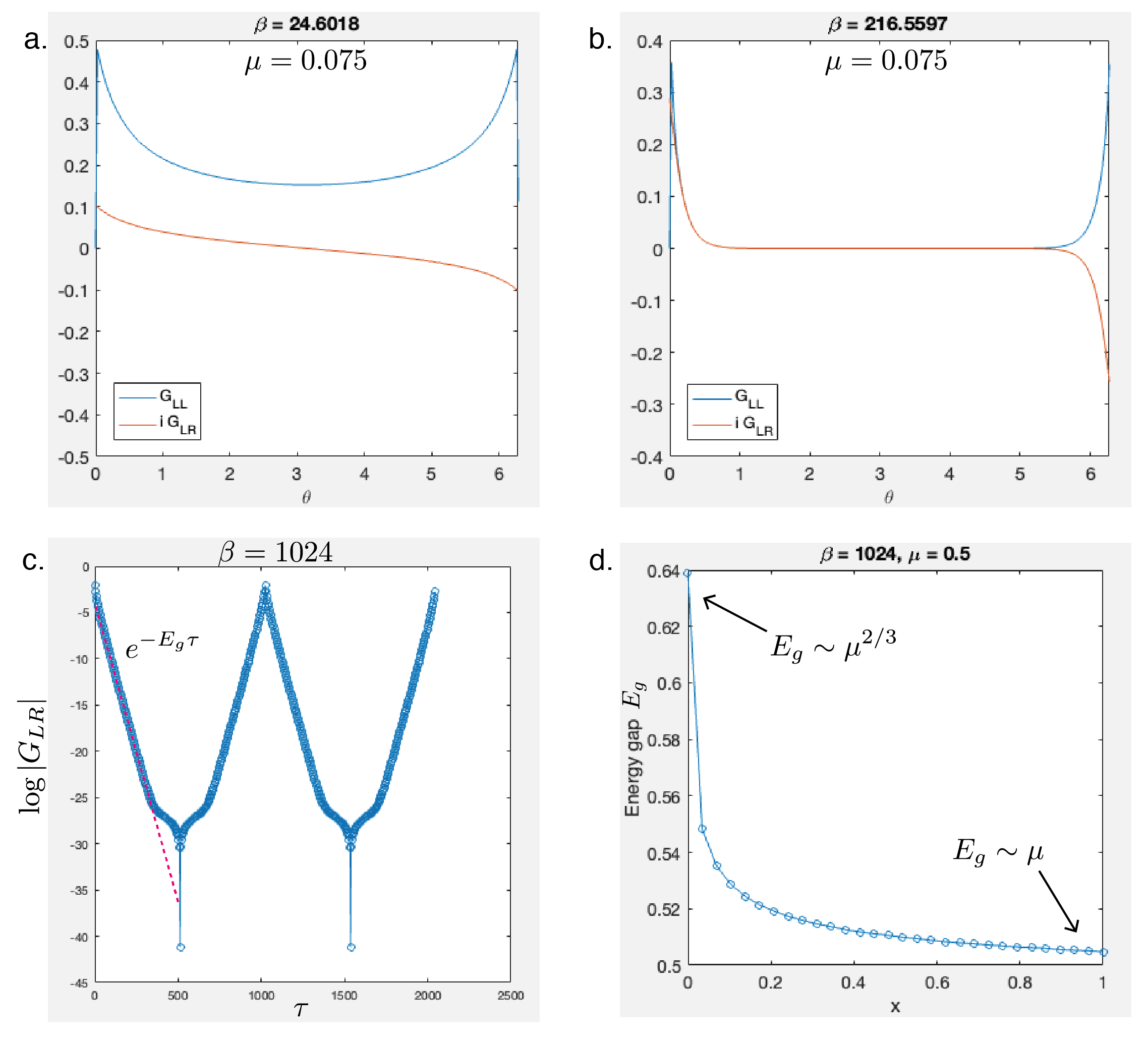}
    \caption{Numerical solutions of the imaginary-time Schwinger-Dyson equations for the two-sided SYK model with Maldacena-Qi coupling. Top row shows the absolute magnitude $\magn{G_{ab}(\theta)}$ of the Green's functions $G_{LL}$ (blue) and $G_{LR}$ (orange) versus $\theta = 2 \pi \tau / \beta$, calculated at $\mu = 0.075$ and $J \beta = 24.6018$ (a) and $J \beta = 216.5597$ (b). Bottom row shows energy gap $E_g$ extracted from Green's functions $G_{LR}$ by fitting a linear slope on a log scale (c). The resulting energy gap is plotted in (d) versus the tuning parameter $x$, where $J_2 = J \sqrt{x}$ and $J_4 = J\sqrt{1-x}$.}
    \label{fig:numericalsyktwosided}
\end{figure}

We obtain self-consistent solutions to these equations using the iterated weighted update procedure described in Appendix~\ref{app:numericalmethods}. The resulting solutions $G_{LL}(\tau)$ and $G_{LR}(\tau)$ are plotted in Fig. \ref{fig:numericalsyktwosided}. At low temperatures, these solutions decay exponentially
\begin{equation}
    \magn{G_{ab}} \approx e^{-E_g \tau}
\end{equation}
with the timescale set by the smallest energy gap $E_g$ in the system. We fit this exponential decay in $\magn{G_{LR}}$ to extract the energy gap $E_g$ for any specific model. In particular, in Fig. \ref{fig:numericalsyktwosided}(d) we consider tuning the model between a purely $q = 4$ model and a purely $q = 2$ model by choosing coupling coefficients $J_2 = J\sqrt{x}$ and $J_4 = J\sqrt{1-x}$ whose ratio is controlled by the tunable parameter $x \in [0,1]$. We find that the energy gap remains open at all points between these two extreme cases. Further, we find that the gap scales like $E_g \sim \mu^{2/3}$ for the pure $q = 4$ case and like $E_g \sim \mu$ for the pure $q = 2$, which agrees with the analysis of Ref. \cite{maldacena2018eternal} and the arguments presented above for $q = 2$.

\begin{figure}
    \centering
    \includegraphics[width=0.8\textwidth]{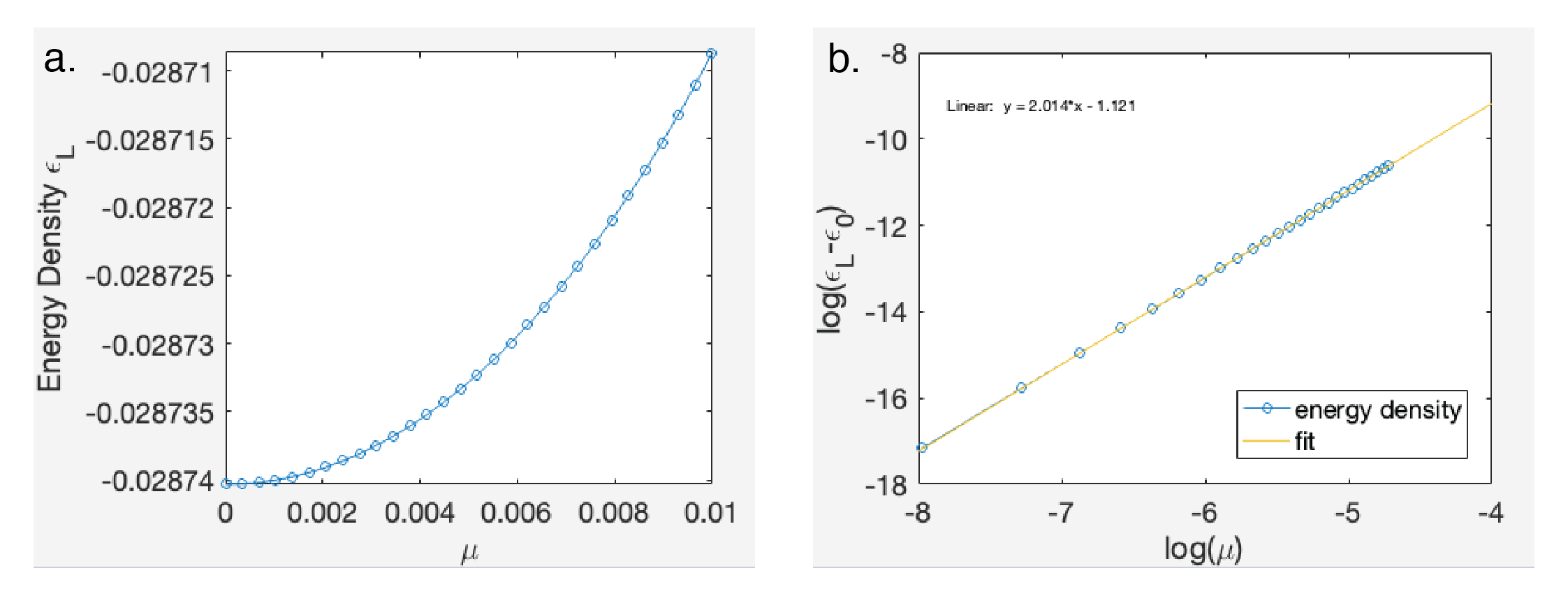}
    \caption{Energy density $\varepsilon_L = \langle H_L \rangle / N$ of the left side of the two-sided SYK model as a function of coupling $\mu$ (a). Plotting on a log-log scale reveals an almost perfectly quadratic dependence on $\mu$ (b).}
    \label{fig:energydensityL}
\end{figure}

Further, we can show that the one-sided energy smoothly approaches its decoupled $\mu = 0$ value as we tune $\mu$ to zero for the pure $q = 4$ model. From the equations of motion, one can readily derive the energy expectation value of the left side \cite{maldacena2018eternal}:
\begin{equation}
    \varepsilon_L = \langle H_L \rangle / N = - \frac{J_4^2}{4} \int_0^{\beta} d \tau G_{LL}(\tau)^4 - \frac{J_4^2}{4} \int_0^{\beta} d \tau G_{LR}(\tau)^4
\end{equation}
similar to Eq. \eqref{eq:sykenergydensity}. The resulting energy density can be calculated directly from the numerical solutions to the Schwinger-Dyson equations. We plot the result in Fig. \ref{fig:energydensityL}. Fitting on a log-log scale reveals a quadratic growth of the energy density $\varepsilon_L = \varepsilon_{L,0} + \alpha \mu^2$ with $\mu$.

\end{document}